%% file: Lecture_Notes_on_TFT.tex
\documentclass[10pt]{article}
\usepackage{fullpage,epsfig,graphics,amsbsy,amssymb,cancel,bbm}
\usepackage{psfrag,color}
\usepackage[breaklinks]{hyperref}
\usepackage{graphicx}
\usepackage{amsfonts}
\usepackage{amssymb,amsmath,amscd}
\usepackage{calligra}
\usepackage{tikz}
\usetikzlibrary{arrows,chains,matrix,positioning,scopes,snakes}

\makeatletter
\tikzset{join/.code=\tikzset{after node path={%
\ifx\tikzchainprevious\pgfutil@empty\else(\tikzchainprevious)%
edge[every join]#1(\tikzchaincurrent)\fi}}}
\makeatother

\tikzset{>=stealth',every on chain/.append style={join},
         every join/.style={->}}
\tikzset{
    >=stealth',
    punkt/.style={
           rectangle,
           rounded corners,
           draw=black, very thick,
           text width=6.5em,
           minimum height=2em,
           text centered},
    pil/.style={
           ->,
           thick,
           shorten <=2pt,
           shorten >=2pt,}
}

\newcommand{\BS}{\boldsymbol}
\newcommand{\BB}{\mathbb}
\newcommand{\SF}{\mathsf}
\newcommand{\FR}{\mathfrak}

\newcommand{\bea}{\begin{eqnarray}}
\newcommand{\eea}{\end{eqnarray}}
\newcommand{\nn}{\nonumber}
\newcommand{\Tr}{\textrm{Tr}}

\newcommand{\sbullet}{\textrm{\tiny{\textbullet}}}
\newcommand{\bra}{\langle}
\newcommand{\ket}{\rangle}

\def\ga{\alpha}
\def\gb{\beta}
\def\gc{\gamma}
\def\gd{\delta}

\DeclareMathAlphabet{\mathpzc}{OT1}{pzc}{m}{it}

\newenvironment{example}[1][Example]{\begin{trivlist}
\item[\hskip \labelsep {\bfseries #1}]}{\end{trivlist}}
\newenvironment{remark}[1][Remark]{\begin{trivlist}
\item[\hskip \labelsep {\bfseries #1}]}{\end{trivlist}}

\newcommand{\qed}{\nobreak \ifvmode \relax \else
      \ifdim\lastskip<1.5em \hskip-\lastskip
      \hskip1.5em plus0em minus0.5em \fi \nobreak
      \vrule height0.5em width0.5em depth0.00em\fi}

\begin{document}
\thispagestyle{empty}
\smallskip
\begin{center}
\Large{\bf Lecture Notes on Topological Field Theory,}\\
\large{\bf Perturbative and Non-perturbative Aspects}
  \\[12mm] \normalsize
{\bf Jian Qiu$^{\dagger}$} \\[8mm]
 {\small\it
 ${}^{\dagger}$I.N.F.N. Dipartimento di Fisica, Universit\`a di Firenze\\
     Via G. Sansone 1, 50019 Sesto Fiorentino - Firenze, Italia\\
 }
\end{center}
\vspace{7mm}

\begin{abstract}
\noindent These notes are based on the lecture the author gave at the workshop 'Geometry of Strings and Fields' held at Nordita, Stockholm. In these notes, I shall cover some topics in both the perturbative and non-perturbative aspects of the topological Chern-Simons theory. The non-perturbative part will mostly be about the quantization of Chern-Simons theory and the use of surgery for computation, while the non-perturbative part will include brief discussions about framings, eta invariants, APS-index  theorem, torsions and finite type knot invariants.
\end{abstract}

\smallskip

\begin{center}
\large\it To my mother
\end{center}

\tableofcontents

\section{Prefatory Remarks}
These notes are based on the lectures the author gave in November 2011, during the workshop 'Geometry of Strings and Fields' at Nordita institute, Stockholm. A couple months back when Maxim Zabzine, who is one of the organizers, asked me to give some lectures about topological field theory, I agreed without much thinking. But now that the lectures are over, I felt increasingly intimidated by the task: the topological field theory has evolved into a vast subject, physical as well as mathematical, and I can only aspire to cover during the lectures a mere drop in the bucket, and there will be nothing original in these notes either.

This note is divided into two parts: the non-perturbative and perturbative treatment of topological field theory (which term will denote the Chern-Simons theory exclusively in this note). I managed to cover most of the first part during the lectures, which included Atiyah's axioms of TFT and quantization of CS theory, as well as some of the most elementary 3-manifold geometry. Certainly, none of these materials are new, in fact, all were well established by the mid nineties of the previous century. Yet, I hope at least this note may provide the reader with a self-contained, albeit somewhat meagre overview; especially, the basics of surgery of 3-manifolds is useful for application in any TFT's.

The second part, the perturbative treatment of CS theory has to be dropped from the lectures due to time pressure. The perturbative treatment is, compared to the non-perturbative treatment, more intrinsically three dimensional, and therefore, by comparing computation from both sides one can get some interesting results.
I need to add here that for people who are fans of localization technique, he will find nothing interesting here. Still some of the topics involved in the second part, such as the eta invariant, the framing and torsion etc are interesting mathematical objects too, and it might just be useful to put up an overview of these materials together and provide the reader with enough clue to grasp the main features (which can be somewhat abstruse for the novices) of a 1-loop calculation and its relation to finite type invariants.

So enough with the excusatory remarks, the Chern-Simons (CS) theory is defined by the path integral
\bea \int DA~ \exp\Big(\frac {ik}{4\pi}\int_{M^3}~\Tr\big[ AdA+\frac23A^3 \big]\Big),~~k\in\BB{Z}\nn\eea
where $A$ is the gauge connection on an oriented closed 3-manifold $M$. The Lie algebra of the gauge group is assumed to be simply laced throughout.
The normalization is such that under a gauge transformation $A\to g^{-1}dg+g^{-1}Ag$, the action shifts by
\bea \int_{M^3}~\Tr\big[ AdA+\frac23A^3 \big]\to \int_{M^3}~\Tr\big[ AdA+\frac23A^3 \big]+\frac13\int_{M^3}~\Tr[(g^{-1}dg)^3.\nn\eea
Together with the factor $1/4\pi$, the second term gives the $2\pi$ times the winding number of the map $g$ and hence drops from the exponential. In particular, if the gauge group is $SU(2)$, then (where the trace is taken in the 2-dimensional representation of $SU(2)$)
\bea\textrm{vol}_{SU(2)}=\frac{1}{24\pi^2}\Tr\big[(g^{-1}dg)^3\big]\nn\eea
is the volume form of $S^3$, normalized so that $S^3$ has volume 1.

The non-perturbative part of the lecture will involve the following topics
1. Atiyah's axiomitiztion of TFT, which provides the theoretical basis of cutting and gluing method. 2.
basics of surgery. 3. quantization of CS theory. 4. some simple sample calculations.

The perturbative part of the lecture will cover
1. The eta invariant, needed to get the phase of the 1-loop determinant. 2. 2-framing of 3-manifolds, which is needed to relate the eta invariant to the gravitational CS term that is inserted into the action as a local counter term. 3. the analytic torsion, which is the absolute value of the 1-loop determinant. 4. the Alexander polynomial, which is related to the torsion of the complement of a knot inside a 3-manifold.

\vspace{.5cm}

\noindent{\it Acknowledgements}:
I would like to thank the organizers of the workshop 'Geometry of Strings and Fields': Ulf Lindstr\"om and Maxim Zabzine for their kind invitation and hospitality during my sojourn in Stockholm, and especially the latter for encouraging me to finish the lecture notes and publish them. I would also like to thank Anton Alekseev, Francesco Bonechi, Reimundo Heluani, Vasily Pestun and Alessandro Tomasiello, whose inquisitiveness helped me to sharpen my understanding of some of the problems. And finally, a cordial thanks to everyone who strained to keep himself awake during my rambling lectures.

\section{Non-perturbative Part}

\subsection{Atiyah's Axioms of TFT}\label{AAoT}
\input{AtiyahAxiom}

\subsection{Cutting, Gluing, Fun with Surgery}\label{sec_CGFwS}
\input{CuttingGluingFunwithSurgery}

\subsection{Quantization of CS-part 1}\label{sec_QoC1}
\input{Quantization_of_CS_part1}

\subsection{Quantization of CS-part 2-the Torus}\label{sec_QoC2}
\input{Quantization_of_CS_torus}

\subsection[Some Simple Calculations]{Some Simple Calculations (Partition Function and the Jones Polynomial)}\label{sec_SCatJP}
\input{SimpleComputation}

\section{Perturbative Part}
In the perturbative computation of the Chern-Simons theory, I shall outline the calculation of the 1-loop determinant and also some Feyman diagrams. The Feynmann diagrams are actually interesting independent of the Chern-Simons theory: these are the finite type invariants of 3-manifolds or knots. I give a simple example to motivate this.
\subsection{Motivation: Finite type Invariants}
\input{Perturb_Motivation}

\subsection{Relating Chord Diagrams to Knot Polynomials}\label{RCDtKP}
\input{CD_Alexander}

\subsection[The Torsion]{The Torsion, Whitehead, Reidemeister-de Rham-Franz and Ray-Singer}\label{TTWRdRFRS}
\input{theTorsion}

\subsection{The eta Invariant}\label{TeI}
One would like to write the eta invariant $\eta_a$ in terms of something more familiar.
Let $a$ be a flat connection over a 3-manifold $Y$, $\theta$ be the trivial connection on $Y$ and $\tt d$ be the dimension of the bundle (which is $|G|$ for our problem, since all fields transform in the adjoint representation). It is shown that the combination $\eta_{a}-\eta_{\theta}$ is independent of the metric (Thm 2.2 \cite{AtiyahPatodiSingerII}), thus the metric dependence of the phase Eq.\ref{phase_det} is encapsulated in $\eta_{\theta}={\tt d}\eta_g$, where $\eta_g$ is purely 'gravitational'. One can obtain more precise information of the difference $\eta_{a}-\eta_{\theta}$ by invoking the Atiyah-Potodi-Singer index theorem.

The eta invariant was introduced in the paper \cite{AtiyahPatodiSingerI} in order to obtain an index theorem for a manifold with boundary. Let $X$ has boundary $Y$, and $D$ be a linear first order differential operator acting from a vector bundle $E$ to another vector bundle $F$ over $X$. And $D$ is assumed to be of the form $D=\partial_u+{\cal A}$ close to the boundary, where $u$ is the inward pointing normal to the boundary and ${\cal A}:\,E|Y\to E|Y$ is elliptic self-adjoint. The index formula takes the following form
\bea \textrm{index}\,D=\int_X~\alpha_0 dx-\frac{h+\eta_{\cal A}}{2}.\label{APS_I}\eea
As there is a boundary, the index must be defined under proper boundary conditions. The major realization of ref.\cite{AtiyahPatodiSingerI} is that the boundary condition should be a \emph{non-local} one, more concretely, close to the boundary, any solution can be expanded in terms of eigen-modes of the operator ${\cal A}$, and the boundary condition sets to zero all modes with eigenvalues greater than or equal to zero. It is essentially this different treatment of plus or negative eigenvalues of ${\cal A}$ that led to the factor $\textrm{sgn}(\lambda)$ in Eq.\ref{eta_fun}. The symbol $h$ above is the multiplicity of zero eigenvalues of ${\cal A}$ on $Y$ and finally $\alpha_0$ is some characteristic polynomial obtained from the constant term in the asymptotic expansion of the heat kernel $e^{-tD^{\dagger}D}-e^{-tDD^{\dagger}}$. For example, if the operator $D$ is $d+d^{\dagger}$, then $\alpha_0=L(p)$ is the Hirzebruch $L$-polynomial, $\eta_{\cal A}$ is the purely gravitational $\eta_g$ above and the index is the signature of the manifold.

For our application, we need to put a (not necessarily flat) bundle $E$ over $X$, and twist the operator by a connection $a$ of the bundle $D_a=d_a+d_a^{\dagger}$, the corresponding ${\cal A}$ equals $L$ defined in Eq.\ref{def_L}. The characteristic polynomial $\alpha_0$ is calculated in ref.\cite{AtiyahBottPatodiHeat} to be
\bea\ga_0=2^{\dim X/2}\textrm{ch}(E)\prod_i\frac{x_i/2}{\tanh{x_i/2}},\nn\eea
At dimension 4, one has
\bea \ga_0=-\pi^{-2}\Tr[F^2]+\frac23 p_1.\label{char_poly}\eea
The calculation of this using the susy QM method is given in the appendix.

The discussion below is from ref.\cite{FreedGompf} (Eq 1.29). Let $X$ be the product $Y\times I$ and the bundle $E$ is induced from the flat bundle over $Y$ and the symbols $\theta$, $a$ and ${\tt d}$ are as in the first paragraph.
One puts the connection $a$ and the trivial connection on the left and right end ($u=0,~1$) of $Y\times I$, with an arbitrary interpolation in between. Of course, the interpolated connection in the bulk is not flat, one can nonetheless apply the index theorem \ref{APS_I}. On the lhs of Eq.\ref{APS_I}, one gets the index of $D$ which is twice the index $I_a$ in Eq.\ref{general_feature}. On the rhs, now that the boundary has two components, we have
$(b^0_Y+\cdots b_Y^3-2{\tt d}\eta_g)/2=(b^0_Y+b_Y^1)-{\tt d}\eta_g$ on the right end and $(\dim H_a^0+\cdots \dim H_a^3+2\eta_{a})/2=(\dim H_a^0+\dim H_a^1)+\eta_a$ on the left. The factor of 2 in front of $\eta$ is because $L_-$, whose $\eta$ we want, is the restriction of $L$ to the odd forms. As for the characteristic polynomial Eq.\ref{char_poly}, the integral of the Pontryagin class gives zero. The curvature squared term can be integrated using the Stokes theorem
\bea \int_{Y\times I}\Tr_{ad}[F^2]=-\int_Y\Tr_{ad}\big(ada+\frac23aaa\big).\nn\eea
The trace taken in the adjoint can be rewritten as a trace over the fundamental, but multiplied by a factor $h$. This is the source of the shift $k\to k+h$ in Eq.\ref{general_feature}. Putting everything together
\bea \frac{i\pi}{4}\eta_a=\frac{ih}{4\pi}CS(a)-\frac{i\pi}{4}\Big((b^0_Y+b_Y^1)+(\dim H_a^0+\dim H_a^1)+2I_a-|G|\eta_g\Big).\nn\eea
This explains $N_{ph}$ in Eq.\ref{general_feature} up to $\eta_g$. In fact this seemingly non-local term can be countered by a gravitational Chern-Simons term $CS_{\ga}(g)$, the subscript $\ga$ is a reminder that this term is defined up to an integer shift and that by fixing a 2-framing $\ga$ of $Y$, one can fix this ambiguity. Atiyah \cite{AtiyahFraming} showed that there exists a canonical framing such that $CS_{\ga}(g)+3\eta_g=0$, this will be reviewed in sec.\ref{2Fo3M}.

\subsection{2-Framing of 3-Manifolds}\label{2Fo3M}
\input{two_framing}

\appendix
\input{Lecture_Notes_Appendix}



\providecommand{\href}[2]{#2}\begingroup\raggedright\endgroup

\end{document}

%% file: AtiyahAxiom.tex
Our goal is to compute the path integral of the Chern-Simons functional over a 3-manifold, and the major tool we shall emply is the surgery, the theoretical basis of which is Atiyah's axiomatization of TFT.

First, let us review a simple quantum mechanics fact. Consider a particle at position $q_0$ at time 0 and position $q_1$ at time $T$, the QM amplitude for this process is
\bea \bra q_1| e^{-iH(q,p)T}|q_0\ket,\nn\eea
where the Hamiltonian is a function of both momentum and coordinate. Then Feynman taught us that this amplitude can also be written as an integration over all the paths starting from $q_0$ and ending at $q_1$ weighted by the action associated with the path
\bea \bra q_1| e^{-iH(q,p)T}|q_0\ket=\int Dq(t)~e^{i\int_0^T(\dot qp-H)}.\label{Feynman_taught_us}\eea
In general, one may be interested in the transition amplitude from an in state $\psi_0$ to an out state $\psi_1$, this amplitude is written as a convolution
\bea \bra \psi_1| e^{-iH(q,p)T}|\psi_0\ket=\int dq_0\int dq_1 \bra\psi_1|q_1\ket \bra q_0|\psi_0\ket~\int Dq(t)~e^{i\int_0^T(\dot qp-H)},\nn\eea
where we simply expand the in and out states into the coordinate basis and the quantities
\bea \psi(q)\stackrel{def}{=}\bra\psi|q\ket\nn\eea
are called the \emph{wave function}s.

Since QM is just a 0+1 dimensional QFT, our 'source manifold' in this case is a segment $M=*\times[0,T]$; if one thinks of this problem somewhat 'categorically', one may say that the QFT assigns to the boundary of the source manifold (which is two disjoint points) the quantum Hilbert space of the system. And to the segment that connects the two boundary components, is assigned a mapping between the two Hilbert spaces, namely the evolution operator $e^{-iHT}$.
The same story can be easily extended to the higher dimensional case: if the source manifold is $\Sigma\times I$, then this time a QFT assigns to the boundary $\Sigma$ a different Hilbert space ${\cal H}_{\Sigma}$ for different $\Sigma$'s.

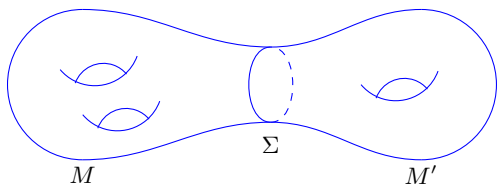
\begin{figure}[h]
\begin{center}
\begin{tikzpicture}
\draw [-,blue] (-2.5cm,-1cm) to [out=0, in=180] (0cm,-0.5cm);
\draw [-,blue] (2cm,-1cm) to [out=180, in=0] (0cm,-0.5cm);
\draw [-,blue] (-2.5cm,1cm) to [out=0, in=180] (0cm,0.5cm);
\draw [-,blue] (2cm,1cm) to [out=180, in=0] (0cm,0.5cm);

\draw [-,blue] (2cm,-1cm) arc (-90:90:1cm);
\draw [-,blue] (-2.5cm,1cm) arc (90:270:1cm);

\node at (2cm,-1.2cm) {\small $M'$};
\node at (-2.5cm,-1.2cm) {\small $M$};
\node at (0cm,-.8cm) {\small $\Sigma$};

\draw [-,blue] (1.2cm,0cm) arc (220:340:.6cm);
\draw [-,blue] (1.4cm,-.18cm) arc (160:40:.4cm);

\draw [-,blue] (-2.5cm,-.4cm) arc (220:340:.6cm);
\draw [-,blue] (-2.3cm,-.58cm) arc (160:40:.4cm);

\draw [-,blue] (-2.8cm,0.2cm) arc (220:340:.6cm);
\draw [-,blue] (-2.6cm,.02cm) arc (160:40:.4cm);

\draw [dashed,blue] (0cm,-.5cm) to [out=0, in=0] (0cm, .5cm);
\draw [-,blue] (0cm,-.5cm) to [out=180, in=180] (0cm, .5cm);
\end{tikzpicture}\caption{Gluing of two manifolds along their common boundary ${\cal M}=M\cup_{\Sigma}M'$}\label{glue_along_bound_fig}
\end{center}
\end{figure}

One can now consider topological field theories, which have the virtue that the Hamiltonian is zero (or BRST-exact, in general), one can formalize what was said above and obtain Atiyah's axioms of TFT \cite{AtiyahTFT}. In the third row of tab.\ref{tab_Atiyah_ax}, P.I. means path integral and B.C. means boundary condition, and $\phi$ is a generic field in the theory and $\phi|_{\Sigma}$ denotes the boundary value of $\phi$, and the wave function $\psi_M$ is a functional of $\phi_{\Sigma}$. In fact, the rhs of the third row is completely analogous to the expansion of a state into a coordinate basis in the QM problem above, and it is just a reformulation of the second row.
\begin{table}[h]
\begin{center}
\begin{tabular}{|c|c|}
  \hline
  2D Riemann surface $\Sigma$ & Hilbert space ${\cal H}_{\Sigma}$ \\
  3D Mfld $M$, s.t. $\partial M=\Sigma$ & a vector or state $\psi_M\in{\cal H}_{\Sigma}$ \\
  P.I. on $M$ with fixed B.C. on $\Sigma$ & wave function~$\psi_M(\phi|_{\Sigma})=\bra \psi_{M}|\phi|_{\Sigma}\ket$  \\
  reversing the orientation on $\Sigma\hookrightarrow M$ & hermitian conjugation \\
  bordism $M$,~$\partial M=\Sigma\sqcup\Sigma'$ & morphism ${\cal H}_{\Sigma}\to {\cal H}_{\Sigma'}$\\
  trivial bordism $M=\Sigma\times I$ & Id \\
  gluing along common boundary $(-M)\cup_{\Sigma}M'$ & inner product $\bra \psi_M|\psi_{M'}\ket$\\
  disjoint union $\Sigma\sqcup \Sigma'$ & tensor product ${\cal H}_{\Sigma}\otimes {\cal H}_{\Sigma'}$\\
  $M$ with $\partial M=\emptyset$ & $Z_M$\\
  \hline
\end{tabular}
\end{center}
\caption{Atiyah's Axiom for 3D TFT, the story can also be extended to $\Sigma$ with punctures}\label{tab_Atiyah_ax}
\end{table}
As for the seventh row, one may refer to fig.\ref{glue_along_bound_fig}.

The implication of such a reformulation is far-fetched. If one is interested in computing the partition function of a TFT on a complicated 3-manifold $M$, then in principle, one can try to break down $M$ into simpler parts $M=(-M_1)\cup_{\Sigma}M_2$, then so long as one possesses sufficient knowledge of $\psi_{M_{1,2}}\in{\cal H}_{\Sigma}$, the partition function is just
\bea Z_M=\bra \psi_{M_1}|\psi_{M_2}\ket.\label{recipe_atiyah}\eea
To put this simple principle into practice, one needs to understand ${\cal H}_{\Sigma}$ which is the subject of sec.\ref{sec_QoC1} and \ref{sec_QoC2}. Before that, we also need to know how to break down a 3-manifold into simpler bits.

%% file: CuttingGluingFunwithSurgery.tex
First, let me give several facts about 3-manifolds (\emph{oriented and closed}),
\begin{itemize}
\item any oriented closed 3-manifold is parallelizable
\item they possess the so called Heegard splitting
\item they can be obtained from performing surgery along some links inside of $S^3$
\item they are null-bordant
\end{itemize}

\vskip 1cm

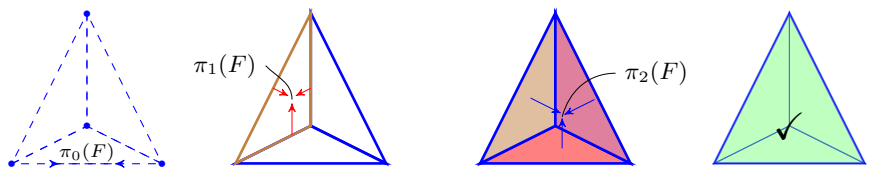
\begin{figure}[h]
\begin{center}
\begin{tikzpicture}
\path[draw, dashed, blue] (1,1) -- (2,1.5)--(3,1)--cycle;
\path[draw, dashed, blue] (1,1) -- (2,1.5)--(2,3)--cycle;
\path[draw, dashed, blue] (2,3) -- (2,1.5)--(3,1)--cycle;
\fill[blue] (1,1) node {$\sbullet$}
 (3,1) node {$\sbullet$}
  (2,3) node {$\sbullet$}
   (2,1.5) node {$\sbullet$};
\draw [>-<, dashed, blue] (1.5,1) to (2.5,1);
\node at (2,.9) [above] {\scriptsize $\pi_0(F)$};
\end{tikzpicture}
\begin{tikzpicture}
\path[draw, blue, line width =1pt] (1,1) -- (2,1.5)--(3,1)--cycle;
\path[draw, blue, line width =1pt] (2,3) -- (2,1.5)--(3,1)--cycle;
\path[draw, brown, line width =1pt] (1,1) -- (2,1.5)--(2,3)--cycle;
\draw [->, red, line width =.3pt] (1.5,2) -- (1.7,1.9);
\draw [->, red, line width =.3pt] (2,2) -- (1.8,1.9);
\draw [->, red, line width =.3pt] (1.75,1.375) -- (1.75,1.8);
\draw [-,black,line width =.3pt] (1.75,1.85) to [out=90, in=-45] (1.4,2.3) node [left] {\small $\pi_1(F)$};
\node at (1,.95) {};
\end{tikzpicture}\hspace{1cm}
\begin{tikzpicture}
\path[draw, blue, line width =1pt, fill=red!50] (1,1) -- (2,1.5)--(3,1)--cycle;
\path[draw, blue, line width =1pt, fill=brown!50] (1,1) -- (2,1.5)--(2,3)--cycle;
\path[draw, blue, line width =1pt, fill=purple!50] (2,3) -- (2,1.5)--(3,1)--cycle;
\draw [->, blue, line width =.3pt] (1.67,1.87) -- (1.67+.4,1.87-.2);
\draw [->, blue, line width =.3pt] (1.68+.41,1.2) -- (1.68+.41,1.6);
\draw [<-, blue, line width =.3pt] (1.71+.41,1.65) -- (1.71+.41+.4,1.65+.2);
\draw [-,black,line width =.3pt] (1.68+.41,1.62) to [out=75, in=180] (2.8,2.2) node [right] {\small $\pi_2(F)$};
\node at (1,.95) {};
\end{tikzpicture}
\begin{tikzpicture}
\path[draw, blue] (1,1) -- (2,1.5)--(3,1)--cycle;
\path[draw, blue] (1,1) -- (2,1.5)--(2,3)--cycle;
\path[draw, blue] (2,3) -- (2,1.5)--(3,1)--cycle;
\path[draw, blue, line width =1pt, fill=green!50, opacity=.5] (1,1) -- (3,1)--(2,3)--cycle;
\node at (2,1.5) {\Large $\checkmark$};
\node at (1,.95) {};
\end{tikzpicture}\caption{Successive extension of sections. The sections are defined on 0-cells, 1-cells, 2-cells and finally 3-cells in the four pictures}\label{obstruct_thy_fig}
\end{center}
\end{figure}

\noindent$\sbullet$\emph{Triviality of $TM$}\\
The key to the first fact is that for orientable 3-manifolds, $w_1=w_2=w_3=0$, for a quick proof using Wu's formula of Stiefel-Witney classes see the blog \cite{FowlerBlog}. That this it is already sufficient to prove the triviality of the tangent bundle is unclear to me, so I will append to it some arguments using some elementary obstruction theory. The class $w_2$ obstructs the lift of the frame bundle $B$ of $M$ from an $SO(3)$ bundle to a $spin(3)=SU(2)$ bundle, now that this classes vanishes, one may assume that the structure group of $B$ is $SU(2)=S^3$ which is 2-connected. One can now try to construct a global section to the frame bundle. One first divides $M$ into a cell-complex, then one defines the section arbitrarily on the 0-cells. This section is extendable to the 1-cells from the connectivity of $SU(2)$, see the first picture of fig.\ref{obstruct_thy_fig}. Whether or not this section is further extendable to the 2-cells and subsequently to 3-cells depends on the obstruction classes
\bea H^2(M;\pi_1(F)),~~~~H^3(M;\pi_2(F)),~~~~~~F=\textrm{fibre}=SU(2).\label{obstruct_class}\eea
Fig.\ref{obstruct_thy_fig} should be quite self-explanatory, the second picture shows the potential difficulty in extending a section defined on 1-cells into the 2-cells: if the the section painted as brown gives a non-trivial element in $\pi_1(F)$ then the extension is impossible. Likewise, the extension in the third picture is impossible if the sections defined on the boundary of the tetrahedra gives a non-trivial element of $\pi_2(F)$. This is the origin of the two obstruction classes in Eq.\ref{obstruct_class}. One may consult the excellent book by Steenrod \cite{Steenrod} part 3 for more details. Since $\pi_{1,2}(SU(2))=0$, a global section exists, leading to the triviality of the frame bundle.

For a more geometrical proof of the parallelizability of 3-manifolds, see also \cite{springerlink:10.1007/BFb0089039} ch.VII.

\vskip1cm
\noindent$\sbullet$ \emph{Heegard splitting}\\
As for the Heegard splitting, it means that a 3-manifold can be presented as gluing two handle-bodies along their common boundaries. A handle body, as its name suggests, is obtained by adding handles to $S^2$ and then fill up the hollow inside. For a constructive proof of this fact, see ref.\cite{Lickorish} ch.12. (indeed, the entire chapter 12 provides most of the references of this section).

\begin{figure}[h]
\begin{center}
\begin{tikzpicture}[scale=.8]
\draw [-,blue](-2,0cm) to [out=-90,in=-90] (2,0cm);
\draw [-,blue](-2,0cm) to [out=90,in=90] (2,0cm);

\draw [-,blue](-1.1,.24cm) to [out=-120,in=-60] (1.1,.24cm);
\draw [-,blue](-1,-.10cm) to [out=80,in=100] (1,-.10cm);

\draw [->,blue](0,1.16cm) to [out=180,in=120] (-.2,.6cm);
\draw [-,blue](-0.2,0.6cm) to [out=-60,in=180] (0,.48cm);
\draw [dotted,blue](0,1.16cm) to [out=-20,in=20] (0,.48cm);

\draw [-,blue](-1.5,0cm) to [out=-80,in=178] (0,-.73cm);
\draw [-,blue](1.5,0cm) to [out=-100,in=2] (0,-.73cm);
\draw [<-,blue](-1.5,0cm) to [out=90,in=90] (1.5,0cm);

\node (b) at (0,1.15cm) [above] {\scriptsize $a$};
\node (a) at (1.5,0cm) [right] {\scriptsize $b$};
\end{tikzpicture}\caption{The smaller circle is the contractible $a$-cycle, and the meridian is the $b$ cycle}\label{fig_ab_cycle}
\end{center}
\end{figure}
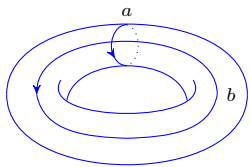
\begin{figure}[h]
\begin{center}
\begin{tikzpicture}[scale=.6]
\draw [rotate = 0, fill=blue!80, opacity=.5](0,0) ellipse (1 and 0.6)
      (0,2.5) ellipse (1 and 0.6)
      (3,0) ellipse (1 and 0.6)
      (3,2.5) ellipse (1 and 0.6);
\node at (0,.6) [above] {\small$a$};
\draw [-,red] (-1,0) -- (-1,2.5)
             (1,0) -- (1,2.5)
             (2,0) -- (2,2.5)
             (4,0) -- (4,2.5);
\node at (-1,1.25) [left] {\small$b$};
\draw [<->,black] (1,1.2) to [out=30, in=210] (2,1.2);
\draw [<->,black] (.4,2) to [out=30, in=150] (2.6,2);
\end{tikzpicture}
\begin{tikzpicture}[scale=.8]
\draw [blue] (0,0) circle (1);
\draw [dashed,blue] (-1,0) to [out=90, in=90] (1,0);
\draw [-,blue] (-1,0) to [out=-90, in=-90] (1,0);
\node at (-1.5,0) {$\Longrightarrow$};
\node at (1.2,0) {$\times$};
\draw [red] (2,0) circle (.6);
\end{tikzpicture}\caption{Gluing two solid tori, $a$ cycle to $a$-cycle and $b$-cycle to $b$-cycle.}\label{fig_glue_id}
\end{center}
\end{figure}
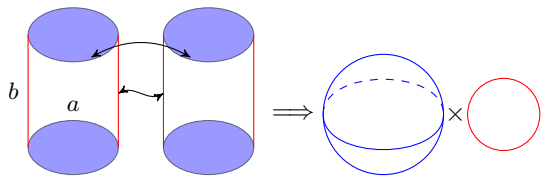
\begin{figure}
\begin{center}
\begin{tikzpicture}[scale=.6]
\draw [rotate = 0,fill=blue!80,opacity=.5](0,0) ellipse (1 and 0.6)
      (0,2.5) ellipse (1 and 0.6);
\draw [rotate = 0,fill=red!100,opacity=.5](3,0) ellipse (1 and 0.6)
      (3,2.5) ellipse (1 and 0.6);
\node at (0,.6) [above] {\small$a$};
\draw [-,red] (-1,0) -- (-1,2.5)
             (1,0) -- (1,2.5);
\draw [-,blue] (2,0) -- (2,2.5)
             (4,0) -- (4,2.5);
\node at (-1,1.25) [left] {\small$b$};
\draw [<->,black] (1,0.8) to [out=0, in=120] (2.6,-0.4);
\draw [<->,black] (.4,1.9) to [out=-30, in=180] (2,1.5);
\end{tikzpicture}
\begin{tikzpicture}[scale=.6]
\draw [blue, fill=blue!60,opacity=.4] (0,0) circle (1.4);
\draw [dashed,blue] (-1.4,0) to [out=90, in=90] (1.4,0);
\draw [-,blue] (-1.4,0) to [out=-90, in=-90] (1.4,0);
\draw [line width=6pt, brown, rotate = 15](0,0) ellipse (.8 and 0.3);
\node at (-2,0) {$\Longrightarrow$};
\node at (1,1) [right] {\small $\cup\{\infty\}$};
\end{tikzpicture}
\caption{Gluing two solid tori after an $S$-dual. We envision the solid blue ball union $\{\infty\}$ as $S^3$, in which there is a brown solid torus embedded.}\label{fig_glue_S}
\end{center}
\end{figure}
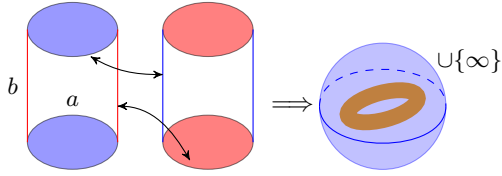

In particular, $S^3$ can be presented as gluing two solid 2-tori through an $S$-dual, or in general two Rieman surfaces of genus $n$ ($n$-arbitrary). To see this, we observe first that if one glues two solid tori as in fig.\ref{fig_glue_id}, one obtains $S^2\times S^1$.
In contrast, if one performs an $S$-dual (exchanging $a$-cycle and $b$-cycle, one obtains fig.\ref{fig_glue_S}, in which one needs to envision the complement of the solid torus (brown) also as a solid torus.
If this is hard for the reader to imagine, there will be an example later in this section that will explain the gluing in a different manner.
\begin{remark}\label{inf_Heegard_gen}
In fact, $S^3$ can be obtained as gluing two handle body of arbitrarily high genus, see the figure 12.5 in ref.\cite{Lickorish}.
\end{remark}
\subsection{Surgery along links}
Let me sketch the idea behind the fact that any 3-manifold can be obtained from $S^3$ after a surgery along a link. One first Heegard splits $M$ into 2 handle bodies $M=M_1\cup M_2$, and we denote $\Sigma=\partial M_{1,2}$. We know from the remark above that one can find a diffeomorphism $h$ on $\Sigma$ such that $S^3=M_1\cup_h M_2$. If the diffeomorphism were extendable to the interior of, say, $M_1$, then one has found a homeomorphism mapping $M$ to $S^3$. Of course, this diffeomorphism is not always extendable, but it can be shown (and not hard to imagine either) that if one remove enough solid tori from $M_1$, then the extension is possible. Consequently, what one obtains after the gluing is $S^3$ minus the tori that are removed. And it is along these solid tori, that one needs to perform surgery to pass from $M$ to $S^3$. For a complete proof, see thm.12.13 of
ref.\cite{Lickorish}.

From this fact, we see that the most important surgeries are those along some links or knots, this is why one can also use knot invariant to construct 3-manifold invariants. But, not all the different surgeries along all the different links will produce different 3-manifolds. There are certain relations among the surgeries, which are called \emph{moves}. The Kirby moves are the most prominent\footnote{Depending on how surgeries are presented (generated), there will be different sets of moves, and hence the complete set of moves go by different names.}; Turaev and Reshetikhin \cite{ReshetikhinTuraevI,ReshetikhinTuraevII} proved that the 3-manifold invariants arising from Chern-Simons theory \cite{Witten_Jones} respects the Kirby moves, and consequently placing these invariants, previously defined through path integral, in a mathematically rigorous framework.

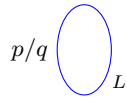
\begin{figure}[h]
\begin{center}
\begin{tikzpicture}[scale=.6]
\draw [rotate = 0](0,0) ellipse (.6 and 1)[blue];
\node at (-.6,0) [left] {\small $p/q$};
\node at (0.4,-0.7) [right] {\scriptsize$L$};
\end{tikzpicture}\caption{A $p/q$ surgery along a link}\label{fig_ration_surgery}
\end{center}
\end{figure}

\begin{figure}[h]
\begin{center}
\begin{tikzpicture}[scale=.8]
\draw [rotate = 0](0,0) ellipse (1 and .6)[blue];
\node at (-1,0) [left] {\small $2/1$};
\node at (0.7,-0.4) [right] {\scriptsize$L$};
\node at (1,0) [right] {\small$:$};
\node at (0,-1) {};
\end{tikzpicture}~~~
\begin{tikzpicture}[scale=.6]
\draw [-,blue](-2,0cm) to [out=-90,in=-90] (2,0cm);
\draw [-,blue](-2,0cm) to [out=90,in=90] (2,0cm);

\draw [-,blue](-1.1,.24cm) to [out=-120,in=-60] (1.1,.24cm);
\draw [-,blue](-1,-.10cm) to [out=80,in=100] (1,-.10cm);

\draw [-,blue](-1.5,0cm) to [out=-80,in=155] (-0.6,-1.13cm);
\draw [dotted,blue](-.6,-1.13cm) to [out=15,in=-125] (-0.2,-0.31cm);
\draw [blue](-0.2,-0.31cm) to [out=-15,in=165] (0.2,-1.13cm);
\draw [dotted,blue](0.2,-1.13cm) to [out=15,in=-125] (0.6,-0.31cm);
\draw [-,blue](1.5,0cm) to [out=-100,in=-30] (0.6,-.28cm);
\draw [<-,blue](-1.5,0cm) to [out=90,in=90] (1.5,0cm);
\node at (2,0) [right] {\small$\Rightarrow$};
\end{tikzpicture}
\begin{tikzpicture}[scale=.6]
\draw [-,blue](-2,0cm) to [out=-90,in=-90] (2,0cm);
\draw [-,blue](-2,0cm) to [out=90,in=90] (2,0cm);

\draw [-,blue](-1.1,.24cm) to [out=-120,in=-60] (1.1,.24cm);
\draw [-,blue](-1,-.10cm) to [out=80,in=100] (1,-.10cm);

\draw [dotted,blue](0,-1.13cm) to [out=145,in=-145] (0,-0.31cm);
\draw [-,blue](0,-1.13cm) to [out=25,in=-25] (0,-0.31cm);
\end{tikzpicture}
\caption{A $2/1$ surgery along a link}\label{fig_21_surgery}
\end{center}
\end{figure}
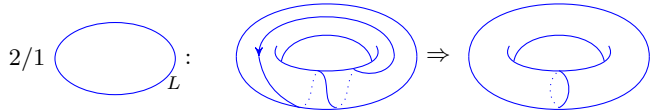

To introduce some of these moves, we first fix a notation, by placing a rational number along a link (fig.\ref{fig_ration_surgery}),
we understand that the tubular neighborhood of $L$ is removed, and reglued in after a diffeomorphism. To describe this diffeomorphsim, we note that before the surgery, $a$ was the contractible cycle, and the diffeomorphism is such that the $pa+qb$ cycle becomes the contractible cycle. For example, the $1/0$ surgery is the trivial surgery, since $a$ was the contractible cycle initially. Fig.\ref{fig_21_surgery} is a 2/1 surgery, so the cycle drawn in the middle panel will become the one on the right after the surgery.

\begin{remark}
The mapping class group of the torus is the $SL(2,\BB{Z})$ matrices
\bea M=\begin{array}{|cc|} p & r \\ q & s\end{array},~~~ps-qr=1.\nn\eea
The $p,~q$ entry are the two numbers that label the rational surgery of the previous paragraph, which determines the diffeomorphism type of the manifold after the surgery. Different choices of $r$ and $s$ will, however, affect the \emph{framing} of the resulting manifold, for example, the matrices
\bea T^p=\begin{array}{|cc|} 1 & p \\ 0 & 1\end{array}\nn\eea
do not change the diffeomorphism type. To see this, one can still refer to the middle panel of fig.\ref{fig_21_surgery} for the case $p=2$. But this time, the cycle drawn in the picture \emph{becomes the $b$-cycle after the surgery} (which is not what is drawn on the right panel). This is also called Dehn twists, namely, one saws open the solid torus along $a$, and reglue after applying $p$ twists. The problem of the framing will be taken up later.
\end{remark}

\begin{figure}[h]
\begin{center}
\begin{tikzpicture}[scale=.8]
\draw[rotate = 0](1.2,0) ellipse (.8 and 1.2)[color=black];
\draw [-<,blue](2,-0.1cm) to [out=90, in=-90] (2,0.1cm);
\draw[white, line width=1.8mm] (.46,-.48cm) arc (190:205:.7cm);
\node at (2,0) [right] {\scriptsize $p/q$};
\draw[rotate = 0](0,0) ellipse (1.2cm and 0.6cm)[color=black];
\draw [->,blue](-0.1,-0.6cm) to [out=0, in=-180] (0.1,-0.6cm);
\draw[white, line width=.8mm] (0.41,0.50cm) arc (170:155:.7cm);
\draw[white, line width=.8mm] (0.52,0.44cm) arc (170:155:.7cm);
\node at (-1.2,0)[right] {\scriptsize $r/s$};
\node at (3.2,0) {\small $\sim$};
\end{tikzpicture}
\begin{tikzpicture}[scale=.8]
\draw[rotate = 0](1.2,0) ellipse (.8 and 1.2)[color=black];
\draw [-<,blue](2,-0.1cm) to [out=90, in=-90] (2,0.1cm);
\node at (2,0) [right] {\scriptsize $p/q-s/r$};
\end{tikzpicture}
\caption{Combining rational surgeries, the slam-dunk move.}\label{fig_comb_rat_surg}
\end{center}
\end{figure}
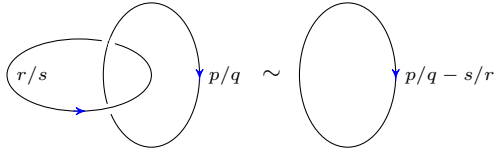

There is in general no obvious way of combining two surgeries, but when two rational surgeries are in the situation depicted as in fig.\ref{fig_comb_rat_surg}, they can be combined. To see this, let $a'$ be the contractible cycle of the left torus, and $a$ the contractible cycle of the right torus and $b$ be the long meridian along the right torus with zero linking number with the left torus. Suppose that after the surgery the combination $xa+yb$ becomes contractible in the right picture, then is must be the linear combination of the two contractible cycles of the two solid tori
\bea xa+yb=\lambda(ra'+sa)+\mu(pa+q(b+a')),\nn\eea
which implies
\bea \Bigg\{\begin{array}{c} \lambda r+\mu q=0 \\ \lambda s+\mu p=x \\ \mu q=y\end{array}\Rightarrow \frac{x}{y}=\frac{p}{q}-\frac{s}{r}.\nn\eea
One defines the matrices
\bea T=\begin{array}{|cc|} 1 & 1 \\ 0 & 1\end{array}~,~~S=\begin{array}{|cc|} 0 & -1 \\ 1 & 0\end{array}~,~~~~(ST)^3=-\bf{1}\label{SL2Z_matrices}\eea
that generate the entire $SL(2,\BB{Z})$.
Notice that the combination
\bea T^pS=\begin{array}{|cc|} p & -1 \\ 1 & 0\end{array}\nn\eea
correspond to the $p/1$ surgery and are called the integer surgeries, and the resulting manifold is exhibited as the boundary of a handle body obtained by gluing 2-handles to the 4-ball along the link $L\subset S^3=\partial B^4$, see ref.\cite{Kirby}.

Then the slam dunk-move tells us that integer surgeries as in fig.\ref{fig_comb_rat_surg} ($q=s=1$) can be composed as one composes matrices
\bea T^pST^rS=\begin{array}{|cc|} p & -1 \\ 1 & 0\end{array}~\begin{array}{|cc|} r & -1 \\ 1 & 0\end{array}=
\begin{array}{|cc|} pr-1 & -p \\ r & -1\end{array}.\nn\eea
And conversely, one can decompose a rational surgery into a product of integer surgeries by writing
\bea p/q=a_r-\frac{1}{\displaystyle a_{r-1}-\frac{1}{\displaystyle \cdots-\frac{1}{a_1}}},\nn\eea
then the $p/q$ surgery can be decomposed as the product $T^{a_r}ST^{a_{r-1}}S\cdots T^{a_1}S$, see ref.\cite{FreedGompf} (marvelous paper!).

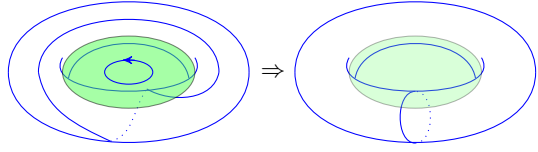
\begin{figure}[h]
\begin{center}
\begin{tikzpicture}[scale=.8]
\draw [-,blue](-2,0cm) to [out=-90,in=-90] (2,0cm);
\draw [-,blue](-2,0cm) to [out=90,in=90] (2,0cm);
\draw [-,blue](-1.1,.24cm) to [out=-120,in=-60] (1.1,.24cm);
\draw [-,blue](-1,-.10cm) to [out=80,in=100] (1,-.10cm);
\draw [-,blue](-1.5,0cm) to [out=-80,in=155] (-0.3,-1.15cm);
\draw [dotted,blue](-.3,-1.15cm) to [out=15,in=-125] (0.3,-.28cm);
\draw [-,blue](1.5,0cm) to [out=-100,in=-30] (0.3,-.28cm);
\draw [-,blue](-1.5,0cm) to [out=90,in=90] (1.5,0cm);
\draw [rotate = 0, fill=green!70, opacity=.5](0,0) ellipse (1.1 and 0.6);
\draw [rotate = 0, blue](0,0) ellipse (.4 and 0.2);
\draw [<-,blue] (-.1,.2) -- (0.1,.2);
\node at (2.4,0) {\small$\Rightarrow$};
\end{tikzpicture}%
\begin{tikzpicture}[scale=.8]
\draw [-,blue](-2,0cm) to [out=-90,in=-90] (2,0cm);
\draw [-,blue](-2,0cm) to [out=90,in=90] (2,0cm);
\draw [-,blue](-1.1,.24cm) to [out=-120,in=-60] (1.1,.24cm);
\draw [-,blue](-1,-.10cm) to [out=80,in=100] (1,-.10cm);
\draw [blue](0,-1.18cm) to [out=180,in=180] (0,-.32cm);
\draw [dotted, blue](0,-1.18cm) to [out=0,in=0] (0,-.32cm);
\draw [rotate = 0, fill=green!50, opacity=.3](0,0) ellipse (1.1 and 0.6);
\end{tikzpicture}%
\caption{Surgery $TS$, which is just adding one Dehn twist on the $b$-cycle.}\label{fig_1_1_surgery}
\end{center}
\end{figure}

There are further equivalence relations amongst surgeries. The following surgery $TS$ in fig.\ref{fig_1_1_surgery} does not change the diffeomorphism type either, but only shifts the 2-framing. Yet it has an interesting effect on the links that pass through the torus hole.
This surgery can be simply summarized as add a Dehn twist along the $b$-cycle, but to help visualize it, one may look at fig.\ref{fig_1_1_surgery}. One cuts open the disc in green, and slightly widen the cut to get an upper and lower disc separated by a small distance. Then one rotates the upper disc by $2\pi$ and recombine it with the lower one. In this process, it is clear that the cycle drawn in the left becomes the $a$-cycle on the right. Equally clear is the fact that any strand that passes through the green disc receives a twist in the process.

Finally, we have examples.
\begin{example}
Performing a 0/1 surgery, which corresponds to $T^0S$, in $S^3$, brings us $S^2\times S^1$, as we have seen before. Since this is a very important concept, we can try to understand it in different ways.

The 3-sphere can be presented as
\bea |z_1|^2+|z_2|^2=1,~~~z_{1,2}\in \BB{C}.\nn\eea
This may be viewed as a torus fibration. Define $t=|z_1|^2$, then we have
\bea &&~S^3\longleftarrow S^1\times S^1\nn\\
&&~\downarrow \nn\\ &&[0,1].\nn\eea
The base is parameterized by $t$, while the fibre is two circles $(te^{i\theta_1},(1-t)e^{i\theta_2})$. Away from $t=0,1$, the two circles are non-degenerate, while if one slides to the $t=0$ end, one shrinks the first $S^1$ and vice versa. Thus one can slice open $S^3$ at $t=1/2$, then to the left and to the right, one has a solid torus each, whose contractible cycles are the first and second circle respectively.
\end{example}

\begin{example} $p/q$ \emph{surgery and the lens space} $L(p,q)$.
A lens space is defined as a quotient of a sphere by a free, discrete group action. Again present $S^3$ as in the previous example, and define
\bea \xi=\exp\frac{2\pi i}{p},\nn\eea
and the group action to be
\bea (z_1,z_2)\to (z_1\xi,z_2\xi^q),~~~\gcd(p,q)=1.\nn\eea
Since $p,\,q$ are coprime, the group action is free, and the quotient is a smooth manifold, which has a picture of torus fibration as the previous example.

Try to convince yourself that the $pa+qb$ cycle to the right of $t=1/2$ corresponds to the contractible cycle to the left, which is exactly what a $p/q$ surgery would do.
\end{example}

%% file: Quantization_of_CS_part1.tex
From the discussion of sec.\ref{AAoT}, especially the quantum mechanics reasoning plus the sixth row of tab.\ref{tab_Atiyah_ax}, one sees that the Hilbert space assigned to a Riemann surface $\Sigma$ can be obtained by quantizing the Chern-simons theory on $\Sigma\times\BB{R}$.

To do this, we explicitly split out the time (the $\BB{R}$) direction, and write
\bea A=A_tdt+\SF{A},\nn\eea
where $\SF{A}$ is a connection on $\Sigma$. The action is written as
\bea S=\frac{k}{4\pi}\Tr \int d^2xdt~\big(A_t(\SF{dA+AA})+\SF{A}\dot{\SF{A}}\big),\nn\eea
where $\SF{d}$ and $\SF{A}$ are the differential and conneciton along $\Sigma$.

In the action, $\SF{A}\dot{\SF{A}}$ corresponds to the $p\dot q$ term in Eq.\ref{Feynman_taught_us}, which signals that the symplectic form for the phase space is
\bea \omega=\frac{k}{4\pi}\Tr \int_{\Sigma} ~\delta\SF{A}\delta \SF{A}.\label{symp_CS}\eea
Furthermore, the field $A_t$ sees no time derivatives and therefore is non-dynamical. It can be integrated out, imposing the flatness constraint on $\SF{A}$,
\bea \SF{F}=\SF{dA+AA}=0.\label{moment_map}\eea

Thus the problem of quantizing Chern-Simons theory is the quantization of flat connection on $\Sigma$ (mod gauge transformation), with the symplectic form Eq.\ref{symp_CS}. It is a typical problem of symplectic reduction, namely, the flat connection is the zero of the constraint functionals
\bea \mu=\int_{\Sigma} \Tr[\epsilon \SF{F}],\nn\eea
where $\epsilon$ is any $\FR{g}$-valued function on $\Sigma$. At the same time, gauge transformations are the Hamiltonian flows of the same $\mu$'s
\bea \delta_{\epsilon}\SF{A}=\{\mu,\SF{A}\},\nn\eea
where the Poisson bracket is taken w.r.t the inverse of the symplectic form Eq.\ref{symp_CS}.

There are some easy results to be harvested from this discussion.

\vskip .5cm

\noindent $\sbullet$ $\Sigma=S^2$\\
Over $S^2$ without puncture, there is no flat connection, other than $\SF{A}=0$, since $S^2$ is compact and simply connected. Thus we conclude the Hilbert space ${\cal H}_{S^2}$ is 1-dimensional.

This alone leads to an interesting formula about connected sums.
\begin{example}
Remove from two 3-manifolds $M,~N$ a solid ball $B^3$, and glue them along their common boundary $S^2$,
the resulting manifold is denoted as
\bea M\#N.\nn\eea
Then the partition function is given by
\bea Z_{M\# N}=\bra \psi_{M\backslash B^3}|\psi_{N\backslash B^3}\ket,~~~\psi_{M\backslash B^3},~\psi_{N\backslash B^3}\in{\cal H}_{S^2}.\nn\eea
But the two vectors are both proportional to $\psi_{B^3}$, as ${\cal H}_{S^2}$ is 1-dimensional. Thus
\bea Z_{M\# N}=\frac{\bra \psi_{M\backslash B^3}|\psi_{B^3}\ket\bra \psi_{B^3}|\psi_{N\backslash B^3}\ket}
{\bra \psi_{B^3}|\psi_{B^3}\ket}=\frac{Z_M\cdot Z_N}{Z_{S^3}}.\nn\eea
In other words, the quantity $Z_M/Z_{S^3}$ is multiplicative under connected sums.
\end{example}

As a variant, we place two punctures on $S^2$, marked with representations\footnote{This situation is best understood in conformal field theory, where punctures labelled by representations are points with vertex operators inserted. In fact, the wave function of CS theory is equivalent to the conformal blocks of the WZW CFT, but in this note I do not plan to go into WZW model. It is however still helpful to have this picture in the back of one's mind.} $R_{1,2}$.

If in the tensor product of $R_1\otimes R_2$, there is no trivial representation, the Hilbert space is empty. Only when $R_2=\bar R_1$, there is one (and only one, due to Schur's lemma) trivial representation, thus the Hilbert space is one dimensional if the two $R$'s are dual, otherwise empty. The former case can be understood as the insertion of a Wilson line that penetrates $S^2$ at the two punctures. This simple observation also leads to a potent formula.
\begin{example}\emph{Verlinde's formula}\\
Consider the situation fig.\ref{fig_verline}, where one Wilson line transforms in the representation $\ga$, and another Wilson-loop of representation $\gb$ encircles the first one.
\begin{figure}[h]
\begin{center}
\begin{tikzpicture}[scale=.8]
\draw [semithick,blue] (0,0) circle (1);
\draw [->,blue](0,-1cm) to [out=60,in=-60] (0,1cm); 
\node at (0,1cm) {\small $\times$};
\node at (-0.1,1cm)[below] {\small $\alpha$};
\node at (0,-1cm) {\small $\times$};
\node at (0.3,-1cm) [below] {\small $\psi_{\ga}$};
\node at (1.4,0cm) {\small $\sim$};
\end{tikzpicture}%
~%
\begin{tikzpicture}[scale=.8]
\draw [semithick,blue] (0,0) circle (1);
%
\draw [-,blue](0.25,-.2cm) to [out=170, in=-80] (-.2,0cm); 
\draw [-,blue](-.2,0cm) to [out=90, in=-170] (0.15,.3cm); 
\draw [-,blue](0.25,-.2cm) to [out=10, in=-100] (.7,.0cm); 
\draw [>-,blue](.7,0cm) to [out=90, in=-10] (0.35,.3cm); 
\draw [->,blue](0,-1cm) to [out=60,in=-60] (0,1cm); 
\node at (0,1cm) {\small $\times$};
\node at (-0.1,1cm)[below] {\small $\alpha$};
\node at (0,-1cm) {\small $\times$};
\node at (-.2,0cm)[left] {\small $\beta$};
\node at (.3,-1cm) [below] {\small $\psi$};
\end{tikzpicture}
\caption{Verlinde's formula}\label{fig_verline}
\end{center}
\end{figure}
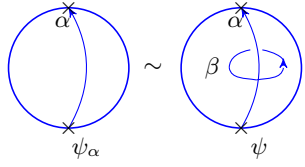
Since each ${\cal H}_{\ga}$ is 1-dimensional, the vector associated with the second configuration in fig.\ref{fig_verline} must be proportional to the one without the Wilson loop $\beta$
\bea \psi=\lambda_{\gb}^{\ga}\psi_{\ga},~~\textrm{no sum over}~\ga.\nn\eea
By gluing these two pictures, one obtains an $S^3$ with a Hopf link, with the two components labelled by $\ga$ and $\gb$. The partition function is given by
\bea Z_{S^3(L(\alpha,\beta))}=\bra\psi_{\ga}|\psi\ket=\bra\psi_{\ga}|\lambda^{\ga}_{\gb}|\psi_{\ga}\ket.\nn\eea

But the same computation can be done by inserting a Wilson line of representations $\ga$ and $\gb$ into the two solid tori of fig.\ref{fig_glue_S}, and after the gluing these two Wilson loops will be linked. Thus the partition function is
\bea Z_{S^3(L(\alpha,\beta))}=\bra\varphi_{\ga}|S|\varphi_{\beta}\ket=S_{\ga\gb}.\nn\eea
From this we get
\bea S_{\ga\gb}=\lambda^{\ga}_{\gb}|\psi_{\ga}|^2,~~S_{\ga0}=\lambda^{\ga}_{0}|\psi_{\ga}|^2=|\psi_{\ga}|^2,\nn\eea
which leads to the formula $S_{\ga\gb}/S_{\ga0}=\lambda^{\ga}_{\gb}$.

This is essentially Verlinde's formula, which laconically says 'the S-matrix diogonalizes the fusion rule'. To see this, we insert two Wilson loops into a solid torus fig.\ref{fig_fusion}, with representation $\ga$, $\gb$.
\begin{figure}
\begin{center}
\begin{tikzpicture}[scale=.6]
\draw [rotate = 0, fill=blue!50, opacity=.5](0,0) ellipse (1 and 0.6);
\draw [rotate = 0, fill=blue!50, opacity=.5](0,2.5) ellipse (1 and 0.6);
\draw [-,blue] (-1,0) -- (-1,2.5)
             (1,0) -- (1,2.5);
\draw [->] (-.3,0) -- (-.3,2.5) node [left] {\scriptsize $\alpha$};
\draw [->] (.3,0) -- (.3,2.5) node [right]{\scriptsize $\beta$};
\end{tikzpicture}
\begin{tikzpicture}[scale=.6]
\draw [rotate = 0, fill=blue!50, opacity=.5](0,0) ellipse (1 and 0.6);
\draw [rotate = 0, fill=blue!50, opacity=.5](0,2.5) ellipse (1 and 0.6);
\draw [-,blue] (-1,0) -- (-1,2.5)
             (1,0) -- (1,2.5);
\draw [->] (0,0) -- (0,2.5) node [left] {\scriptsize $\gamma$};
\node at (-1.8,1) {\small $=N_{\alpha\beta}^{\gamma}$};
\end{tikzpicture}\caption{Fusion of two Wilson loops}\label{fig_fusion}
\end{center}
\end{figure}
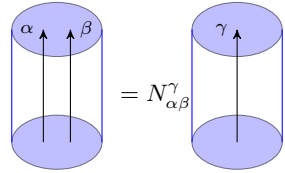
The two Wilson loops can be fused according to
\bea \varphi_{\ga}\otimes\varphi_{\gb}=N_{\ga\gb}^{\gc}\varphi_{\gc}.\nn\eea
The fusion coefficient can be computed using by looking at fig.\ref{fig_S_fusion}.
\begin{figure}[h]
\begin{center}
\begin{tikzpicture}[scale=.8]
\draw [->,blue] (-.2,0) -- (-.2,1.2) node [above] {\scriptsize $\alpha$};
\draw [->,blue] (.2,0) -- (.2,1.2) node [above] {\scriptsize $\beta$};
\node at (.6,.8) {\small $\stackrel{S}{\longrightarrow}$};
\end{tikzpicture}
\begin{tikzpicture}[scale=.8]
\draw [->,blue] (0,0) -- (0,1.2) node [above] {\scriptsize $\alpha$};
\draw [rotate = 0, color=blue](0,0.6) ellipse (.4 and 0.2);
\node at (.6,.6) {\scriptsize $\delta$};
\node at (-1.2,.6) {\small $\sum_{\delta}S^{\beta}_{~\delta}$};
\node at (1.2,.7) {\small $\longrightarrow$};
\end{tikzpicture}
\begin{tikzpicture}[scale=.8]
\draw [->,blue] (0,0) -- (0,1.2) node [above] {\scriptsize $\alpha$};
\draw [dotted, rotate = 0, color=blue](0,0.6) ellipse (.4 and 0.2);
\node at (.6,.6) {\scriptsize $0$};
\node at (-1.2,.6) {\small $\sum_{\delta}S^{\beta}_{~\delta}\lambda^{\alpha}_{\delta}$};
\node at (1.2,.7) {\small $\stackrel{S^{-1}}{\longrightarrow}$};
\end{tikzpicture}
\begin{tikzpicture}[scale=.8]
\draw [->,blue] (0,0) -- (0,1.2) node [above] {\scriptsize $\gamma$};
\node at (-1.2,.6) {\small $\sum_{\delta}S^{\beta}_{~\delta}\lambda^{\alpha}_{\delta}S^{\gamma}_{~\alpha}$};
\end{tikzpicture}\caption{Using S-dual to compute the fusion coefficients}\label{fig_S_fusion}
\end{center}
\end{figure}
And the last formula says that $S$ diagonalizes the matrix $(N_{\ga})^{\gc}_{~\gb}$.
\end{example}

%% file: Quantization_of_CS_torus.tex
To be able to perform surgery along a knot, one needs the knowledge of quantization of CS on a torus, which requires some serious work. Let me first flash review the geometric quantization.
\subsection{Geometrical Quantization}\label{GQ}
For a K\"ahler manifold $M$ with complex structure $J$ and K\"ahler form $\omega$, one has a recipe for the pre-quantization. recall that the prequantization assigns a function $f\in C^{\infty}(M)$ an operator $\hat f$ such that the Poisson bracket is mapped to the commutator as follows
\bea i\widehat{\{f,g\}}=[\hat f,\hat g].\nn\eea
The recipe is as follows. Suppose the K\"ahler form is in the integral cohomology class $i\omega/2\pi\in H^2(M,\BB{Z})$, then one can always construct a complex line bundle $L$ over $M$ with curvature $\omega$. Let the covariant derivative of $L$ be $\nabla$, then $[\nabla_{\mu},\nabla_{\nu}]=-i\omega_{\mu\nu}$. Define the following operator
\bea \hat h=iX_h^{\mu}\nabla_{\mu}+h,~~~h\in C^{\infty}(M),~~X_h^{\mu}=(\partial_{\nu}h)(\omega^{-1})^{\nu\mu}\nn\eea
that acts on the sections of $L$ (the second term above acts by multiplication). One easily checks that
\bea [\hat h,\hat g]=-[\nabla_{X_h},\nabla_{X_g}]+2i\{h,g\}=i\omega(X_h,X_g)-\nabla_{X_{\{h,g\}}}+2i\{h,g\}=-\nabla_{X_{\{h,g\}}}+i\{h,g\}=i\widehat{\{h,g\}}.\nn\eea
Thus the procedure of pre-quantization is more or less standard.
To complete the quantization, one needs to choose a polarization. In the case $M$ is affine, one can use the complex structure to split the 'momenta' and 'coordinates'. For example, one demands that the wave function be the holomorphic sections of the bundle $L$, which is
\bea (-2\nabla_{\bar i}+x^i)\psi=0,~~~\psi\in\Gamma(L).\nn\eea
There is no general recipe for the choice of polarization.

\begin{example}
As an exercise, we consider quantizing the torus parameterized by the complex structure $\tau$, with symplectic form
\bea \omega=\frac{k}{2\pi}dx\wedge dy=\frac{ik}{4\pi\tau_2}dz\wedge d\bar z,~~k\in\BB{Z},~~x,y\in [0,1],~~z=x+\tau y.\nn\eea
After quantization, we had better find $k$ states since the integral of $\omega$ over the torus is $k$.

If we choose $x$ to be the coordinate, then its periodicity forces $y$, the momentum, to be quantized
\bea y=\frac{n}{k},~~n\in \BB{Z} .\nn\eea
At the same time $y$ itself is defined mod $\BB{Z}$, thus we have exactly $k$ states
\bea y=\frac{0}{k},\frac{1}{k},\cdots \frac{k-1}{k}.\nn\eea
The corresponding wave function in the momentum basis is the periodic delta function
\bea \psi_n(y)=\delta^P(y-\frac{n}{k})=\sum_{m\in\BB{Z}}\delta(y-\frac{n}{k}-m)=\sum_{m\in\BB{Z}}\exp\Big(2\pi i(y-\frac{n}{k})m\Big).\label{wav_fun_real}\eea

Now we would like to do the same exercise with the K\"ahler quantization. The line bundles are uniquely fixed by $c_1$, which is related to the Hermitian metric by
\bea c_1=\frac{i}{2\pi}\bar\partial\partial\log h.\nn\eea
Since one would like to have $c_1=\omega$, one may choose
\bea h=\exp\Big(-\frac{\pi k}{\tau_2}z\bar z\Big).\label{metric_L_bundle0}\eea
And as the norm of a section $||s||^2=|s|^2h$ must be a doubly periodic function on the torus, one roughly knows that he would be looking for the holomorphic sections amongst the theta functions. In fact we can derive this, by taking the wave function in the real polarization and find the canonical transformation that takes one from
$(x,y)$ to $(z,\bar z)$. Let
\bea G=-i\pi\Big(-\tau y^2+2yz+\frac{iz^2}{2\tau_2}\Big),\nn\eea
be the generating function, which depends on the old momentum $y$ and new coordinate $z$. It satisfies the usual set of conditions of a canonical transformation
\bea \partial_yG=-2i\pi x,~~~\partial_zG=\frac{\pi\bar z}{\tau_2}.\nn\eea
To convert the wave function Eq.\ref{wav_fun_real} to the complex setting, one applies the kernel
\bea \psi_{n,k}=\int_{-\infty}^{\infty}dy~e^{kG}\psi_n=\sqrt{\frac{1}{-i\tau k}}\exp\big(\frac{\bar\tau\pi k}{2\tau_2\tau}z^2\big)
\sum_m\exp \Big(-\frac{i\pi}{k\tau}m^2+\frac{2\pi i}{\tau} mz - \frac{-2i\pi}{k}mn\Big),\nn\eea
where the phase of $-i\tau$ is taken to be within $(-\pi/2,\pi/2)$ since $\tau_2\gneq0$. Apply the Poisson re-summation,
\bea \psi_{n,k}(\tau,z)=\exp\Big(\frac{\pi kz^2}{2\tau_2}\Big)\sum_m\exp \Big(i\pi k\tau(m-\frac{n}{k})^2+2i\pi kz(m-\frac{n}{k})\Big)=\exp\Big(\frac{\pi kz^2}{2\tau_2}\Big)\theta_{n,k}(\tau;z).\label{wav_fun_cplx}\eea
The level $k$ theta function is defined as
\bea &&\theta_{n,k}(\tau;z)=\sum_m\exp \Big(i\pi k\tau(m-\frac{n}{k})^2+2i\pi kz(m-\frac{n}{k})\Big)\label{theta_n_k},\\
&& \theta(\tau;z+a+b\tau)=\exp\Big(-i\pi b^2\tau-2i\pi bz\Big)\theta(\tau;z),~~~a,b\in\BB{Z}.\nn\eea
One further combines the first factor in Eq.\ref{wav_fun_cplx} with the metric Eq.\ref{metric_L_bundle0} to conclude that the holomorphic sections of the line bundle are theta functions at level $k$, with the standard metric
\bea h=\exp\Big(\frac{\pi k}{\tau_2}(z-\bar z)^2\Big).\label{metric_L_bundle}\eea
Thus the wave function is the theta function with Hermitian structure given in Eq.\ref{metric_L_bundle}. This toy model will be used later.
\end{example}

\subsection{Naive Finite Dimensional Reasoning}\label{NFDR}
Back to the Chern-Simons theory problem, the flat connections on a torus is fixed by the image of the map
\bea \BB{Z}\oplus\BB{Z}=\pi_1(T^2)\to G,\nn\eea
and since the fundamental group of $T^2$ is abelian, so should be the image. Thus we can choose the image, in other words, the holonomy, to lie in the maximal torus of $G$. Take the holonomy round the two cycles $z\to z+m+n\tau$ to be
\bea \textrm{hol}=\exp2\pi i(m{u}+n{v}).\label{holonomy}\eea
This means that one can, up to a gauge transformation, put $\SF{A}$ to be a constant, valued in the Cartan sub-algebra $\FR{h}$ (CSA)
\bea \SF{A}=a{u}+b{v},~~~{u}=\vec u\cdot\vec H,~{v}=\vec v\cdot\vec H,~~\vec H\in \FR{h}.\label{finite_reduction}\eea
where I have also used $a$ and $b$ to denote the 1-forms representing the $a$ and $b$ cycle.
The symplectic form for this finite dimensional problem, which is inherited from Eq.\ref{symp_CS} is
\bea \omega=k\omega_0,~~~\omega_0=\Tr [du\wedge dv].\label{symp_CS_finite}\eea
At this stage, the residual gauge transformation is the Weyl group $W$, which preserves the maximal torus. Both $u$ and $v$ are periodic, shifting them by the root lattice in Eq.\ref{holonomy} affects nothing, since $\exp(2\pi i \ga\cdot \vec H)=1$.
\begin{example}
Since we will be dealing with the root and weight lattice quite a lot in the coming discussions, it is useful to have an example in mind. Let me just use $A_{n-1}=\FR{su}(n)$ as an illustration (and I will also assume throughout the notes that the Lie algebra is simply laced).

Pick a set of $n$ orthonormal basis $e_i$ for the Euclidean space $\BB{R}^n$.
\bea e_i=(0,\cdots ,\mathop{1}_{i\,th},\cdots,0).\nn\eea
The set of positive roots of $\FR{su}(n)$ is $\{e_i-e_j|i<j\}$, while the simple roots are those $j=i+1$, and it is clear that the simple roots $\alpha_i=e_i-e_{i+1}$, $i=1\cdots n-1$ satisfy $\angle(\alpha_{i-1},\alpha_i)=120^{\circ}$.

The \emph{root lattice} $\Lambda^R$ is the lattice generated by the simple roots. The \emph{weight lattice} $\Lambda^W$ is a lattice such that\footnote{I follow notation of ref.\cite{Gepner:1986wi}: $(\ga,\gb)$ is the inner product, while $\bra\ga,\gb\ket=2(\ga,\gb)/(\gb,\gb)$. Note the notation used in refs.\cite{Jeffrey,Rozanskytrivial} is opposite.}
\bea \gb\in\Lambda^W~\Rightarrow~\bra \gb,\ga\ket=\frac{2(\gb,\ga)}{(\ga,\ga)}\in\BB{Z},~~\forall \ga\in\Lambda^R\label{weight_lattice}.\eea
where $(\sbullet,\sbullet)$ is the invariant metric.
When the Lie algebra is simply laced and the simple roots normalized to be of length $\sqrt2$, the weight lattice becomes the dual lattice of the root lattice.

The highest weight vector for the two common irreducible representations are listed
\bea \textrm{fundamental}:&& \frac{1}{n}(n-1,-1,\cdots,-1),\nn\\
\textrm{adjoint}:&& (1,0,\cdots,0,-1).\nn\eea
The Weyl vector $\rho$ is half the sum of positive roots
\bea\rho=\frac{1}{2}(n-1,n-3,\cdots -n+3, -n+1).\nn\eea
The Weyl reflection off of a root $\ga$ is
\bea \beta\to \beta-2\frac{(\beta,\alpha)}{(\alpha,\alpha)}\ga=\gb-\bra\gb,\ga\ket\ga,\nn\eea
Thus the Weyl reflection
of $e_i-e_{i+1}$ off of $e_{i+1}-e_{i+2}$ is $e_i-e_{i+2}$, i.e. a permutation $e_{i+1}\to e_{i+2}$, so we see that the Weyl group for $\FR{su}(n)$ is just the symmetric group $s_n$.

The product $\Delta=e^\rho\prod_{\alpha>0}(1-e^{-\alpha})$ is called the \emph{Weyl denominator}, and we have the important Weyl denominator formula
\bea \Delta=e^\rho\prod_{\alpha>0}(1-e^{-\alpha})=\sum_{w\in W}(-1)^we^{w(\rho)}.\label{Weyl_denominator}\eea
where the sum is over the Weyl group. The sign $(-1)^w$ is + if $w$ is decomposed into an even number of Weyl reflections and $-$ otherwise.
Note that $e^{\ga}$ can be thought of as a function on the Cartan subalgebra $\FR{h}$
\bea e^{\ga}(\gb)\stackrel{def}{=}e^{(\ga,\gb)},~~~~\gb\in\FR{h}.\nn\eea
The dimension of a representation of weight $\mu$ is
\bea d_{\mu}=\prod_{\ga>0}\frac{(\rho+\mu,\ga)}{(\rho,\ga)}\label{dimension_formula}.\eea
And the second Casimir is
\bea C_2(\mu)=\frac{1}{2}\big((\rho+\mu)^2-\rho^2\big)\label{Casimir_formula}.\eea
The dual Coxeter number is the second Casimir of the adjoint representation.

For $\FR{su}(n)$, we can explicitly compute the Weyl denominator.
Let us introduce $x_i = e^{e_i/2}$,  then $e^{\rho}$ is a product
\bea e^{\rho}=(x_1)^{n-1}(x_2)^{n-3}\cdots (x_{n-1})^{-n+3} (x_n)^{-n+1},~~~{\rm where}~~x_i = e^{e_i/2}.\nn\eea
Since a Weyl group in this case permutes the $x_i$'s, the sum over Weyl group is equivalent to the Van de Monde determinant
\bea \Delta&=&\sum_{\sigma\in S_n}{\rm sgn}(\sigma)(x_{\sigma(1)})^{n-1}(x_{\sigma(2)})^{n-3}\cdots (x_{\sigma(n-1)})^{3-n} (x_{\sigma(n)})^{1-n}\nn\\
&=&\det\left|\begin{array}{ccc}
         x_1^{n-1} &\cdots & x_1^{1-n} \\
         \cdots & \cdots &\cdots \\
         x_n^{n-1} &\cdots & x_n^{1-n}\end{array}\right|=\prod_i x^{1-n}_i\cdot\prod_{i<j}(x_i^2-x_j^2)\nn\\
&=&\prod_i x^{1-n}_i\cdot\prod_{i<j}x_ix_j\cdot\prod_{i<j}(\frac{x_i}{x_j}-\frac{x_j}{x_i})=\prod_{i<j}(\frac{x_i}{x_j}-\frac{x_j}{x_i})=\prod_{\alpha>0}(e^{\alpha/2}-e^{-\alpha/2}),\nn\eea
leading to the Weyl denominator formula Eq.\ref{Weyl_denominator}.

For those familiar with matrix models, $x_i$ are the eigen-values of an $SU(n)$ matrix, and $\Delta$ is the Jacobian $[dU]=\Delta^2 d^nx$.
\end{example}

In Eq.\ref{finite_reduction}, the two variables are valued in the dual of the Cartan subalgebra, and they are periodic.
Thus if one takes $u$ in Eq.\ref{finite_reduction} as the coordinate, the momentum $v$ must be quantized, leading to $v\in \Lambda^W/k$. Taking into account the Weyl group and also the periodicity of $v$ itself, the Hilbert space is given by
\bea \frac{\frac1k\Lambda^W}{W\ltimes \Lambda^R}.\label{naive_hilbert}\eea
This description of the Hilbert space coincide with the integrable representation of the affine Lie algebra at level $k$, and the group given by the semi-direct product of the Weyl group and translation $W\ltimes k\Lambda^R$ is the affine Weyl group.

Putting aside the Weyl invariance in Eq.\ref{naive_hilbert}, the reduced finite dimensional phase space $\Lambda^W/k \Lambda^R$ is also a torus, and its quantization is a straight forward generalization of the toy example in sec.\ref{GQ}, with the obvious replacement
\bea \BB{Z}\to \Lambda^R,~~~\frac{\BB{Z}}{k}\to \frac{\Lambda^{W}}{k},\nn\eea
and the resulting section for the prequantum line bundle is similar to the expression Eq.\ref{theta_n_k}, with $n$ replaced by weights and the summation over the integers replaced with a summation over the root lattice
\bea &&\theta_{\gc,k}(\tau;u)=\sum_{\ga\in\Lambda^R}\exp\Big(i\pi k\tau\left|\ga+\frac{\gc}{k}\right|^2+2\pi ik\big( u,\big(\ga+\frac{\gc}{k}\big)\big)\Big),\label{Weyl_theta}\eea
which is the definition of the Weyl theta function. One can further demand the theta function to be invariant or anti-invariant under the Weyl group. Since we have
\bea \theta_{\gc,k}(\tau;w(u))=\theta_{w(\gc),k}(\tau;u),\nn\eea
we let
\bea \theta^+_{\gc,k}(\tau;u)=\sum_{w\in W}~\theta_{w(\gc),k}(\tau,u),~~~~\theta^-_{\gc,k}(\tau;u)=\sum_{w\in W} (-1)^w~\theta_{w(\gc),k}(\tau,u)\nn\eea
be the invariant and anti-invariant Weyl theta function. Thus the wave function for our finite dimensional phase space Eq.\ref{naive_hilbert} is the former $\theta^+_{\gamma,k}$.

The naive reasoning is almost correct, the quantum correction merely accomplishes the well-known shift in Eq.\ref{naive_hilbert}
\bea \frac{\frac1k\Lambda^W}{W\ltimes \Lambda^R}\Rightarrow \frac{\frac1{k+h}(\Lambda^W+\rho)}{W\ltimes \Lambda^R}.\label{well_knownn_shift}\eea

\subsection{Quillen's Determinant Line-Bundle}
The shift above comes from a determinant factor, in the process of using a gauge transformation to put the gauge field into the constant form Eq.\ref{finite_reduction}. There are quite many ways to understand this shift, from current algebra in CFT, from holomorphic anomaly etc\footnote{During my conversation with Reimundo Heluani, he showed me that this can also be understood from some constructions due to Ben-Zvi and Frenkel \cite{Ben-ZviFrenkel}, but it is far byond my range of cognizance; I give the reference nonetheless.}. But let me only present the one I understand.

The discussion above gives us a candidate for the pre-quantum line bundle and its holomorphic sections as the wave functions. Now let us look at the inner product of the wave function
\bea \bra\psi|\psi'\ket= \int_{\FR{M}}~(\omega_0)^{r/2} h\psi^*\psi',\nn\eea
where $h$ is the hermitian metric of the line bundle, $r$ is the rank of the gauge group and $\omega_0$ is the symplectic form on $\FR{M}$ given in Eq.\ref{symp_CS_finite}, whose top power serves as the volume form of $\FR{M}$. Recall how we obtained $\omega_0$, for any flat connection on $T^2$ we preform a gauge transformation and make the connection a constant valued in CSA, naturally, in this process one would expect something similar to the Fadeev-Popov determinant term when gauge fixing Yang-Mills theory. One can get a clearer view of this determinant as follows.

As already mentioned in sec.\ref{sec_QoC1}, the moduli space of flat connection can be obtained as a K\"ahler reduction (the first part of sec.\ref{sec_QoC1})
\bea \FR{M}=\mu^{-1}(0)/\FR{G},\nn\eea
where $\FR{G}$ is the group of gauge transformations. Using the complex structure on $\Sigma$, there is a natural notion of the complexification $\FR{G}_c$ of $\FR{G}$, and the above K\"ahler reduction is diffeomorphic to
\bea \FR{M}=\FR{A}/\FR{G}_c\label{gauge_trans_complex}\eea
by a theorem of Mumford Sternberg and Guillemin \cite{GuilleminSternberg}. At the infinitesimal level, one can understand this equivalence by looking at the decomposition of the tangent space of $\FR{A}$ at $\mu^{-1}(0)$ as follows. Let $\mu_{\epsilon}=\int\Tr[\epsilon F]$, then $T\FR{G}$ is spanned by $X_{\mu_{\epsilon}}$ for all $\epsilon$, while \bea JX_{\mu_{\epsilon}}\cdot\mu_{\eta}=\bra\textrm{grad}\mu_{\epsilon},\textrm{grad}\mu_{\eta}\ket,\nn\eea
where $\bra\sbullet,\sbullet\ket$ denotes the inverse metric and we have used $J\omega^{-1}=g^{-1}$ for a K\"ahler manifold. This means that if $X_{\mu_{\epsilon}}$ generates a gauge transformation then $JX_{\mu_{\epsilon}}$ is necessarily orthogonal to $T\mu^{-1}(0)$, thus
\bea T\FR{A}\big|_{\mu^{-1}(0)}=T\FR{G}\oplus JT\FR{G}\oplus T\FR{M}.\nn\eea
We can use this decomposition to give a description of $T\FR{M}$ and also the volume form on $\FR{M}$.

Consider the gauge transformation in the complex setting, which fits into the complex
\bea &&\underbrace{\Omega^{(0,0)}(\Sigma,ad)}\stackrel{\bar\partial_A}{\longrightarrow}\underbrace{\Omega^{(0,1)}(\Sigma,ad)},\label{gauge_complex}\\
&&\hspace{.8cm}V^0 \hspace{2.2cm} V^1\nn.\eea
where $A$ is \emph{not} necessarily a flat connection. The vector space $V^1$ is the (0,1) tangent direction of the space of connections at $A$.
When $A$ is flat, the kernel and cokernel of $\bar\partial$ are
\bea \ker\bar\partial_A=H_{\bar\partial_A}^{0,0}(\Sigma,ad),~~~\textrm{coker}\bar\partial_A=H_{\bar\partial_A}^{0,1}(\Sigma,ad),\nn\eea
and both cohomology groups can be related to their real cousins $H_{d_A}^{0}(\Sigma,ad)$ and $H_{d_A}^{1}(\Sigma,ad)$ by means of the Hodge decomposition for a K\"ahler manifold (for which $\Delta_{d_A}$ and $\Delta_{\bar\partial_A}$ are proportional to each other).
Now one has the following relation for the determinant line bundle
\bea \wedge^{top}(V^0)^*\otimes \wedge^{top}(V^1)\sim \wedge^{top}(\ker\bar\partial_A)^*\otimes  \wedge^{top}(\textrm{coker}\,\bar\partial_A)\otimes\wedge^{top}\big((\ker\bar\partial_A)^{\perp}\big)^*\otimes  \wedge^{top}(\textrm{img}\,\bar\partial_A).\label{factor_det}\eea
For $A$ flat $H_{\bar\partial_A}^{0,0}(\Sigma,ad)$ is the complexification of a real bundle $H_{d_A}^{0}(\Sigma,ad)$, so its top power is a trivial complex line bundle (at any rate, if the genus $g\geq2$, $H_{d_A}^{0}(\Sigma,ad)$ vanishes). On the other hand $H_{\bar\partial_A}^{0,1}(\Sigma,ad)$ represents the (0,1) tangent direction of $\FR{M}$. Thus one can view the last two factors in the rhs of Eq.\ref{factor_det} as describing the Fadeev-Popov determinant incurred from modding out the complexified gauge transformation Eq.\ref{gauge_trans_complex}. Therefore the naive volume form $(\omega_0)^{r/2}$ of $\FR{M}$ must be corrected by the factor
\footnote{I am using $\det$ for the both the determinant and for the determinant line bundle.}
\bea H=\wedge^{top}\big((\ker\bar\partial_A)^{\perp}\big)^*\otimes  \wedge^{top}(\textrm{img}\,{\bar\partial}_A)={\det}'\bar\partial_A,\label{det_line_bundle}\eea
where the prime in ${\det}'$ denotes the exclusion of zero-modes. However, one often includes this factor as the correction to the wave function and hermitian metric instead of to the volume form. Note that $H$ is formally a trivial bundle, as the map $\bar\partial_A$
provides the trivialization away from the kernel, but the need to regulate an infinite product spoils the triviality.

We now take a closer look at the determinant line bundle Eq.\ref{det_line_bundle}.

\subsubsection{Abstract Computation of the Quillen Bundle}
The computation here is largely from ref.\cite{Quillen}. Since certain manipulations in the original paper simply eludes me and I have written down the following exposition according to my own take, the reader is admonished to check the original paper so that I may not be accused of wilfully misleading the reader.

Consider a \emph{smooth} vector bundle $E$ over a Riemann surface $\Sigma$, and let $\Omega^{p,q}(\Sigma,E)$ be the $E$-valued $(p,q)$-forms. Let $\alpha$ be a $(0,1)$-connection and define
\bea D=d\bar z(\bar\partial+\alpha),\nn\eea
which is the Cauchy Riemann operator since it defines a holomorphic structure on $\Gamma(E)$.
Further let \bea V^0=\Omega^{(p,q)},~~V^1=\Omega^{(p,q+1)}.\nn\eea
Then formally, the determinant line bundle ${\det}'D$ is
\bea {\det}'D=\wedge^{top}(V^0/\ker D)^*\otimes\wedge^{top} \textrm{img}\,D.\nn\eea
The \emph{Quillen} metric for sections of ${\det}'D$ is defined as
\bea \bra s,s\ket=\int_{\Sigma} |s|^2({\det}'{\Delta})^{1/2},~~~~\Delta=D^{\dagger}D.\nn\eea
It is convenient to write the determinant as
\bea {\det}'{\Delta}=e^{-\zeta'(0)},~~~\zeta(s)=\sum_{\lambda\neq 0} \lambda^{-s}.\nn\eea
The curvature of a complex line bundle is given by
\bea(\clubsuit)~~~~ \delta_{\bar\ga}\delta_{\ga}\log{\det}'\Delta=-\delta_{\bar\ga}\delta_{\ga}\zeta'(0)=-\delta_{\bar\ga}\delta_{\ga}\partial_s\Tr'[\Delta^{-s}]\big|_{s=0},\nn\eea
where the trace is taken over the non-zero eigen-space.

We now apply the first variation $\delta_{\ga}$
\bea \clubsuit&=&\delta_{\bar \ga}\partial_s\diamondsuit\big|_{s=0},\nn\\
\diamondsuit&=&-\delta_{\ga}\Tr'[\Delta^{-s}]=s\Tr'[\Delta^{-s-1}D^{\dagger}\delta_{\ga}D]\nn\\
&=&s\Tr'[\Delta^{-s}D^{-1}\delta_{\ga}D]=\frac{s}{\Gamma(s)}\int_0^{\infty}\Tr'[e^{-t\Delta}D^{-1}\delta_{\ga}D]t^{s-1},\nn\eea
where we have used the fact that $\Delta^{-1}D^{\dagger}$ is the Hodge theoretic inverse of $D$.

First of all, when $t\gneq0$, the coincidence limit $\bra z|e^{-t\Delta}D^{-1}|z\ket$ is finite, so the expression above makes sense except at the lower limit of the integral. Secondly, as $s/\Gamma(s)$ vanishes as fast as $s^2$, one need only keep the singular part in the factor
\bea\spadesuit=\int_0^{\infty}\Tr'[e^{-t\Delta}D^{-1}\delta_{\ga}D]t^{s-1},\nn\eea
and likewise any term that is finite at the $s=0$ limit will be discarded in the subsequent manipulation.

The integral converges for large $t$ and any $s$, thus we can write cut the integral off at $t=1$. For easy reference I record also the asymptotic expansion of the heat kernel $e^{-t\Delta}$ for small $t$
\bea \bra x |e^{-t\Delta}|y\ket=\frac{1}{(4\pi t)^{\textrm{dim/deg}}}e^{-d^2(x,y)/(4t)}\Big[c_0(x,y)+c_1(x,y)t^{1/\deg}+c_2(x,y)t^{2/\deg}+\cdots\Big],\label{aymp_heat_kernel}\eea
where $d(x,y)$ is the geodesic distance between $x$, $y$ and $c_i(x,y)$ are non-singular at $x=y$, $c_0(x,x)=1$. Now apply the variation $\delta_{\bar\ga}$
\bea \delta_{\bar \ga}\spadesuit=\int_0^1\Tr'[e^{-t\Delta}(\delta_{\bar \ga}D^{-1})\delta_{\ga}D]t^{s-1},\nn\eea
where even though $\delta_{\bar \ga}$ could have hit $e^{-\Delta}$, this variation is only going to give us something finite as $s\to 0$. Formally, $\delta_{\bar \ga}D^{-1}$ should be zero as $D$ is defined without reference to $\bar \ga$, yet as we shall see, due to the need to regulate the infinite sum, $D^{-1}$ develops an anomalous $\bar\ga$ dependence.

Let us work out the kernel $\bra x|D^{-1}|y\ket$ at small separation, which is the only region of interest to us. In the short distance limit, one has the expansion of $D^{-1}$ as
\bea G(x,y)&=&\bra x|D^{-1}|y\ket=\frac{dy}{x-y}\big[A+B(y)(x-y)+B'(y)(\bar x-\bar y)+\cdots\Big],\nn\\
D^{-1}\psi(x)&=&\int_y G(x,y)\psi(y),\nn\eea
where $\psi\in V^1$. Demanding $DD^{-1}=\delta^2(x-y)$ fixes $B'$
\bea G(x,y)=\frac{dy}{\pi(x-y)}\big[1+B(y)(x-y)-\alpha(y)(\bar x-\bar y)+\cdots\Big],\nn\eea
but leaves $B$ free. To fix also $B$, we impose the gauge invariance, namely, if $\psi^{\Omega}=\Omega\psi$ is the gauge transformation of $\psi$, then
\bea  (D^{-1}\psi)^{\Omega}\psi(y)&=&D^{-1}\psi^{\Omega}\nn\\
 G(x,y)&=&\frac{dy}{\pi(x-y)}\big[1-\alpha^*(y)(x-y)-\alpha(y) (\bar x-\bar y)+\cdots\Big].\label{expansion_kernel}\eea
Indeed, using $\alpha^{\Omega}=\Omega^{-1}\bar\partial \Omega+\Omega^{-1}\alpha \Omega$ one may check that
\bea D^{-1}\Omega(y)&=&\int_y~\frac{dy}{\pi(x-y)}\Big(\Omega(x)+\partial_x\Omega(x)(y-x)+\bar\partial_x\Omega(x)(\bar y-\bar x)+\cdots\nn\\
&&\hspace{2cm}-\alpha^*(y)(x-y)\Omega(y)-\alpha(y) (\bar x-\bar y)\Omega(y)+\cdots\Big)\psi(y),\nn\\
&=&\int_y\frac{\Omega(x)dy}{\pi(x-y)}\big(1-(\alpha^{\Omega})^*(y)(x-y)-\alpha^{\Omega}(y) (\bar x-\bar y)\Omega(y)+\cdots\big)\psi(y)=\Omega D^{-1}\psi.\nn\eea
From Eq.\ref{expansion_kernel}, one sees that $\delta_{\bar\ga}D^{-1}$ is finite in the coincidence limit, as a result one may set $e^{-t\Delta}=1$ in the expression of $\delta_{\bar \ga}\spadesuit$
\bea \delta_{\bar \ga}\spadesuit=\Tr[\delta_{\alpha^*}D^{-1}\delta_{\ga}D]\int_0^{1}t^{s-1}.\nn\eea
Finally one arrives at,
\bea -\delta_{\bar\ga}\delta_{\ga}\zeta'(0)=-\frac1\pi\int_{\Sigma}\Tr_E[\delta\alpha^*\delta{\ga}],\label{Quillen}\eea
where up to the last step the trace is taken in the infinite dimensional space $V^0$, while the last trace is taken in $E$.

Eq.\ref{Quillen} shows clearly a conflict between holomorphicity and gauge invariance, indeed ${\det}'\Delta$ should naively be the norm squared of a holomorphic function \bea {\det}'\Delta\stackrel{?}{=}|{\calligra det}'D|^2,\nn\eea
which would lead to a trivial first chern-class. Instead, one can rewrite
\bea {\det}'\Delta=e^{-||\ga-\ga_0||^2}|{\calligra det}'(D)|^2,\label{temp_11}\eea
where I have again used ${\calligra det}'D$ to denote a holomorphic function of $\ga$ defined in the above manner.

Now apply Eq.\ref{Quillen} to our problem Eq.\ref{gauge_complex}, we have
\bea c_1(H)\sim\int_{\Sigma}~\Tr_{ad}[\delta \SF{A}\delta \SF{A}]=h^{\vee}\int_{\Sigma}~\Tr_{f}[\delta \SF{A}\delta \SF{A}],\nn\eea
where $\Tr_f$ is in the fundamental representation and $h^{\vee}$ is the dual Coxeter number. Since we restrict ourselves to simply laced groups, we have $h=h^{\vee}$.
This explains roughly the shift in Eq.\ref{well_knownn_shift}.

\subsubsection{Concrete Computation of the Quillen Bundle at $g=1$}
Now we would like to compute the determinant ${\det}'\Delta_A$ explicitly on the torus. We put the connection $\ga$ in the previous section to be a flat connection, namely, we specify that holonomy as in Eq.\ref{finite_reduction}
\bea A_{\bar z}=ud\bar z,~~~~u\in\FR{h}\nn\eea
and effectively we have reduced our problem to the case of a $U(1)$ bundle. Indeed, for a field $\phi\in V^0=\Omega^{(0,0)}(\Sigma,ad)$, one may decompose
\bea \phi=\sum_{H_a\in \FR{h}}\phi^aH_a+\sum_{\ga>0}(\phi^{\ga}t_{\ga}+\phi^{-\ga}t_{-\ga}),\nn\eea
where $\FR{h}$ is the CSA and $\alpha>0$ are the positive roots. In this decomposition, $\phi^h$ have trivial holonomy, while $\phi^{\ga}$ has holonomy
\bea  \phi^{\ga}(z+m+\tau n)=\phi^{\ga}e^{m\bra u_1,\ga\ket}e^{n\bra u_2,\ga\ket}\nn\eea
without any mixing with each other.

Now consider a $U(1)$ bundle over a torus, with holonomy $u=u_1-\tau u_2$, with determinant ${\det}'\Delta_u$, then the determinant we would like to compute for the gauge theory problem is the product
\bea {\det} '\Delta_A=\prod_{\ga>0}\big|{\det}'\Delta_{\bra u,\ga\ket}\big|^2\cdot\big({\det}'\Delta_{0}\big)^r,\label{det_assemble}\eea
where the last factor comes from $\phi^h$ and $r$ is the rank of the gauge group.

We see that we just need to compute ${\det}'\Delta_u$ for a $U(1)$ bundle, which is left to the appendix, the result is Eqs.\ref{det_non_zero}, \ref{det_zero}
\bea
{\det}'\Delta_u&=&\exp\big(\frac{\pi}{2\tau_2}(u-\bar u)^2\big)\big|{\calligra det}'\bar\partial_u\big|^2,\nn\\
{\calligra det}'\bar\partial_u&=&\exp\big(i\pi u+\frac{i\pi\tau}{6}\big)\prod_{n=-\infty}^{\infty}\Big(1-\exp2\pi i\big(|n|\tau-\epsilon_nu\big)\Big)\nn\\
{\det}'\Delta_0&=&\tau_2\big|{\calligra det}'\bar\partial\big|^2,\nn\\
{\calligra det}'\bar\partial&=&\exp\big(\frac{i\pi\tau}{6}\big)\prod_{n\neq 0}\big(1-\exp(2\pi i|n|\tau)\big)\nn,\eea
where $u=u_1-\tau u_2$ and $0<u_2<1$. One should compare the first equation with Eq.\ref{temp_11}.

{\color{black}Assembling all these together according to Eq.\ref{det_assemble}, one first sees that the metric is corrected by
\bea \Big(\tau^r_2\cdot\exp\big(\frac{\pi}{2\tau_2}\sum_{\ga>0}(u-\bar u,\ga)^2\big)\Big)^{1/2}=\exp\Big(\frac{r}{2}\log\tau_2+\frac{h\pi}{2\tau_2}(u-\bar u)^2\Big).\nn\eea}
If one uses this metric to compute the Chern-class, one recovers the abstract computation Eq.\ref{Quillen}!

Secondly, one  sees that the wave function is corrected by (the square root of)
\bea &&{\calligra det}'(\bar\partial+\SF{A})=\big({\calligra det}'\bar\partial\big)^r\prod_{\alpha>0}{\calligra det}'\bar\partial_{\bra\ga,u\ket}{\calligra det}'\bar\partial_{\bra-\ga,u\ket}\nn\\
&=&\exp\Big(\frac{i\pi\tau}{6}(r+\sum_{\ga>0}2)\Big)\prod_{\alpha>0}\Big\{e^{2i\pi\ga}(1-e^{-2i\pi \alpha})^2\prod_{n>0}
\big(1-e^{2i\pi \alpha}q^n\big)^2\big(1-e^{-2i\pi \alpha}q^n\big)^2\Big\}\prod_{n>0}(1-q^n)^{2r},\nn\eea
where $q=e^{2i\pi\tau}$ and also to avoid clutter, I have written $\ga$ instead of $(\ga,u)$ on the exponentials. One can rewrite the product as
\bea {\calligra det}'(\bar\partial+\SF{A})&=&\tilde\Pi(\tau,u)^2,\nn\\
\tilde\Pi(\tau,u)&=&q^{|G|/24}\Delta\Big\{\prod_{\alpha>0}\prod_{n>0}
\big(1-e^{2i\pi \alpha}q^n\big)\big(1-e^{-2i\pi \alpha}q^n\big)\Big\}\prod_{n>0}(1-q^n)^{r},\nn\eea
where $\Delta$ is the Weyl denominator.

Here I need to invoke the \emph{MacDonald identity}, which I do not know how to prove
\bea \theta^-_{\rho,h}(\tau,u)=\tilde\Pi(\tau,u)\label{Macdonald},\eea
where $\theta^-$ was defined in Eq.\ref{Weyl_theta}.

But we can look at the two sides of Eq.\ref{Macdonald} at lowest order of $q$, the rhs simply gives $q^{|G|/24}\Delta$, while the double sum over $\Lambda^R$ and $W$ in $\theta^{-}_{\rho,h}$ can be rewritten as
\bea &&\sum_{w,\ga}(-1)^w\exp\big({i\pi\tau h|\alpha+\rho/h|^2+2i\pi w(h\alpha+\rho)}\big)=\sum_{w,\ga}(-1)^w q^{\frac{h}{2}|\ga|^2+\bra\ga,\rho\ket+\frac{|G|}{24}}e^{2i\pi w(h\alpha+\rho)},\nn\eea
where we have used Freudenthal's strange formula
\bea \frac{|\rho|^2}{2h}=\frac{|G|}{24}.\label{Freudenthal}\eea
In this expression, the lowest power possible for $q$ is when $\ga=0$, and we have
\bea q^{\frac{|G|}{24}}\sum_{w,\ga}(-1)^w e^{2i\pi w(\rho)}=q^{\frac{|G|}{24}}\Delta,\nn\eea
as is given by the Weyl denominator formula Eq.\ref{Weyl_denominator}.

To summarize, we have seen that the corrected wave function is the product of the Weyl theta functions $\theta^-_{\rho,h}\theta^+_{\gamma,k}$. One can expand this function
as the level $k+h$, weight $\rho+\gamma$ anti-invariant theta functions $\theta^-_{\gamma+\rho,k+h}(\tau;u)$; and the inner product is given by
\bea \bra \theta^-_{\ga+\rho,k+h},\theta^-_{\gb+\rho,k+h}\ket=\int \Big(\frac{dud\bar u}{2i\tau_2}\Big)^r
e^{\frac12r\log\tau_2+\frac{(h+k)\pi}{2\tau_2}(u-\bar u)^2}(\theta^-_{\ga+\rho,k+h})^*\theta^-_{\gb+\rho,k+h},\nn\eea
where $u=u_1-\tau u_2$ and $u_{1,2}$ are integrated in one fundamental cell of the root lattice.

The inner product for a general theta function at level $k$ is
\bea \bra \theta_{\ga,k},\theta_{\gb,k}\ket=\int \Big(\frac{dud\bar u}{2i\tau_2}\Big)^r
e^{\frac12r\log\tau_2+\frac{k\pi}{2\tau_2}(u-\bar u)^2}(\theta_{\ga,k})^*\theta_{\gb,k},\label{inner_product}\eea
The theta functions are orthogonal under this pairing. To see this,
one looks at
\bea \theta^*_{\gd,k}\theta_{\lambda,k}=\sum_{\ga,\gb\in\Lambda^R}\exp\Big\{i\pi k\tau(\ga+\lambda/k)^2-i\pi k\bar\tau(\gb+\gd/k)^2+2i\pi k (u,\ga+\lambda/k)-2i\pi k (\bar u,\gb+\gd/k)\Big\}.\nn\eea
Performing the $u_1$ integral in this expression will force $\ga+\lambda/k=\gb+\gd/k$, which is fulfilled only when $\lambda=\gd$. Assuming this one has
\bea\bra\theta_{\lambda,k},\theta_{\lambda,k}\ket&=&\int d^ru_2~e^{\frac12r\log\tau_2-2k\pi \tau_2u_2^2}\sum_{\ga\in\Lambda^R}\exp\Big\{-2\pi k\tau_2(\ga+\lambda/k)^2+4\pi k \tau_2(u_2,\ga+\lambda/k)\Big\}\nn\\
&=&\sum_{\ga\in\Lambda^R}\int d^ru_2~\tau^{r/2}_2\exp\big\{-2\pi k\tau_2(u_2-\ga-\lambda/k)^2\big\}\nn\\
&=&\int_{\BB{R}^r} d^ru_2~\tau^{r/2}_2\exp\big\{-2\pi k\tau_2(u_2-\lambda/k)^2\big\}=\big(\frac{1}{2k}\big)^{r/2}.\nn\eea
In the last step, I have combined the integral over one fundamental cell of $\Lambda^R$ and the sum over shifts of $\Lambda^R$ into an integration over $\BB{R}^r$.

In the end one gets the inner product
\bea\bra\theta_{\ga,k},\theta_{\gb,k}\ket=\big(\frac{1}{2k}\big)^{r/2}\delta_{\ga,\gb},~~~~~\ga,\gb\in \Lambda^W/k\Lambda^R.\label{orthogonality}\eea
The orthogonality of $\theta^-$ is similar
\bea\bra\theta^-_{\ga,k},\theta^-_{\gb,k}\ket=|W|\big(\frac{1}{2k}\big)^{r/2}\delta_{\ga,\gb},~~~~~\ga,\gb\in \textrm{ALC}_k,\label{orthogonality_anti}\eea
where $\textrm{ALC}_k$ is the Weyl alcove to be introduced shortly.

Now we discuss what are the independent set of weights that labels $\theta^-_{\gc+\rho,k+h}$. By definition, the theta function $\theta^-_{\gc,k}$
is anti-invariant under the Weyl group, and invariant under the shift $\gamma\to\gamma+k\Lambda^R$, as can be seen from Eq.\ref{Weyl_theta}.
To choose a representative $\gc$ in the orbit of the affine Weyl group $W\ltimes k\Lambda^R$, one can demand $\gc$ to be in the fundamental region of the Weyl group, which is the closure of the fundamental Weyl chamber, defined as
\bea \textrm{FWC}:~~\{\gc\in\Lambda^W|(\gc,\ga)>0,~\forall\, \textrm{simple root}~\ga\}.\label{weyl_chamber}\eea
Taking the closure $\overline{\textrm{FWC}}$ allows $\geq0$ in the definition. In fact, one can restrict $\gc$ to be in FWC. The reason is that on the boundary of FWC, there must exist a simple root $\ga$ s.t. $(\gc,\ga)=0$, thus the Weyl reflection off of $\ga$ leaves $\gc$ invariant $\sigma_{\ga}(\gc)=\gc$. But $\sigma_{\ga}(\gc)$ and $\gc$ contribute to the sum with opposite signs, which means $\theta^-_{\gc,k}=0$ if $\gc$ is on the boundary of FWC.

To fix the shift symmetry, we define first the fundamental Weyl $k$-alcove
\bea \textrm{ALC}_k=\{\gb\in \overline{\textrm{FWC}}|(\gb,\ga_h)<k\},\label{weyl_alcove}\eea
where $\ga_h$ is the highest root. Taking the closure $\overline{\textrm{ALC}}_k$ allows $\leq$ in the above definition.
Now one fixes the shift symmetry by demanding $\gc$ to lie in the interior $\textrm{ALC}_k$.

Thus $\theta^-_{\gb+\rho,k+h}$ is labelled by the $\gb+\rho$ in the interior of the $k$-alcove.
Furthermore, we notice that the quadratic Casimir of a representation with weight $\gb$ is
\bea C_2(\gb)=\frac12\big(|\gb+\rho|^2-|\rho|^2\big).\label{quad_casim}\eea
Since $h^{\vee}$ is the quadratic Caimir of the adjoint representation it is given by
\bea h^{\vee}=\frac12\big(|\ga_h|^2+2(\ga_h,\rho)\big)=(\ga_h,\rho)+1.\nn\eea
This shows that if $\gb\in\overline{\textrm{ALC}}_k$, then
\bea (\gb+\rho,\ga_h)\leq k+h-1,\nn\eea
that is, $\gb+\rho \in \textrm{ALC}_{(k+h)}$.
Finally we conclude that the independent weights that label $\psi_{\gb,k}$ are
\bea \overline{\textrm{ALC}}_k=\{\gb\in \overline{\textrm{FWC}}|(\gb,\ga_h)\leq k\}.\label{integrable_rep}\eea
In particular, there is only one vector in $\textrm{ALC}_h$.
\begin{example} For $SU(2)$, we have the weights
\bea \gb=\frac n2(1,-1),~~~0\leq n\leq k.\nn\eea
And in the language familiar to us, we have representations of spin
\bea \textrm{spin}=0,~\frac12,~1,\cdots,~\frac k 2,\label{integrable_SU2}\eea
a total of $k+1$ states.
\end{example}

In practice, it is more convenient to shuffle the theta functions $\theta^-_{\rho,h}$ to the metric, and let the wave function be the ratio
\bea ch_{\gc,k}(\tau,u)=\frac{\theta^-_{\gc+\rho,k+h}(\tau,u)}{\theta^-_{\rho,h}(\tau,u)}\label{wave_function},\nn\eea
which is actually the basis of the character functions of the affine Lie algebra with highest weight $\gamma$ and level $k$. The reason for doing this is: 1. the character is a modular function while the theta functions are only modular forms, 2. the K\"ahler quantization depends on the complex structure, and one can identify the Hilbert space obtained for different complex structures by using a certain flat connection, under which the above character function is parallel. For details see ref.\cite{AxelrodPietraWitten}, sec.5.b.

\subsection{Surgery Matrices}
With the explicit form of the wave functions, one can see how the Hilbert space respond to the $SL(2,\BB{Z})$ matrices.

The $T$ surgery as defined in Eq.\ref{SL2Z_matrices} acts on the wave function by a shift $\tau\to \tau+1$. Note that $\ga\cdot\gamma$ is an integer and for simply laced groups the norm $|\ga|^2$ is normalized to be 2, so the effect of $T$ is simply a phase
\bea T\theta_{\gc,k}(\tau;u)=\exp\big(\frac{i\pi}{k}|\gamma|^2\big)\theta_{\gc,k}(\tau,u).\nn\eea
Thus $T$ acts on the wave function as
\bea T\psi_{\gc,k}=\exp\big(\frac{i\pi}{k+h}|\gamma+\rho|^2-\frac{i\pi}{h}\rho^2\big)=\exp\big(\frac{2i\pi}{k+h}C_2(\gc)-\frac{2i\pi c}{24} \big)\label{T_matrix},\eea
where $c$ is
\bea c=\frac{\dim G\,k}{k+h},\label{central_charge}\eea
it is also the central charge of the WZW model.

The $S$ duality acts as $\tau\to-1/\tau$ and $u\to u/\tau$, one can use a Possoin re-summation to restore $\tau$ to the numerator in front of the quadratic term in the exponential. In the process of re-summation, the sum over the root lattice $\Lambda^R$ becomes the sum over the dual lattice, which we replace with the weight lattice $\Lambda^W$ for simply laced groups
\bea \theta_{\gc,k}(-1/\tau;u/\tau)=\big(\frac{-i\tau}{k}\big)^{r/2}\Big(\frac{\textrm{vol}\Lambda^W}{\textrm{vol}\Lambda^R}\Big)^{1/2}
\sum_{\gb\in \Lambda^W}\exp\Big(\frac{i\tau\pi}{k}\gb^2+\frac{i\pi k}{\tau}u^2-\frac{2i\pi}{k}\gc\cdot\gb+2i\pi u\cdot\gb\Big),\nn\eea
where $r$ is the rank of the Lei algebra. We shall also break the sum
\bea \sum_{\gb\in\Lambda^W}=\sum_{\gb^*\in\Lambda^W/k\Lambda^R}\sum_{\ga\in k\Lambda^R},\nn\eea
leading to
\bea \theta_{\gc,k}(-1/\tau;u/\tau)=\big(\frac{-i\tau}{k}\big)^{r/2}\Big(\frac{\textrm{vol}\Lambda^W}{\textrm{vol}\Lambda^R}\Big)^{1/2}
\exp\big(\frac{i\pi k}{\tau}u^2\big)\sum_{\gb^*\in \Lambda^W/k\Lambda^R}\exp\big(-\frac{2i\pi}{k}\gc\cdot\gb^*\big)
\theta_{\gb^*,k}(\tau,u).\nn\eea
Using this and the orthogonality relation Eq.\ref{orthogonality}, one can derive
\bea \bra S\theta_{\gd,k},S\theta_{\gc,k}\ket&=&\big(\frac{1}{2k}\big)^{r/2}k^{-r}\Big(\frac{\textrm{vol}\Lambda^W}{\textrm{vol}\Lambda^R}\Big)
\sum_{\gb^*\in \Lambda^W/k\Lambda^R}\exp\big(\frac{2i\pi}{k}(\gd-\gc)\cdot\gb^*\big)\nn\\
&=&\big(\frac{1}{2k}\big)^{r/2}k^{-r}\Big(\frac{\textrm{vol}\Lambda^W}{\textrm{vol}\Lambda^R}\Big)
\Big(\frac{\textrm{vol}\,k\Lambda^R}{\textrm{vol}\Lambda^W}\Big)\delta_{\gd,\gc}=\bra \theta_{\gd,k},\theta_{\gc,k}\ket.\nn\eea
We obtain the important fact that the inner product defined in Eq.\ref{inner_product} is \emph{modular invariant}, and equivalently $S$ acts unitarily w.r.t to the inner product.

The corresponding transformation for $\theta^-$ is
\bea \theta^-_{\gc,k}(-1/\tau;u/\tau)=\big(\frac{-i\tau}{k}\big)^{r/2}\Big(\frac{\textrm{vol}\Lambda^W}{\textrm{vol}\Lambda^R}\Big)^{1/2}\exp\big(\frac{i\pi k}{\tau}u^2\big)\sum_{\gb^*\in C_k}\sum_{w\in W}(-1)^w\exp\big(-\frac{2i\pi}{k}w(\gc)\cdot\gb^*\big)
\theta^-_{\gb^*,k}(\tau,u),\nn\eea
where $C_k$ is the interior of the Weyl $k$-alcove defined in Eq.\ref{weyl_alcove}.

The function $\theta_{\rho,h}^-$ is special under modular transformation
\bea \theta^-_{\rho,h}(-1/\tau;u/\tau)=\big(\frac{-i\tau}{h}\big)^{r/2}\Big(\frac{\textrm{vol}\Lambda^W}{\textrm{vol}\Lambda^R}\Big)^{1/2}\exp\big(\frac{i\pi h}{\tau}u^2\big)\sum_{w\in W}(-1)^w\exp\big(-\frac{2i\pi}{h}w(\rho)\cdot\rho\big)
\theta^-_{\rho,h}(\tau,u),\nn\eea
as there is only one vector $\rho$ in the $h$-alcove. Thus $\theta_{\rho,h}^-$ transforms with a mere multiplicative factor
\bea \theta^-_{\rho,h}(-1/\tau;u/\tau)=\big(\frac{-i\tau}{h}\big)^{r/2}\Big(\frac{\textrm{vol}\Lambda^W}{\textrm{vol}\Lambda^R}\Big)^{1/2}\exp\big(\frac{i\pi h}{\tau}u^2\big)\Delta\big(-\frac{2i\pi}{h}\rho\big)\theta^-_{\rho,h}(\tau,u),\nn\eea
where the Weyl denominator formula is used. One should be able to evaluate the factor involving $\Delta$ directly, but in ref.\cite{KacsPeterson}, an indirect method is used. Notice that one can fix $\Delta(-2i\pi\rho/h)$ up to a positive multiplicative factor
\bea \Delta(-\frac{2i\pi}{h}\rho)=(-2i)^{(|G|-r)/2}\prod_{\ga>0}\sin\big(\frac{\pi}{h}(\ga,\rho)\big),\nn\eea
the factor involving sinus is positive as all the arguments in the sinus are within $(0,\pi)$. Combine this with the fact that $S$ acts unitarily
\bea \bra\theta_{\rho,h},\theta_{\rho,h}\ket=\bra S\theta_{\rho,h},S\theta_{\rho,h}\ket\nn\eea
one can fix the $\Delta$ term as
\bea i^{(r/2-|G|)}h^{r/2}\Big(\frac{\textrm{vol}\Lambda^R}{\textrm{vol}\Lambda^W}\Big)^{1/2}.\nn\eea
Thus the formula for the $S$ matrix is
\bea S_{\ga,\gb}=\frac{i^{(|G|-r)/2}}{(k+h)^{r/2}}\left|\frac{\textrm{vol}\Lambda ^W}{\textrm{vol}\Lambda ^R}\right|^{1/2}\sum_w(-1)^w\exp\big(-\frac{2i\pi}{k+h}( w(\ga+\rho),\gb+\rho)\big)\label{S_matrix}.\eea
which in the case of $G=SU(2)$, we have
\bea \ga+\rho=\frac{a}{2}(1,-1)+\frac12(1,-1),\nn\eea
and the Weyl relection exchanges 1 and $-1$. Thus
\bea S_{a,b}&=&\frac{i}{(k+2)^{1/2}}\left|\frac{1}{2}\right|^{1/2}\Big(\exp\big(-\frac{2i\pi}{k+2}\frac{(a+1)(b+1)}{2}\big)-\exp\big(\frac{2i\pi}{k+2}\frac{(a+1)(b+1)}{2}\big)\Big)\nn\\
&=&\sqrt{\frac{2}{k+2}}\sin\big(\frac{\pi(a+1)(b+1)}{k+2}\big)\label{S_matrix_SU2}.\eea
Recalling Eq.\ref{integrable_SU2}, one can easily show directly that $S_{a,b}$ is unitary.

%% file: SimpleComputation.tex
With this much information, one can compute the partition function of Chern-Simons theory with gauge group $SU(2)$.

Physically, the states $\psi_{\gc,k}$ in the Hilbert space can be described by placing a Wilson-line inside the torus along the b-cycle (the $\tau_2$ direction) transforming in the representation of weight $\gc$. Hence the partition function on $S^3$ for $SU(2)$ CS theory at level $k$ is given by the matrix element
\bea Z_{S^3}=\bra\psi_{0,k}|S|\psi_{0,k}\ket=S_{00}=\sqrt{\frac{2}{k+2}}\sin\big(\frac{\pi}{k+2}\big).\label{part_fun_S3}\eea

\begin{figure}
\begin{center}
\begin{tikzpicture}[scale=.6]
\draw [blue,fill=blue!40,opacity=.4] (0,0) circle (1.4);
\draw [dashed,blue] (-1.4,0) to [out=90, in=90] (1.4,0);
\draw [-,blue] (-1.4,0) to [out=-90, in=-90] (1.4,0);
\draw [brown, rotate = 15](0,0) ellipse (1 and 0.5);
\node at (1,1) [right] {\small$\cup\{\infty\}$};
\end{tikzpicture}
\caption{An unknot embedded in $S^3$}\label{fig_one_W_loop}
\end{center}
\end{figure}
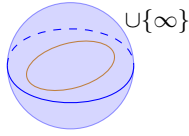

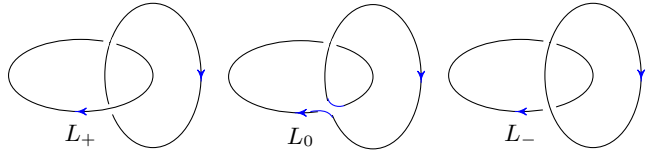
\begin{figure}[h]
\begin{center}
\begin{tikzpicture}[scale=.8]
\draw[rotate = 0](1.2,0) ellipse (.8 and 1.2)[color=black];
\draw [-<,blue](2,-0.1cm) to [out=90, in=-90] (2,0.1cm);
\draw[white, line width=1.8mm] (.46,-.48cm) arc (190:205:.7cm);

\draw[rotate = 0](0,0) ellipse (1.2cm and 0.6cm)[color=black];
\draw [-<,blue](-0.1,-0.6cm) to [out=0, in=-180] (0.1,-0.6cm);
\draw[white, line width=.8mm] (0.41,0.50cm) arc (170:155:.7cm);
\draw[white, line width=.8mm] (0.52,0.44cm) arc (170:155:.7cm);

\node at (0,-1) {\small$L_+$};
\end{tikzpicture}~~
\begin{tikzpicture}[scale=.8]
\draw[rotate = 0](1.2,0) ellipse (.8 and 1.2)[color=black];
\draw [-<,blue](2,-0.1cm) to [out=90, in=-90] (2,0.1cm);

\draw[rotate = 0](0,0) ellipse (1.2cm and 0.6cm)[color=black];
\draw [-<,blue](-0.1,-0.6cm) to [out=0, in=-180] (0.1,-0.6cm);
\draw[white, line width=.8mm] (0.41,0.50cm) arc (170:155:.7cm);
\draw[white, line width=.8mm] (0.52,0.44cm) arc (170:155:.7cm);

\draw[white, line width=10pt] (.46,-.40cm) arc (190:210:.7cm);

\draw[blue] (.53,-.68) to [out=120, in=10] (.16,-.58);
\draw[blue] (.45,-.36cm) to [out=-80, in=200] (.72,-0.48);

\node at (0,-1) {\small$L_0$};
\end{tikzpicture}~~
\begin{tikzpicture}[scale=.8]
\draw[rotate = 0](0,0) ellipse (1.2cm and 0.6cm)[color=black];
\draw [-<,blue](-0.1,-0.6cm) to [out=0, in=-180] (0.1,-0.6cm);
\draw[white, line width=.8mm] (0.41,0.50cm) arc (170:155:.7cm);
\draw[white, line width=.8mm] (0.52,0.44cm) arc (170:155:.7cm);
\draw[white, line width=1.8mm] (.46,-.48cm) arc (190:205:.7cm);

\draw[rotate = 0](1.2,0) ellipse (.8 and 1.2)[color=black];
\draw [-<,blue](2,-0.1cm) to [out=90, in=-90] (2,0.1cm);

\node at (0,-1) {\small$L_-$};
\end{tikzpicture}\caption{Un-skeining the Hopf link}\label{fig_Hopf}
\end{center}
\end{figure}

One can also compute the expectation value of Wilson lines. If there is only one Wilson line of spin 1/2 trivially embedded in $S^3$, see fig.\ref{fig_one_W_loop}.
Then the expectation value is
\bea \bra\bigcirc\ket=\frac{Z_{L_{1,0}}}{Z_{S^3}}=\frac{S_{10}}{S_{00}}=q^{1/2}+q^{-1/2},~~~~q=\exp\frac{2i\pi}{k+2}.\label{Wilson_loop_norm}\eea
In CS theory, the expectation value of a knot, every component of which is labelled by a representation, is invariant under isotopy of the knot, so long as no two strands cross each other during the isotopy. This is the field theory construction of knot invariants in the ground breaking paper \cite{Witten_Jones}.
One may compute the associated invariant of a Hopf link (the first picture in fig.\ref{fig_Hopf}), labelled by spin 1/2
\bea \bra \textrm{Hopf}\ket=\frac{\bra \psi_{1,k}|S|\psi_{1,k}\ket}{S_{00}}=\frac{S_{11}}{S_{00}}
=(q^{1/2}+q^{-1/2})(q+q^{-1}).\label{Z_Hopf}\eea

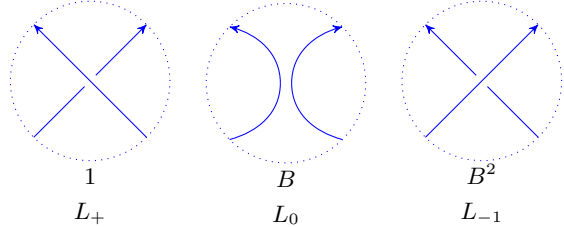
\begin{figure}[h]
\begin{center}
\begin{tikzpicture}[scale=.75]
\draw [->,blue] (-1,-1)--(1,1);
\draw [-,line width=6pt,draw=white] (1/2,-1/2)--(-1/2,1/2);
\draw [->,blue] (1,-1)--(-1,1);
\draw [dotted,blue] (0,0) circle (1.414);
\node (a) at (0,-2cm) [above]{\small $1$};
\node (b) at (0,-2cm) [below]{\small $L_+$};
\end{tikzpicture}
~~~%
\begin{tikzpicture}[scale=.75]
\draw [<-,blue] (-1,1) to [out=-15,in=90] (-.1,0);
\draw [-,blue] (-.1,0) to [out=-90,in=15] (-1,-1);
\draw [<-,blue] (1,1) to [out=195,in=90] (.1,0);
\draw [-,blue] (.1,0) to [out=-90,in=165] (1,-1);
\draw [dotted,blue] (0,0) circle (1.414);
\node (a) at (0,-2cm) [above]{\small $B$};
\node (b) at (0,-2cm) [below]{\small $L_0$};
\end{tikzpicture}
~~~%
\begin{tikzpicture}[scale=.75]
\draw [->,blue] (1,-1)--(-1,1);
\draw [-,line width=6pt,draw=white] (1/2,1/2)--(-1/2,-1/2);
\draw [->,blue] (-1,-1)--(1,1);
\draw [dotted,blue] (0,0) circle (1.414);
\node (a) at (0,-2cm) [above]{\small $B^2$};
\node (b) at (0,-2cm) [below]{\small $L_{-1}$};
\end{tikzpicture}
\caption{Applying braiding operation to obtain skein relations}\label{fig_skein}
\end{center}
\end{figure}

As the knot gets more complicated, a direct computation becomes harder, yet one can instead try to obtain the so called skein relation of a knot. The fig.\ref{fig_skein}
shows the vicinity of a point where two strands are close. If there is a relation of type
\bea L_++\lambda L_0+ \kappa L_{-}=0,\nn\eea
then one may use this relation iteratively to untie a knot until it becomes a collection of non-linking unknots.
Kauffman pointed out that the skein relation above in conjunction with the normalization of one unknot $\bra \bigcirc\ket$ completely fixes a knot invariant. Of course not all choices $\lambda$ $\kappa$ are consistent, using the result of Kauffman (see e.g ref.\cite{Kauffman_CRM_workshop}), the following is a necessary condition
\bea \frac{\lambda^2\kappa}{(\kappa^2-1)^2}=-1.\label{Kauffman}\eea

\begin{remark}
What one should do seriously to obtain the skein relation is to quantize CS theory on a sphere with 4 marked points. The Techm\"uller space of such a space is 2 dimensional, namely, the cross ratio $r$ of the position of the two points on the sphere. Then one can identify the Hilbert space at different $r$ by using a flat connection, the Knizhnik-Zamolodchikov connection. This then tells one what happens to the Hilbert space when two punctures, say, 3 and 4, revolve around each other. It is clear from fig.\ref{fig_KZ} that when 3 and 4 revolve around each other counter clockwise by $\pi$ one passes from $L_+$ to $L_0$, another $\pi$ takes one to $L_-$. The corresponding integral of the KZ connection gives one the coefficients $A,B$ and $C$.
\begin{figure}
\begin{center}
\begin{tikzpicture}[scale=.8]
\draw [blue, fill=blue!40,opacity=.3] (0,0) circle (1.4);
\draw [dashed,blue] (-1.4,0) to [out=90, in=90] (1.4,0);
\draw [-,blue] (-1.4,0) to [out=-90, in=-90] (1.4,0);
\draw [brown] (-0.7,-1) to [out=80,in=-120] (-0.1,-0.1);
\draw [brown] (0.1,0.1) to [out=60,in=-100] (0.7,1);
\draw [brown] (0.7,-1) to [out=100,in=-80] (-0.7,1);
\node at (-0.7,-1.2) {\small $1$};
\node at (0.7,-1.2) {\small $2$};
\node at (0.7,1.2) {\small $3$};
\node at (-0.7,1.2) {\small $4$};
\end{tikzpicture}\caption{Sphere with 4 marked points}\label{fig_KZ}
\end{center}
\end{figure}
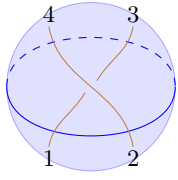
The KZ connection can also be derived in the context of conformal field theory see ref.\cite{Moore:1988uz}, and this is how the skein relation was obtained in ref.\cite{Witten_Jones}. But as I am not going to touch WZW model in these notes, I will try to get the skein relation by stretching our existing knowledge a little, and I offer my apology for the sloppiness in the discussion.
\end{remark}

We know that two Wilson loops parallel to each other can be fused into one Wilson loop of spin 0 or 1 (provided $k\geq2$, otherwise there is only the spin 0 state). And the $T$ matrix which adds one twist to the Wilson loop is given by a pure phase
\bea T=q^{C_2(r)}e^{\frac{2i\pi c}{24}},\nn\eea
where $C_2(r)$ is the quadratic Casimir of the relevant representation. The phase factor involving the central charge is due to the framing change from the surgery, and has nothing to do with the knot, so only the first phase factor matters. We have the table
\bea \begin{array}{c|c|c|}
& \textrm{spin}~1 & \textrm{spin}~0 \\
\hline
        2\pi & q^{C_2(2)-C_2(1)} & q^{C_2(0)-C_2(1)} \\
\hline  \pi &  q^{(C_2(2)-C_2(1))/2} & -q^{(C_2(0)-C_2(1))/2}
\end{array},\nn\eea
where $C_2(2)=2$, $C_2(1)=3/4$ and $C_2(0)=0$. And I have inserted a minus sign to the lower right entry, which come from the fact that spin 0 is the anti-symmetric combination of the two spin 1/2.
Thus we know that in this basis, the braiding matrix is diagonal
\bea B=\begin{array}{|cc|}
         -q^{-3/4} & 0 \\
         0 & q^{1/4} \end{array}.\nn\eea
And in a general basis, $B$ satisfies its own characteristic equation
\bea B^2-B\Tr[B]+\det B=0~\Rightarrow B^2-B(q^{1/4}-q^{-3/4})-q^{-1/2}=0.\nn\eea
Applying this equation to $L_+$ one gets the relation
\bea L_--L_0(q^{1/4}-q^{-3/4})-L_+q^{-1/2}=0,\label{skein_relation}\eea
which is the skein relation we seek. Note that we have now $\lambda=q^{3/4}-q^{-1/4}$ and $\kappa=-q^{1/2}$, and they satisfy the condition Eq.\ref{Kauffman}.

Next we use this relation to recompute the Hopf link as a check. Apply the skein relations to fig.\ref{fig_Hopf}
Amongst the three knots, we already know the value of the last one, it is
\bea L_-=\bra\bigcirc\ket^2=(q^{1/2}+q^{-1/2})^2.\nn\eea
The middle one is topologically an unknot, however, it has one more (left hand) twist compared to the unknot, thus its value is
\bea L_0=q^{-3/4}\bra\bigcirc\ket=q^{-3/4}(q^{1/2}+q^{-1/2}).\nn\eea
The value of $L_0$ can also be computed as the matrix element
\bea L_0=\frac{\bra\psi_0|ST^{-1}|\psi_{1}\ket}{\bra\psi_0|ST^{-1}|\psi_{0}\ket},\nn\eea
since the surgery $ST^{-1}$ exactly produces an unknot with one left hand twist.

From these information and the skein relation one has
\bea L_+=q^{1/2}\Big(L_--L_0(q^{1/4}-q^{-3/4})\Big)=q^{3/2}+q^{1/2}+q^{-1/2}+q^{-3/2},\nn\eea
agreeing with the direct computation Eq.\ref{Z_Hopf}.

One sees from the above manipulation that, the knot invariant directly defined from the CS theory is sensitive to the twisting, or in other words, the framing of the knot. One can define a different knot invariant by manually removing this framing dependence
\bea q^{3/2}L_--q^{3/4}L_0(q^{1/4}-q^{-3/4})-L_+q^{-1/2}=0,\label{skein_relation_Jones}\eea
where we have inserted the phase $q^{3/2},~q^{3/4}$ in front of $L_-$ and $L_0$ to offset the twisting incurred by applying the braiding $B^2$ and $B$. this new skein relation is the skein relation of the Jones  polynomial.

\subsection{Reparation of Omission}
There is no discussion of Chern-Simons theory without mentioning also the Wess-Zumino-Witten model. So before leaving the non-perturbative first part of the notes, let me just give some references, and most of the features of quantization problem discussed can be understood from the WZW point of view. Needless to say, the references given here is not meant to be complete.

First of all, the Hilbert space of CS theory equivalent to the conformal blocks of the WZW model. In ref.\cite{Elitzur:1989nr}, the wave function of CS theory was written as a path integral of the gauged WZW model, and besides, the quantization problem is worked out in detail for the disc (with or without punctures), the annulus and the torus, which contains all the discussion of sec.\ref{NFDR}. In ref.\cite{Witten:Hol_fact_WZW}, many features of the wave function explained in the tome \cite{AxelrodPietraWitten} was understood neatly from the formal manipulations of the path integral of WZW model. In ref.\cite{Moore:1988uz}, the KZ connection was derived and it was shown that the conformal blocks satisfy the axioms of the so-called modular tensor category, the latter category can be used to construct abstract knot and 3-manifolds invariants, for this one may see the book \cite{BakalovKirillov}. The shift by $h$ of $k$ that appeared prevalently can be alternatively understood as certain anomaly computation (the determinant of a functional change of variable) as done in refs.\cite{Gawedzki:1988hq,Gawedzki:1988nj}. Or it can be understood from the current algebra of WZW model \cite{Knizhnik:1984nr}. In ref.\cite{Gepner:1986wi}, the independent set of wave functions Eq.\ref{integrable_rep} correspond to the primary fields in the integrable representations, and in particular, it was shown that the non-integrable ones decouple by using the Ward identity of WZW model. And finally, the book by Kohno \cite{Kohno} is quite good.

%% file: Perturb_Motivation.tex
The following fig.\ref{Hopf_Link_fig} is a Hopf link embedded in $\BB{R}^3$; a basic topological invariant is its linking number, which is just how many times (counted with sign) one loop penetrates the other.
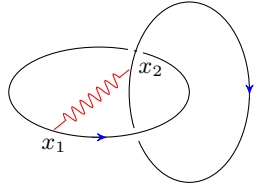
\begin{figure}[h]
\begin{center}
\begin{tikzpicture}
\draw[rotate = 0](1.2,0) ellipse (.8 and 1.2)[color=black];
\draw [-<,blue](2,-0.1cm) to [out=90, in=-90] (2,0.1cm);
\draw[white, line width=1.8mm] (.46,-.48cm) arc (190:205:.7cm);

\draw[rotate = 0](0,0) ellipse (1.2cm and 0.6cm)[color=black];
\draw [->,blue](-0.1,-0.6cm) to [out=0, in=-180] (0.1,-0.6cm);
\draw[white, line width=.8mm] (0.41,0.50cm) arc (170:155:.7cm);
\draw[white, line width=.8mm] (0.52,0.44cm) arc (170:155:.7cm);

\draw[snake=coil, line before snake=2mm, line after snake=1mm, segment aspect=0, segment length=4pt, color=red] (-.60,-0.50cm) -- (0.4,0.3cm);
\node (a) at (-0.6,-.5cm)[below] {\small $x_1$};
\node (b) at (0.4,0.3cm) [right]{\small $x_2$};
\end{tikzpicture}\caption{A Hopf link}\label{Hopf_Link_fig}
\end{center}
\end{figure}
One can give a simple formula for this linking number, which is
\bea L(K_1,K_2)=\int_{K_1}~\int_{K_2}~G(x_1,x_2),~~~\textrm{where}~~~ G(x_1,x_2)=\frac{(\vec x_1-\vec x_2)\times d(\vec x_1-\vec x_2) \times d(\vec x_1-\vec x_2)}{|\vec x_1-\vec x_2|^3}.\label{linking_number}\eea
The propagator $G(x_1,x_2)$, which is the wiggly line in the figure, is actually the inverse of the de Rham differential in $\BB{R}^3$
\bea d^{-1}\psi(x)=\int_y G(x,y)\psi(y),~~~~\psi\in\Omega^{\sbullet}(\BB{R}^3).\nn\eea
This formula Eq.\ref{linking_number} calculates the degree of the Gauss map
\bea K_1\times K_2\to S^2:~~~(x_1,x_2)\to \frac{\vec x_1-\vec x_2}{|\vec x_1-\vec x_2|}\in S^2.\label{gauss_map}\eea
because one clearly sees that $G(x,y)$ is the pull back of the volume form by the Gauss map above.

The integral Eq.\ref{linking_number} is invariant under deformations of the two circles so long as the strands do not cut through each other. In physics, this is known as the 'Amp\`ere's circuital law':
\bea \oint B\cdot dl =I.\nn\eea
Imagine that the right circle is a wire, then the integral of the propagator over $K_2$ gives the magnetic field excited by the current at point $x_1$, and a further integral over $x_1$ gives back the current, which is an invariant.

\begin{figure}[h]
\begin{center}
\begin{tikzpicture}[scale=0.9]
\draw [-,blue] (0:0cm) -- (0:1.25cm);
\draw [-,blue] (0:1.25cm) -- (0:2.5cm);

\draw [-,blue](0:2.5cm) arc (45:90:1.7675cm);
\draw [-,blue](0:0cm) arc (135:90:1.7675cm);

\draw [-,blue](0:2.5cm) arc (-45:-90:1.7675cm);
\draw [-,blue](0:0cm) arc (225:270:1.7675cm);

\node (p) at (0,0) {};
\node (q) at (2.5,0) {};
\draw (p) node [left] {\scriptsize{1}}
 (q) node [right] {\scriptsize{2}};
\end{tikzpicture}\caption{The simplest invariant}\label{fig_cocycle}
\end{center}
\end{figure}
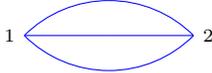

One can form invariants for 3-manifolds by using similar techniques. One lets $G(x,y)$ be the inverse of $d$ for a general 3-manifold, and form integrals like the following
\bea \int_{M}\int_{M} G(x_1,x_2)^3,\nn\eea
where the two points $x_{1,2}$ are integrated over the entire manifold $M$. This integral can be represented by a Feynman graph fig.\ref{fig_cocycle}
Naturally, to form the inverse $d^{-1}$ one needs to introduce some extraneous data such as the metric, yet the integral given above is independent of these extra data, and hence, are 3-manifold invariants. In fact, this integral is the Casson-Walker invariant.

Of course, if one takes an arbitrary Feynman graph, or a collection of Feynman graphs, and form the integral as above, there is no guarantee that the result is going to be independent of the metric. One can use a powerful tool, the graph complex, to single out the invariant combinations, but I shall not go into this subject.

\begin{figure}[h]
\begin{center}
\begin{tikzpicture}[scale=.8]
\draw [semithick,blue] (0,0) circle (1);
\draw [->,blue] (0,1) arc (90:180:1cm);
\draw [-,blue](-.717,.717cm)--(-.5,-.866cm);
\draw [->,blue](-.717,.717cm)--(-.717/2-.5/2,.717/2-.866/2cm);
%
%
\node (p) at (-.717,.717cm) [left] {\small $1$};
\node (q) at (-.5,-.866cm) [left] {\small $2$};
\node (g) at (0,-1) [below] {\small $\Gamma_0$};
\end{tikzpicture}~~
\begin{tikzpicture}[scale=.8]
\draw [semithick,blue] (0,0) circle (1);
\draw [->,blue] (0,1) arc (90:180:1cm);
\draw [-,blue](-.717,.717cm)--(-.5,-.866cm);
\draw [->,blue](-.717,.717cm)--(-.717/2-.5/2,.717/2-.866/2cm);
\draw [-,blue](.5,.866cm)--(.5,-.866cm);
\draw [->,blue](.5,.866cm)--(.5,-.2cm);
\node (p) at (-.717,.717cm) [left] {\small $1$};
\node (q) at (-.5,-.866cm) [left] {\small $2$};
\node (s) at (.5,.866cm) [right] {\small $4$};
\node (r) at (.5,-.866cm) [right] {\small $3$};
\node (g) at (0,-1) [below] {\small $\Gamma_4$};
\end{tikzpicture}~~
\begin{tikzpicture}[scale=.8]
\draw [semithick,blue] (0,0) circle (1);
\draw [->,blue] (0,1) arc (90:180:1cm);
\draw [-,blue](-.717,.717cm)--(.717,-.717cm);
\draw [->,blue](-.717,.717cm)--(.2,-.2cm);
\draw [-,blue](-.866,-.5cm)--(.866,.5cm);
\draw [->,blue](-.866,-.5cm)--(0.4,0.217cm);
\node (p) at (-.717,.717cm) [left] {\small $1$};
\node (q) at (-.866,-.5cm) [left] {\small $2$};
\node (r) at (.717,-.717cm) [right] {\small $3$};
\node (s) at (.866,.5cm) [right] {\small $4$};
\node (g) at (0,-1) [below] {\small $\Gamma_5$};
\end{tikzpicture}~~
\begin{tikzpicture}[scale=.8]
\draw [semithick,blue] (0,0) circle (1);
\draw [->,blue] (0,1) arc (90:180:1cm);
\draw [-,blue](-.717,-.717cm)--(0,0cm);
\draw [->,blue](-.717,-.717cm)--(-.2,-.2cm);
\draw [-,blue](.717,-.717cm)--(0,0cm);
\draw [->,blue](.717,-.717cm)--(.2,-.2cm);
\draw [-,blue](0,1cm)--(0,0cm);
\draw [->,blue](0,1cm)--(0,0.5cm);
\node (p) at (0,0cm) [left] {\small $1$};
\node (q) at (-.717,-.717cm) [left] {\small $2$};
\node (r) at (.717,-.717cm) [right] {\small $3$};
\node (s) at (0,1cm) [above] {\small $1$};
\node (g) at (0,-1) [below] {\small $\Gamma_6$};
\end{tikzpicture}~~
\begin{tikzpicture}[scale=.8]
\draw [semithick,blue] (0,0) circle (1);
\draw [->,blue] (0,1) arc (90:180:1cm);
\draw [->,blue](0,-1cm)--(0,-.5cm);
\draw [->,blue](0,1cm)--(0,.5cm);
\draw [-<,blue](0,-.5cm) to [out=135,in=-90] (-2/7,0cm);
\draw [-,blue](-2/7,0cm) to [out=90,in=-135] (0,0.5cm);
\draw [-<,blue](0,-.5cm) to [out=45,in=-90] (2/7,0cm);
\draw [-,blue](2/7,0cm) to [out=90,in=-45] (0,0.5cm);
\node (p) at (0,0.5cm) [left] {\small $1$};
\node (q) at (0,-.5cm) [above] {\small $2$};
\node (r) at (0,1cm) [above] {\small $1$};
\node (s) at (-.1,-.85cm) [ ] {\small $2$};
\node (g) at (0,-1cm) [below] {\small $\Gamma_7$};
\end{tikzpicture}~~
\caption{Examples of some simple chord diagrams} \label{ex_gph_fig}
\end{center}
\end{figure}
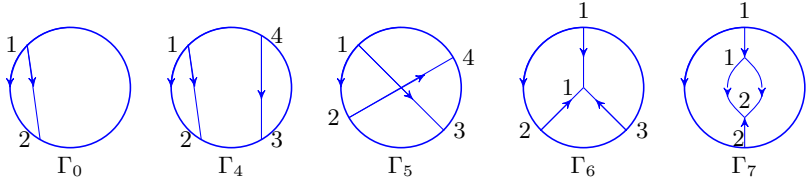

In case of a knot embedded in some ambient 3-manifold (usually taken to be $S^3$), one can use similar strategies to construct knot invariants.
One takes some propagators and strings them onto the knot (meaning the ends are integrated along the entire knot), or one can string the ends of the propagators together and integrate the position of the vertex over the whole 3-manifold. This can again be conveniently depicted by Feynman graphs, as in fig.\ref{ex_gph_fig}, thse graphs are known as the \emph{chord diagrams}.
In these pictures, the big circle represents the knot, of course, one understands that this circle is embedded non-trivially in the 3-manifold.

To summarize, we see that some simple integrals involving propagators and Feynman diagrams are interesting in their own right, without reference to the Chern-Simons theory. Of course, their invariance properties are best understood in the context of perturbation expansion of CS theory, as they can be extracted from the contribution of the saddle point coming from the trivial background connection to the perturbation expansion of CS.

\subsection{Perturbation Expansion, an Overview}
The perturbation expansion in QFT is basically an infinite dimensional Gaussian integral. One expands around the saddle points, which are the solution to the classical
equation of motion. The expansion series then consist of a determinant factor at one loop and Feynamn diagrams at higher loops. Duh, easier said than done!

The solution to the eom of CS theory are the flat connections $a$ on $M$, $da+aa=0$. To perform the expansion
one splits the gauge field into
\bea A\to a+A.\label{splitaA}\eea
One then has to gauge fix the action, which I shall skip, since it is quite standard. 
Define $d_a$ and its adjoint $d_a^{\dagger}$
\bea d_a\omega=d\omega+[a,\omega],~~~d_a^{\dagger}\omega_{(p+1)}=(-1)^{pd+1}*d_a*\omega_{(p+1)}.\nn\eea
The BRST-gauge fixed action is
\bea S_{CS-brst}=S_{CS}(a)+\Tr\int_M~Ad_aA+\frac{2}{3}AAA+2\big(bd_a^{\dagger}A-\bar cd_a^{\dagger}d_ac-\bar cd_a^{\dagger}[A,c]\big),\label{BRST_G_fixed}\eea
and the BRST transformation is
\bea \gd \bar c=b,~~\gd b=0,~~\gd A=d_ac+[A,c],~~\gd c=-\frac12[c,c],\label{BRST_rule}\eea

To help organize the bosonic kinetic term, define an operator $L$
\bea L\omega_{(p)}=(-1)^{p(p-1)/2}(*d_a-(-1)^pd_a*)\omega_{(p)},\label{def_L}\eea
and its restriction to the odd forms
\bea L_-\omega_{(p)}=(-1)^{p(p-1)/2}(*d_a+d_a*)\omega_{(p)}.\nn\eea
The kinetic term for the  bosonic fields (the gauge field $A$ which a 1-form and the Lagrange multiplier $b$ which is a 3-form) can be organized as
\bea S_{kin-bos}=\int_M~(b+A)*L_-(b+A).\nn\eea
We also have
\bea L_-^2(b+A)=-d_a*d_a*b+(*d_a*d_a-d_a*d_a*)A=d_ad_a^{\dagger}b+\{d_a^{\dagger},d_a\}A=\{d_a^{\dagger},d_a\}(b+A).\nn\eea
The kinetic term for the fermions is just the operator $d_a^{\dagger}d_a$.
To the lowest order of perturbation theory, we just have the ratio of the two determinants
\bea \frac{\det d_a^{\dagger}d_a}{\sqrt{\det L_-}},\label{Ray-Singer}\eea
but this is interestingly the most messy part.

First of all, the absolute value of the above determinant is equal to the Ray-Singer analytic torsion, which is explained in sec.\ref{TTWRdRFRS}. Secondly, since one has to take a square root, one has to also define the phase carefully. Consider the following 1-dimensional integral, which is defined by rotating the contour $\pm45^{\circ}$ depending on the sign of $\lambda$.
\bea \int_{\BB{R}}dx~e^{i\lambda x^2}=\bigg\{\begin{array}{c}
                                             e^{i\pi/4},~~~\lambda>0 \\
                                             e^{-i\pi/4},~~~\lambda<0 \end{array}\nn\eea
So the phase is formally equal to
\bea N_{ph}=\frac{\pi}{4}\sum_{\lambda\neq 0}\textrm{sgn}(\lambda),\label{phase_det}\eea
where the summation runs over all the non-zero eigenvalues of $L_-$. This infinite sum has to be regularized, the following regularization is the most elegant
\bea \eta_a(s)=\sum_{\lambda\neq 0}|\lambda|^{-s}\textrm{sgn}(\lambda),\label{eta_fun}\eea
which is called the \emph{eta-invariant}.

For continuity, we finish the story first. Round a saddle point given by a flat connection $a$, the one loop correction is written as (see ref.\cite{Jeffrey} 5.1) 
\bea Z_{1-\textrm{loop}}&=&\sum_a\,Z_a,\nn\\
Z_a&=&\frac{1}{G_a}\Big(\frac{1}{k+h}\Big)^{(\dim H_a^0-\dim H_a^1)/2}\frac{N_{ph}}{\tau_a}\exp\Big(\frac{i(k+h)}{4\pi}CS(a)\Big),\label{general_feature}\\
\textrm{where}&& N_{ph}=\exp\Big(-\frac{i\pi}{4}|G|(1+b^1)-\frac{i\pi}{4}(2I_a+\dim H^0_a+\dim H^1_a)\Big).\nn\eea
{\color{black}In particular when $H^0_a\neq0$, then this is a signal that the holonomy of $a$ is in a subgroup of $G$ only and there can be non-trivial constant gauge transformations that leaves $a$ invariant; this centralizer is denoted $G_a$ above}.

One may compare this result to the calculation done using surgery in sec.\ref{sec_SCatJP}. For $G=SU(2)$ and $S^3$, the only flat connection is the trivial one $a=\theta$ and $\dim H_{\theta}^1=0$ $\dim H_{\theta}^0=3$, so the above formula would predict the asymptotic behavior
\bea Z_{\theta}\sim k^{-3/2},\nn\eea
and this is in accordance with Eq.\ref{part_fun_S3}.
\begin{remark}
In general situation, the comparison of the perturbative result and the surgery result is not as straightforward. One notices that the in surgery coefficient Eq.\ref{S_matrix}, the factor $k+h$ appears in the \emph{denominator} on the exponential, while in Eq.\ref{general_feature} there is the $k+h$ factor \emph{multiplying} the classical action $CS(a)$. So to facilitate the comparison, one needs to resort to a special kind of Poisson re-summation to flip the $1/(k+h)$ in the surgery coefficient to $k+h$, see also the discussion in ref.\cite{Jeffrey}, sec.5. In the example immediately above, we are only spared this step because we are expanding around a trivial connection $\theta$ and $CS(\theta)=0$.
\end{remark}

Sections \ref{TTWRdRFRS}, \ref{TeI} and \ref{2Fo3M} will be devoted to the understanding of $N_{ph}$.

%% file: CD_Alexander.tex
The main computation we would like to do here is the computation of the diagrams of fig.\ref{ex_gph_fig} for reasons already stated. The strategy is similar to the one adopted in the appendix for the computation of the determinant ${\calligra det}'\bar\partial$, namely, one first adds a non-trivial holonomy to the differential forms and then carefully tune the holonomy to zero. The computation of the determinant with a non-trivial holonomy is much simpler than otherwise. The central result is that these diagrams can be written as the second derivative of the Alexander polynomial Eq.\ref{alex_perturb}, and hence are knot invariants (of course, one can argue the invariance of these diagrams in completely different ways, such as by means of the graph complex \cite{Bar-Natan_I}).

To start, one includes a Wilson loop operator in the path integral
\bea \int DA~W~\exp\big(\frac{ik}{4\pi}\int_M~\frac{1}{2}\big( \eta_{ab}A^adA^b+\frac{i}{3}f_{abc}A^aA^bA^c\big),~~~~~W=\Tr_{\mu}\Big[\BB{P}\exp\big(-\small{\textrm{$\oint$}}_K~A\big)\Big],\nn\eea
where $M$ is a \emph{rational homology sphere} and $K$ is the knot embedded therein. Symbol $\BB{P}$ signifies the path ordering and the trace is taken in a representation with weight $\mu$. There are two ways to proceed at this point: 1. one puts $W$ on the exponential and finds the saddle points of $We^{CS}$ together, thus the saddle point are connections with non-trivial holonomy.  2. one treats $W$ as perturbation, and does perturbation around the saddle point of the action $e^{CS}$ alone. As one expands $W$ into powers of $A$, the perturbation will involve Feynman diagrams like those of fig.\ref{ex_gph_fig}.

\begin{remark}Let me make a final remark concerning the Feynman diagram calculation. Once we have obtained the 1-loop determinant, for further loop expansion, it is more convenient to integrate out the $b$ field in Eq.\ref{BRST_G_fixed} enforcing $d_a^{\dagger}A=0$ and reorganize the fields into a \emph{superfield}
\bea \BS{A}=c+\theta^{\mu}A_{\mu}-\frac12 \theta^{\mu}\theta^{\nu}(d^{\dagger}_a\bar c)_{\mu\nu},\nn\eea
where $\theta^{\mu},~\mu=1,1,3$ are Grassman variables that transoform like $dx^{\mu}$ on $M$. One can check that the gauge fixed action Eq.\ref{BRST_G_fixed} can be recast as
\bea S_{CS-brst}=S_{CS}(a)+\Tr\int_M d^3xd^3\theta~\big(\BS{A}D_a\BS{A}+\frac{2}{3}\BS{AAA}\big),~~~~D_a=\theta^{\mu}(\partial_{\mu}+[a_{\mu},\cdot])\nn\eea
with $D_a^{\dagger}\BS{A}=0$ imposed. In this way, the propagator is indeed the super propagator
\bea&& \bra \BS{A}^c(x,\theta),\BS{A}^d(y,\zeta)\ket=\frac{-4i\pi}{k}G(x,\theta;y,\zeta)\eta^{cd},\nn\\
\textrm{where}&& d^{-1}\psi=\int d^3y~G(x,dx;y,dy)\psi(y),~~~~\psi\in \Omega^{\sbullet}(M).\label{super_prop}\eea
This superfield technique was used in ref.\cite{Axelrod:1991vq} to compute the 2-loop diagram fig.\ref{fig_cocycle} directly.
\end{remark}

\subsubsection{First Approach}
We start with the first approach. For the fully non-abelian case, it is not easy to find the saddle point, yet for our purpose it suffices to look for a saddle point at which the flat connection is reducible, i.e. the holonomy can be brought into the maximal torus of $G$. In this way, our problem is reduced to finding a solution to the \emph{abelian} CS theory with a Wilson loop. In this case, one can drop the path ordering in the Wilson loop operator and write it as
\bea W=\Tr_{\mu}\Big[\exp\big(-\small{\textrm{$\oint$}}_K~A\big)\Big],~~~\textrm{where}~A=A^aH_a,~~H_a\in \textrm{CSA},~~a=1\cdots \textrm{rk} G.\nn\eea

\begin{figure}[h]
\begin{center}
\begin{tikzpicture}[scale=1]
\draw [-,blue] (.06,-.06) to [out=-60, in=90] (.6,-1) arc (0:-180:.6)
to [out=90, in=-90] (.6,1) arc (0:180:.6) to [out=-90, in=120] (-.12,.12);
\draw [-,red] (.16,0) to [out=-60, in=90] (.7,-1) arc (0:-180:.7)
to [out=90, in=-140] (-.12,0) to [out=50, in=-90] (.5,1) arc (0:180:.5) to [out=-90, in=120] (-.02,0.18);
\end{tikzpicture}~~
\caption{The 'blackboard framing' of a knot.}\label{fig_framing_knot}
\end{center}
\end{figure}
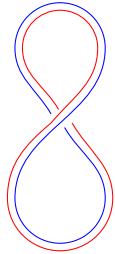

From the discussion of sec.\ref{NFDR}, especially, the shift in Eq.\ref{well_knownn_shift}, we know that the 'effective weight' of the holonomy is $\mu+\rho$, and of course $k$ is shifted by $h$ as well.\footnote{I admit that the reasoning here has a gap, and it is made even more bizarre when one sees that to remove the Wilson loop one must take $\mu=-\rho$! Recall that in sec.\ref{sec_QoC2}, the last two factors of Eq.\ref{factor_det} are the 'Fadeev-Popov' determinant. The determinant was calculated in that section and its effect is exactly the two shifts just mentioned, even though the setting there is the K\"ahler quantization. With some work, one probably could do a better job in explaining this.}
It is more convenient to label the vectors in the representation by their weights $|\mu_i\ket$, where $\mu_0=\mu$ is the highest weight and $\mu_i<\mu_0$. In this basis the Wilson line is written as
\bea W=\sum_{\mu_i}\bra\mu_i|e^{-\oint_KA}|\mu_i\ket=\sum_{\mu_i} e^{-\oint_KA^a\bra \mu_i|H_a|\mu_i\ket}=\sum_{\mu_i} e^{-\oint_KA\cdot(\mu_i+\rho)}.\nn\eea
Now one can easily solve for the flat connection. Let $G(x,y)$ be the inverse of the de Rham differential in $M$ as in Eq.\ref{super_prop},
then the flat connection can be written as
\bea a(x)=\eta^{-1}(\mu_i+\rho)\frac{-4i\pi}{k+h}\int_{K_y} G(x,y),\label{explicit_sol}\eea
and the CS term evaluated at this connection is
\bea CS(a)=\frac{2i\pi(\mu_i+\rho)^2}{(k+h)}\int_{K_x}\int_{K_y}G(x,y)=\frac{2i\pi(\mu_i+\rho)^2}{(k+h)}\nu_K.\nn\eea
where the double integral
\bea \nu_K=\int_{K_x}\int_{K_y}G(x,y).\label{self_linking}\eea
is just the self-linking number of the knot, in analogy with the case Eq.\ref{linking_number}. However, this number is ambiguous, since the propagator $G(x,y)$ is not well-defined when $x=y$. So one needs to push the $y$ integral slightly off the knot. Looking at fig.\ref{fig_framing_knot}, one integrates $x$ along the blue line and $y$ along the red. This pushing off is equivalent to giving a \emph{framing of the knot}. The framing in fig.\ref{fig_framing_knot} is called the 'blackboard framing'.
It is instructive to obtain this factor $\nu_K$ in our setting carefully, which is done in ref.\cite{KirkKlassen}, but for our purpose we will just leave it as it is since it will cancel out eventually.

Having obtained the CS term evaluated at the saddle point, the next task is the one-loop determinant. The fluctuations $A$ in Eq.\ref{splitaA} as well as the other fields can be decomposed as
\bea \phi=\phi^aH_a+\phi^{\ga}E_{\ga}+\phi^{-\ga}E_{-\ga},~~~~~\phi=\{A,b,\bar c,c\},\nn\eea
where $a$ ranges over the CSA and $\ga$ ranges over the positive roots. The fields $\phi^{\pm\ga}$ are charged under the flat connection and the torsion for these fields
is given by the Alexander polynomial (reviewed in sec.\ref{KCatAP}). To work this out, one needs to find the generator for the infinite cyclic subgroup of the first homology group of the knot complement, this generator is characterized by the fact that it has intersection number 1 with the Seifert surface. Let us name the $a$ and $b$ cycles of the tubular neighbourhood of the knot as $C_1$ and $C_2$. If $M$ were $S^3$, then $C_2$ is contractible in the knot complement, but in general, let us assume that $pC_1+qC_2$ is contractible\footnote{We remark that this information is also encoded in the propagator $G(x,y)$, that is, if one integrates Eq.\ref{explicit_sol} along the cycle $pC_1+qC_2$, one must get an integer multiple of $2\pi$}. The Seifert surface can be chosen to be bounded by this same cycle and the generator we seek is \bea rC_1+sC_2,~~~rq-ps=1.\nn\eea
So the holonomy of a field $\phi^{\ga}$ is
\bea t=\exp\big(\frac{2i\pi\bra\mu_i+\rho,\ga\ket}{(k+h)}\frac{1-sn}{q}\big),~~~n\in\BB{Z}.\nn\eea
Any integer $n$ serves to guarantee a trivial holonomy along $pC_1+qC_2$, but if one insists that by letting $\mu_i+\rho=0$ (removing the Wilson loop), the flat connection be trivial, one must set $n=0$.

The combinatorial calculation of the torsion by Milnor \cite{MilnorAlexander} (given in the appendix) shows
$\tau_{\ga}=A_K(t)/(1-t)$ with $t$ given above. If one normalizes the torsion so that $A(t)$ is real and $\tau$ is 1 for an unknot embedded in $S^3$, then the torsion is
\bea \tau_{\ga}=\frac{A_K(\exp\frac{2i\pi\bra\mu_i+\rho,\ga\ket}{q(k+h)})}{2\sin(\frac{\pi\bra\mu_i+\rho,\ga\ket}{q(k+h)})}.\nn\eea
One of course gets one such factor for each root $\prod_{\ga>0}|\tau^{-1}_{\ga}|^2$.

In contrast, the fields that are in the CSA are neutral, and their torsion is claimed in ref.\cite{Rozanskytrivial} Eq.(2.19) to be
$|H_1(M,\BB{Z})|$-the number of elements in the first homology group (for reasoning see ref.\cite{RozanskySaleurST} sec.6.3 and \cite{RozanskySaleurU11}).
To summarize, the contribution of the trivial connection to the perturbation series is given by (Eq.(2.27) of ref.\cite{Rozanskytrivial})
\bea &&{\cal Z}^{\textrm{tr}}_{\mu_i+\rho}=\big(2(k+h)\,\big|H_c(M,\BB{Z})\big|\big)^{-\textrm{rk}_G/2}\exp\big(\frac{i\pi(\mu_i+\rho)^2}{k+h}\nu_K\big)\nn\\
&&\hspace {4cm}\cdot\prod_{\ga>0}\frac{2\sin\big(\frac{\pi\bra\mu_i+\rho,\ga\ket}{q(k+h)}\big)}{A_K\big(\exp\frac{2i\pi\bra\mu_i+\rho,\ga\ket}{q(k+h)}\big)}
\exp\Big(i\sum_{n=1}^{\infty}\big(\frac{2\pi}{k+h}\big)^nS_{n+1}(\frac{\ga}{k+h})\Big)\label{appch_1}.\eea
The superscript ${\cal Z}^{\textrm{tr}}$ is a reminder that the expansion is around a connection which, if one sets the weight $\mu_i+\rho$ to zero, is the trivial connection on $M$.

In particular, we will need the ratio
\bea \frac{{\cal Z}^{\textrm{tr}}_{\mu+\rho}}{{\cal Z}^{\textrm{tr}}_{\rho}}=\exp\Big(\frac{i\pi\nu_K}{k+h}((\mu+\rho)^2-\rho^2\big)\Big)
\cdot\prod_{\ga>0}\frac{\sin\big(\frac{\pi\bra\mu+\rho,\ga\ket}{q(k+h)}\big)}{\sin\big(\frac{\pi\bra\rho,\ga\ket}{q(k+h)}\big)}
\frac{A_K\big(\exp\frac{2i\pi\bra\rho,\ga\ket}{q(k+h)}\big)}{A_K\big(\exp\frac{2i\pi\bra\mu+\rho,\ga\ket}{q(k+h)}\big)}+{\cal O}(k^{-3})\nn\eea
The product can be simplified using $A_K(1)=1$ and $A_K'(1)=0$
\bea &&\prod_{\ga>0}\frac{\bra\mu+\rho,\ga\ket}{\bra\rho,\ga\ket}\Big(1+\frac{\pi^2}{q^2(k+h)^2}\big(\bra\mu+\rho,\ga\ket^2-\bra\rho,\ga\ket^2\big)
\big(2A_K''(1)-\frac16\big)\Big)+{\cal O}(k^{-3})\nn\\
&=&d_{\mu}+\frac{d_{\mu}\pi^2}{q^2(k+h)^2}\big(2A_K''(1)-\frac16\big)\sum_{\ga>0}\big(\bra\mu+\rho,\ga\ket^2-\bra\rho,\ga\ket^2\big)\nn\\
&=&d_{\mu}+\frac{d_{\mu}\pi^2}{q^2(k+h)^2}\big(2A_K''(1)-\frac16\big)\sum_{\ga>0}h\big((\mu+\rho)^2-\rho^2\big)\nn.\eea
The ratio up to $k^{-2}$ is thus
\bea \frac{{\cal Z}^{\textrm{tr}}_{\mu+\rho}}{{\cal Z}^{\textrm{tr}}_{\rho}}
=\exp\Big(\frac{i\pi\nu_K}{k+h}((\mu+\rho)^2-\rho^2\big)\Big)d_{\mu}\Big(1+\frac{2\pi^2}{q^2(k+h)^2}\big(2A_K''(1)-\frac16\big)C_2(ad)C_2(\mu)\Big)\label{ratio_1}.\eea
Note that in getting the ratio, we have not taken into account the last exponent in Eq.\ref{appch_1}, whose effect is easily seen to be of order $k^{-3}((\mu+\rho)^2-\mu^2)$.

\subsubsection{Second Approach}
Next, we use the second approach mentioned earlier, namely, we treat the Wilson loop as perturbation and expand around a saddle point corresponding to the trivial flat connection of the entire $M$. The computation are given by Feynman diagrams; the set of diagrams that are not connected to the Wilson loop can be factored out, since these diagrams give merely $Z^{\textrm{tr}}$ without any Wilson loops in it. To lowest orders, diagrams that are connected to the Wilson loop are as in fig.\ref{ex_gph_fig}.

Observe that the first diagram $\Gamma_0$ gives
\bea \Gamma_0&\to&\frac12\Tr[T^{\ga}T_{\ga}]\frac{4i\pi}{k}b_{\Gamma_0}=\frac12C_2(\mu)d_{\mu}\cdot \frac{4i\pi}{k}b_{\Gamma_0},\nn\\
b_{\Gamma_0}&=&\int_0^1 dt \int_{t-1}^t ds G(x(t),x(s)),\nn\eea
where $s$ is taken mod 1, $C_2(\mu)$ is the representation labelled by the highest weight $\mu$ and $d_{\mu}$ is the dimension. The term $b_{\Gamma_0}$ is once again the integral form of the self-linking number, which must be regulated to fix the ambiguity. We can see that in fact this diagram \emph{exponentiates}, this was first observed by Bar-Natan \cite{Bar-Natan95}. To illustrate this point, we compute the diagrams $\Gamma_{4,5,6,7}$
\bea -\frac{1}{2}\Tr[T^{\alpha}T_{\alpha}T_{\beta}T^{\beta}]b_{\Gamma_4}-\frac{1}{4}
\Tr[T^{\alpha}T^{\beta}T_{\alpha}T_{\beta}]b_{\Gamma_5}-\frac{i}{3}f_{\alpha\beta\gamma}
\Tr[T^{\alpha}T^{\beta}T^{\gamma}]b_{\Gamma_6}-\frac{1}{4}f_{\alpha\beta\gamma}f^{\alpha\beta\gamma}\frac{C_2(\mu)d_{\mu}}{|G|}b_{\Gamma_7}.\label{LieAlgWght}\eea
I have dropped the common factor $(4i\pi/k)^2$. After some algebra, we can rewrite the above into two combinations
\bea
\Big(\frac{4i\pi}{k}\Big)^{-2}\textrm{Eq}.\ref{LieAlgWght}=\frac{d_{\mu}C_2(ad)C_2(\mu)}{2}\cdot\big(\frac{1}{4}b_{\Gamma_5}+\frac{1}{3}b_{\Gamma_6}-\frac{1}{2}b_{\Gamma_7}\big)
-\frac{d_{\mu} C_2^2(\mu)}{4}\cdot \big(b_{\Gamma_5}+2b_{\Gamma_4}\big)~,\label{two_linear_comb}\eea
\begin{remark}
The two linear combinations are gauge invariant separately. There is a powerful tool to analyze the gauge invariance of a combination of graphs, called the graph complex. Namely, one can endow the Feynman graphs with a differential complex structure. If any linear combination of graphs is closed w.r.t the differential, then the Feynam integral given by these graphs is gauge invariant. The rule for the graph differential is listed in ref.\cite{cov_wght}, the version of proof (closest in notation) can be found in ref.\cite{WilsonLoop}. I do not plan to going into this topic in this note, but one may refer to the original papers by Kontsevich \cite{Kontsevich:formal,Kontsevich:feynman} and also ref.\cite{ConantVogt} for a more careful exposition.
\end{remark}

I want to show now that the second combination in Eq.\ref{two_linear_comb} \emph{is the square of $\Gamma_0$}. The $b_{\Gamma_{4,5}}$ term is given by
\bea&& b_{\Gamma_4}=\int_0^1 dt_1 \int_{t_1-1}^{t_1} dt_2\int_{t_1-1}^{t_2} dt_3 \int_{t_1-1}^{t_3} dt_4~G(x(t_1),x(t_2)) G (x(t_4),x(t_3)),\nn\\
&&b_{\Gamma_5}=\int_0^1 dt_1 \int_{t_1-1}^{t_1} dt_2\int_{t_1-1}^{t_2} dt_3 \int_{t_1-1}^{t_3} dt_4~G(x(t_1),x(t_3)) G (x(t_4),x(t_2)),\nn\eea
where as usual $t_{2,3,4}$ are taken mod 1. One can see the crucial relation
\bea b_{\Gamma_5}+2b_{\Gamma_4}=-\frac12 b_{\Gamma_0}^2\label{exponentiation}\eea
by writing down $b_{\Gamma_0}^2$ and dividing up the integration range.

\begin{figure}[h]
\begin{center}
\begin{tikzpicture}[scale=.8]
\draw [semithick,blue] (0,0) circle (1);
\draw [->,blue] (0,1) arc (90:195:1cm);
\draw [->,blue](-.717,.717cm)--(.717,-.717cm);
\node  at (-.717,.717cm) [left] {\small $1$};
\node  at (.717,-.717cm) [right] {\small $4$};
\draw [<-,blue](-.717,-.717cm)--(.717,.717cm);
\node  at (-.717,-.717cm) [left] {\small $3$};
\node  at (.717,.717cm) [right] {\small $6$};
\draw [->,blue](-.953,.3cm)--(0.953,.3cm);
\node  at (-.953,.3cm) [left] {\small $2$};
\node  at (.953,.3cm) [right] {\small $5$};
\node  at (0,-1) [below] {\small $\Gamma_{9}$};
\end{tikzpicture}~~
\begin{tikzpicture}[scale=.8]
\draw [semithick,blue] (0,0) circle (1);
\draw [->,blue] (0,1) arc (90:180:1cm);
\draw [->,blue](-.717,.717cm)--(-.717,-.717cm);
\node  at (-.717,.717cm) [left] {\small $1$};
\node  at (-.717,-.717cm) [left] {\small $2$};
\draw [->,blue](-.5,-.866cm)--(.5,-.866cm);
\node  at (-.5,-.866cm) [below] {\small $3$};
\node  at (.5,-.866cm) [below] {\small $4$};
\draw [<-,blue](0,1cm)--(0.717,-.717cm);
\node  at (0,1cm) [above] {\small $6$};
\node  at (.717,-.717cm) [right] {\small $5$};
\node  at (0,-1) [below] {\small $\Gamma_{10}$};
\end{tikzpicture}~~
\begin{tikzpicture}[scale=.8]
\draw [semithick,blue] (0,0) circle (1);
\draw [->,blue] (0,1) arc (90:180:1cm);
\draw [->,blue](-.717,.717cm)--(-.717,-.717cm);
\node  at (-.717,.717cm) [left] {\small $1$};
\node  at (-.717,-.717cm) [left] {\small $2$};
\draw [->,blue](0,1cm)--(0,-1cm);
\node  at (0,1cm) [above] {\small $6$};
\node  at (0.2,-.7cm) {\small $3$};
\draw [<-,blue](.717,.717cm)--(.717,-.717cm);
\node  at (.717,.717cm) [right] {\small $5$};
\node  at (.717,-.717cm) [right] {\small $4$};
\node at (0,-1) [below] {\small $\Gamma_{11}$};
\end{tikzpicture}~~
\begin{tikzpicture}[scale=.8]
\draw [semithick,blue] (0,0) circle (1);
\draw [->,blue] (0,-1) arc (270:315:1cm);
\draw [->,blue](-.717,.717cm)--(-.717,-.717cm);
\node  at (-.717,.717cm) [left] {\small $1$};
\node  at (-.717,-.717cm) [left] {\small $3$};
\draw [->,blue](0,1cm)--(0,-1cm);
\node  at (0,1cm) [above] {\small $6$};
\node  at (0.2,-.7cm) {\small $4$};
\draw [->,blue](-1,0cm)--(1,0cm);
\node  at (-1,0cm) [left] {\small $2$};
\node  at (1,0cm) [right] {\small $5$};
\node at (0,-1) [below] {\small $\Gamma_{12}$};
\end{tikzpicture}~
\begin{tikzpicture}[scale=.8]
\draw [semithick,blue] (0,0) circle (1);
\draw [->,blue] (0,1) arc (90:180:1cm);
\draw [->,blue](-.717,.717cm)--(0,-1cm);
\node  at (-.717,.717cm) [left] {\small $1$};
\node  at (.2,-.7cm) {\small $3$};
\draw [->,blue](0,1cm)--(-0.717,-0.717cm);
\node  at (0,1cm) [above] {\small $6$};
\node  at (-0.717,-.717cm) [left] {\small $2$};
\draw [->,blue](.717,-.717cm)--(.717,.717cm);
\node  at (.717,-.717cm) [below] {\small $4$};
\node  at (.717,.717cm) [above] {\small $5$};
\node at (0,-1) [below] {\small $\Gamma_{13}$};
\end{tikzpicture}~~
\caption{The 3-loop diagrams that contribute to the self linking number} \label{fig_3_loop}
\end{center}
\end{figure}
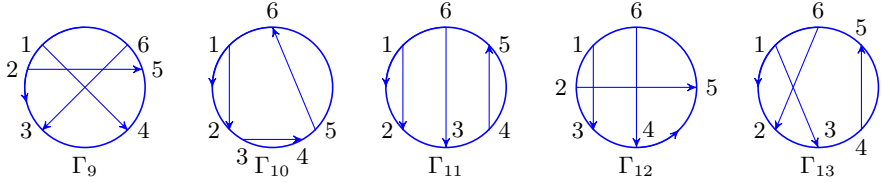

Proceeding to the next order, fig.\ref{fig_3_loop} contains only part of the 3-loop diagrams.
There will be a gauge invariant combination
\bea d_{\mu}C^3_2(\mu)\cdot\Big(\frac{4i\pi}{k}\Big)^3\Big(-\frac{1}{3}b_{\Gamma_{9}}+\frac{1}{3}b_{\Gamma_{10}}-\frac{1}{2}b_{\Gamma_{11}}-\frac12b_{\Gamma_{12}}+b_{\Gamma_{13}}\Big)
\label{3_loop_expon},\nn\eea
it should clear to the reader by now how to write down $b_{\Gamma}$. And I need to add that the total contribution of, say, $\Gamma_{12}$ is not just $-1/2d_{\mu}(4i\pi C_2(\mu)/k)^3$, rather, $\Gamma_{12}$ participates in several gauge invariant combinations. The absolute normalization of Eq.\ref{3_loop_expon} is determined by $\Gamma_{10}$, which appears only in this particular combination.

We again observe that Eq.\ref{3_loop_expon} is
\bea \textrm{Eq}.\ref{3_loop_expon}=d_{\mu}C^3_2(\mu)\cdot\Big(\frac{4i\pi}{k}\Big)^3\frac{1}{48}b^3_{\Gamma_0}.\nn\eea
The general pattern can be established by an induction, and one concludes that one can factor out the self-linking number term
\bea d_{\mu}\exp\big(\frac{2i\pi C_2{\mu}}{k}b_{\Gamma_0}\big)=d_{\mu}\exp\big(\frac{2i\pi C_2{\mu}}{k}\nu_K\big)\nn\eea
and consider only diagrams without isolated chords. Then the expectation value of the Wilson loop can be written as
\bea \bra W\ket=d_{\mu}\exp\Big(\frac{2i\pi}{k}C_2(\mu)\nu\Big)\bigg(1+\frac{C_2(ad)C_2(\mu)}{2}\Big(\frac{4i\pi}{k}\Big)^2
\big(\frac{1}{4}b_{\Gamma_5}+\frac{1}{3}b_{\Gamma_6}-\frac{1}{2}b_{\Gamma_7}\big)+\cdots\bigg),\label{ratio_2}\eea
where the terms $\cdots$, like the second term, consist of diagrams with no isolated chords.

One can equate the result Eq.\ref{ratio_1} with that of Eq.\ref{ratio_2}, and get
\bea \frac{1}{4}b_{\Gamma_5}+\frac{1}{3}b_{\Gamma_6}-\frac{1}{2}b_{\Gamma_7}=-\frac{1}{2q^2}\big(A_K''(1)-\frac{1}{12}\big).\label{alex_perturb}\eea
This is Eq.2.25 of ref.\cite{Rozanskytrivial}. Note that all the Lie algebra related factors cancel out, for otherwise we are in trouble.

From the explicit result of the diagrams in fig.\ref{ex_gph_fig}, one can proceed further and compute the Casson-Walker invariant fig.\ref{fig_cocycle}.
In fact, it is not really the actual value of fig.\ref{fig_cocycle} that interests us, rather it is how the integral responds to a surgery that is more interesting, since this is a measure of 'how revealing' a particular 3-manifold invariant is. What one does is to use a surgery (e.g an $r/s$ surgery) to remove the Wilson loop calculated earlier, and thereby obtaining how fig.\ref{fig_cocycle} responds to such a surgery. If I were to review this, I would merely be copying pages of formulae already well-written in ref.\cite{Rozanskytrivial}, from which I shall desist.

%% file: theTorsion.tex
The torsion is a more refined invariant compared to the homology of a chain complex, especially if the chain complex is acyclic (with vanishing homology). For example, when two manifolds $X$ and $Y$ are homotopy equivalent, then no homology nor homotopy group is able to distinguish them, yet one can define a torsion in this situation that measures roughly how twisted is the homotopy between $X$ and $Y$. The torsion was used to good effect to classify the lens manifolds by Reidemeister. In a sense, the torsion is to homology what Chern-Simons form is to Chern class. There are different ways of defining the torsion, but all are similar enough for a physicist. Amongst the different definitions, the Whitehead torsion takes value in the so called Whitehead group of the fundamental group, the second uses a representation of $\pi_1$ in the orthogonal matrices and the last one is entirely analytical. I am not going to worry anymore about the difference in these definitions, they are enumerated in the section title purely for visual effect.

Consider an acyclic chain complex
\bea 0\to A_0 \mathop{\rightleftharpoons}^{d_0}_{k_1} A_1 \mathop{\rightleftharpoons}^{d_1}_{k_2} A_2\cdots \mathop{\rightleftharpoons}^{d_{n-1}}_{k_n}A_n\to 0,\nn\eea
where $d$ is the differential and $k$ is the chain contraction\footnote{In general an acyclic chain complex need not be contractible, but as we work over a field, the chain complex is projective. A projective acyclic chain complex is contractible} satisfying $\{d,k\}=1$, and $k$ can be chosen to satisfy $k^2=0$. Choose for each $A_i$ a basis $e_i^a,~a=1\cdots\textrm{rk}A_i$, since $A_0$ injects into $A_1$, the map $d_0$ induces a volume element for $A_1/A_0$; equally as $A_1/A_0$ injects into $A_2$, the map $d_1$ induces a volume element for $A_2/A_1/A_0$ ..., and finally, we have a volume element for $A_{n-1}/A_{n-2}/\cdots /A_0$ that differs from the chosen one for $A_n$ by a number (or an element of the Whitehead group, as the case might be), denoted as
\bea \Delta=A_n/A_{n-1}/\cdots /A_0\label{torsion_Milnor}.\eea
This number is the Whitehead torsion associated with this chosen basis (usually the problem itself suggests a preferred basis).

To present the torsion in a slightly different way, one forms a matrix $d+k$ that acts on the even $A_{2i}$
\bea \begin{array}{|ccccc|}
       d_0 & k_2 & 0 & \cdots & \cdots \\
       0 & d_2 & k_4 & 0 & \cdots \\
       0 & 0 & d_4 & k_6 & 0 \\
       \cdots & \cdots & \cdots & \cdots & \cdots \\
       0 & \cdots & 0 & d_{2p-2} & k_{2p}
     \end{array}\,\begin{array}{|c|}
                    A_0 \\
                    A_2 \\
                    A_4 \\
                    \cdots \\
                    A_{2p}
                  \end{array}\label{big_matrix}\eea
where $p=\lfloor n/2\rfloor$. The determinant of this matrix is the torsion we seek.

To see this, take the simplest example when $n=2$, and pick bases
\bea \underbrace{e_1,e_2,\cdots e_{r_0}}_{\dim 0},~~~\underbrace{f_1,f_2,\cdots f_{r_1}}_{\dim 1},~~~\underbrace{g_1,g_2,\cdots g_{r_2}}_{\dim 2}.\nn\eea
then the matrix $d_0$ is $r_1\times r_0$, defined as
\bea d_0 e_i=(d_0)^j_{~i}f_j\nn\eea
and similarly $k_2$ is $r_1\times r_2$. Clearly $r_0+r_2=r_1$. The juxtaposed matrix $[d_0,k_2]$ looks like the first one of fig.\ref{fig_d_plus_k}.
\begin{figure}[h]
\begin{center}
\begin{tikzpicture}[scale=.8]
\draw[blue, fill=blue!30, opacity=.5] (0,0) rectangle (.6,2);
\node at (0.3,1) {\small $d_0$};
\draw[blue, fill=blue!50, opacity=.5] (0.64,0) rectangle (2,2);
\node at (1.3,1) {\small $k_2$};
\node at (.3,2.2) {\small $r_0$};
\node at (1.3,2.2) {\small $r_2$};
\node at (-.2,1) {\small $r_1$};
\node at(2.4,1) {\small $\Rightarrow$};
\end{tikzpicture}
\begin{tikzpicture}[scale=.8]
\draw[blue] (0,1.4) rectangle (.6,2);
\node at (0.3,1.7) {\small $1$};
\draw[blue] (0,0) rectangle (.6,1.36);
\node at (0.3,0.7) {\small $0$};
\draw[blue] (0.64,1.4) rectangle (2,2);
\node at (1.3,1.7) {\small $0$};
\draw[blue, fill=blue!50, opacity=.5] (0.64,0) rectangle (2,1.36);
\node at (1.3,0.7) {\small $k'_2$};
\node at (.3,2.2) {\small $r_0$};
\node at (1.3,2.2) {\small $r_2$};
\node at (-.2,.7) {\small $r_2$};
\end{tikzpicture}\caption{Pictorial description of the matrix $d+k$ for dimension 2}\label{fig_d_plus_k}
\end{center}
\end{figure}
To compute the determinant of this matrix, one can use two matrices $M_0$ $M_1$ to changes basis at dimension 0 and 1 so that $d_0$ in basis $e'$ and $f'$ becomes the standard form $M_1^{-1}d_0M_0=1$, namely: $e_i'=(M_0)^j_{~i}e_j$ and $f'=(e_1',\cdots,e'_{r_0},f'_1,\cdots f'_{r_2})$.
In this basis, $k_2$ is necessarily of the shape indicated in the figure. Thus the determinant is
\bea \det{(d+k)}=\det{k'_2}\det{M_1}\det{M^{-1}_0}.\nn\eea
While a direct application of the definition Eq.\ref{torsion_Milnor} of the torsion goes as
\bea (f_1f_2\cdots f_{r_1})/(e_1e_2\cdots e_{r_0})&=&\det{M_1^{-1}}\det{M_0}(f_1'f_2'\cdots f_{r_2}')\nn\\
(g_1g_2\cdots g_{r_2})/(f_1f_2\cdots f_{r_1})/(e_1e_2\cdots e_{r_0})&=&\det{M_1}\det{M^{-1}_0}\big(\det{d_1'}\big)^{-1}=\det{M_1}\det{M^{-1}_0}\det{k_2'},\nn\eea
which of course agrees with the direct calculation for $\det{(d+k)}$. Instead of Eq.\ref{big_matrix}, one can equally use the matrix
\bea \begin{array}{|ccc|}
       k_1 & 0 & \cdots \\
       d_1 & k_3 & \cdots \\
       \cdots & \cdots & \cdots \end{array}\,
       \begin{array}{|c|}
                    A_1 \\
                    A_3 \\
                    \cdots \end{array}\label{big_matrix_dual},\eea
whose inverse is the torsion.

Using this presentation of the torsion, it is easy to understand the dependence of the torsion on $k$. A change of $k$ is $d$ exact, $\gd k=[l,d]$ for some $l$ that shifts the degree down by 2 and satisfies $[l,k]=0$. Thus the new torsion is computed with the matrix Eq.\ref{big_matrix} composed from the right and from the left with the following upper-triangular matrices
\bea L_{even}=\begin{array}{|cccc|} 1 &  -l_2 & \cdots &\cdots \\
                                                            0 & 1 &  -l_4 & \cdots \\ \vdots & \vdots & \vdots & \vdots \\
                                                   \vdots & \vdots & 1 &  -l_{2p} \\
                                                    \cdots & \cdots & 0  & 1\end{array}~~~~
L_{odd}=\begin{array}{|cccc|} 1 &  l_3 & \cdots &\cdots \\
                                                            0 & 1 &  l_5 & \cdots \\ \vdots & \vdots & \vdots & \vdots \\
                                                   \vdots & \vdots & 1 &  l_{2p-1} \\
                                                    \cdots & \cdots & 0  & 1\end{array}\nn\eea
which clearly gives
\bea \det((d+k+\delta k))=\det(L_{odd}(d+k)L_{even})=\det(d+k).\nn\eea

There is one more property of the torsion that we shall need later, let
\bea 0\to A'_{\sbullet}\to A_{\sbullet}\to A''_{\sbullet}\to0\nn\eea
be an exact sequence of acyclic chain complexes, then their torsion satisfy
\bea \Delta(A)=\Delta(A')\Delta(A'').\label{additivity_torsion}\eea
This is proved in ref.\cite{MilnorTorsion} Thm.3.1 (for the Whitehead torsion, but the proof is easily transferable to other versions of the torsion).

When the complex has non-trivial homology, one can still define the torsion, in several ways. One can choose a representative for the (co)homology group $H_i(A)$ within $A_i$ and the torsion is dependent upon the choice of basis of $A_i$ and also the representative, see ref.\cite{MilnorTorsion} sec.3. This is quite easily done in the case of Riemannian manifold, where one can choose the harmonic form as a representative of cohomology, which will be done in the next paragraph. Or one can change the coefficient ring, which will be dealt with in sec.\ref{KCatAP}.

In the case of Riemannian manifold, one can define an analytic analogue of the above torsion \cite{RaySingerAnalytic}.
The metric gives a Hodge dual and a pairing; the pairing gives us a preferred orthornomal basis for each $A_p$, and different choices differ by the determinant of an $SO$ matrix hence drops. One can choose an explicit splitting of $A_p$ using Hodge decomposition. Let
$\phi^I_p$ be an orthonormal basis of the co-exact component of $A_p$, such that they are eigen-vectors of the Laplacian with eigenvalue $-(\lambda_p^I)^2$
\bea d^{\dagger}\phi_p^I=0,~~~\square\phi_p^I=-(\lambda_p^I)^2\,\phi_p^I,~~~~\lambda_p^I\neq0,~~I=1\cdots r_p.\nn\eea
One can split the basis for $A_p$ as
\bea \{\frac{1}{\lambda_{p-1}^I}d\phi_{p-1}^I, \phi_p^J, h\},\nn\eea
where the first set of the basis is $d$-exact, the middle set $d^{\dagger}$-exact and $h$ is the harmonic modes.

In this basis, the torsion is easily seen to be\footnote{There are no separate names for $\Delta$ and $\log\Delta$, both are called torsion; I will use $\tau$ for the latter one}
\bea \tau=\log\Delta=\sum^{n-1}_{p=0}(-1)^{n-p}\log\prod_{I=1}^{r_p}\lambda^I_p.\nn\eea
One can reorganize this sum to a better form. Notice that
\bea \det\square_p=\prod_{I=1}^{r_p}\big(-(\lambda_p^I)^2\big)\prod_{I=1}^{r_{p-1}}\big(-(\lambda_{p-1}^I)^2\big),\nn\eea
from which one can check that
\bea (-1)^p\log\prod_{I=1}^{r_p}\lambda_p^I=\sum_{q=p+1}^n (-1)^q\log{\det}^{1/2}(-\square_q),\nn\eea
and the torsion can be rewritten as a sum involving eigen-values of the Laplacian
\bea \tau&=&(-1)^n\sum^{n-1}_{q=0}(-1)^{q}\log\prod_{I=1}^{r_q}\lambda^I_q
=(-1)^n\sum^{n-1}_{q=0}\sum_{p=q+1}^n(-1)^{q}\log{\det}^{1/2}(-\square_p)\nn\\
&=&(-1)^n\sum_{p=1}^np(-1)^{p}\log{\det}^{1/2}(-\square_p).\label{torsion_better}\eea

\begin{example}
We have yet to explain why is the expression in Eq.\ref{Ray-Singer} the torsion defined here. This is quite easy, since one may use the Hodge decomposition to define a chain contraction
\bea k=\frac{d_a^{\dagger}}{\square},~~~~~\square=\{d_a,d_a^{\dagger}\}.\nn\eea
Plugging this into Eq.\ref{big_matrix_dual}
\bea \begin{array}{|cc|}
       \square^{-1}d_a^{\dagger} & 0 \\
       d_a & \square^{-1}d_a^{\dagger}\end{array}\,\begin{array}{|c|}
                                                 A \\
                                                 b \end{array}\nn.\eea
Using Hodge duality to relate the eigenvalues of even forms with odd forms, one can see that the determinant gives the inverse squared of Eq.\ref{Ray-Singer}.
We also observe that the deformation of $k$ can be written as
\bea \delta k=[d_a,l],~~~l=\frac1{\square}d_a^{\dagger}\delta d_a^{\dagger}\frac{1}{\square}.\nn\eea
\end{example}

\begin{example}\label{ex_flat_bundle}
\emph{Flat $U(1)$-bundle over the circle}, specified by the holonomy
\bea s(1)=s(0)e^{2\pi i u},~~~u\in\BB{R}\backslash\BB{Z}.\label{U1_bundle_circle}\eea
Since the transition function can be chosen to be a constant, the covariant derivative is just $d$, and this $d$ is \emph{invertible}.

The differential complex is two levelled $0\to A_0\to A_1\to 0$, for which we fix a basis
\bea \phi_0^m,~~\frac{1}{u+m}d\phi_0^m,~~~~\textrm{where}~\phi_0^m=\frac{1}{\sqrt{2\pi}}e^{i\theta(u+m)},~~m\in\BB{Z}.\nn\eea
The torsion is obtained by applying Eq.\ref{torsion_better} (w.o.l.g. one can assume $|u|<1$)
\bea\Delta=\big(u^2\prod_{n\neq0}(n^2-u^2)\big)^{1/2}=u\big(\prod_{1}^{\infty}n^2\big)\prod_{1}^{\infty}(1-\frac{u^2}{n^2}).\nn\eea
Apart from a multiplicative constant (which can also be regulated using the zeta function), the product is equal to $\sin u$.
\end{example}

\begin{example}\emph{the Alexander polynomial}\\
Consider a knot $K$ embedded in $S^3$, let $X$ be the infinite cyclic covering of the complement $S^3\backslash K$. The torsion of the complex $C_{\sbullet}(X)$ is given by
\bea \Delta(X)=\frac{A(t)}{1-t}.\nn\eea
where $t$ is the infinite cyclic generator of $\pi_1(S^3\backslash K)$; it can be chosen as a loop having linking number 1 with $K$. Furthermore $A(t)$ is called the Alexander polynomial. The sketch of the proof is given in the appendix.
\end{example}

%% file: two_framing.tex
The framing of a 3-manifold is a trivialization of its frame bundle.
As listed in the sec.\ref{sec_CGFwS}, orientable closed 3-manifolds are parallelizable, but there are different homotopy classes of trivialization of its tangent bundle.   Two different trivializations is easily seen to differ by a map (homotopic to) $M\to SO(3)$. If we further assume that the two trivializations give the same spin structure, then the framing difference is characterized by a map (homotopic to) $M\to SU(2)$, which is given by the degree of the map. To see the latter statement, one uses a similar argument as in sec.\ref{sec_CGFwS}: as $SU(2)$ is 2-connected, one can first homotope any map $\phi:~M\to SU(2)$ so that the 0-cells are mapped to $e$, the identity element of $SU(2)$. There is no obstruction to extending the homotopy to 1-cells and 2-cells either. Thus without loss of generality, we can assume that the $\phi$ maps the 2-skeleton of $M$ to $e$. But $M$ with its 2-skeleton pinched is equivalent to
\bea M/M_2\sim\vee S^3,\nn\eea
a number of $S^3$ touching at one point. Thus homotopy class of the map $\phi$ is determined by the degree of the map only.

The maps $M\to SO(3)$, however, cannot always be lifted to a map $M\to SU(2)$, but by using 2-framing instead, one can get round the obstruction (see also ref.\cite{KirbyMelvin} for different definitions of framings). The 2-framing is a trivialization of twice the tangent bundle $2T_M$ \emph{as a $spin(6)$-bundle}. To understand this, one notice that the diagonal map $\Delta$
\bea SO(3)\stackrel{\Delta}{\to} SO(3)\times SO(3)\to SO(6)\nn\eea
has a lift to $spin(6)$, and furthermore the map can be deformed into the $spin(3)\subset spin(6)$ subgroup. The reason for this is basically that a map $\BB{R}P^3\to \BB{R}P^3$ of degree 2 can be lifted to $S^3$.

Since now we are considering $2T_M$ as a $spin(6)$ bundle, two different trivializations $\ga$ and $\gb$ differ by a map (I will omit the word 'homotopic to' from now on)
\bea \phi_{\ga\gb}:~~~M\to spin(6).\nn\eea
And as the successive quotients $SU(4)/SU(3)\,, SU(3)/SU(2)$ are all 3-connected, the map $\phi_{\ga\gb}$ can be homotoped to the $SU(2)$ subgroup. Thus, we are back to the scenario of the first paragraph, namely, the 2-framings are characterized by the degree.

\begin{figure}[h]
\begin{center}
\begin{tikzpicture}
\draw[rotate = 0](-1,0) ellipse (.4 and .6);
\draw[rotate = 0](1,0) ellipse (.4 and .6);
\draw[-] (-1,.6) -- (1,.6);
\draw[-] (-1,-.6) -- (1,-.6);
\node at (-1,-.8) {\small $0$};
\node at (1,-.8) {\small $1$};
\node at (-.4,.8) {\small $\overbrace{~~~~~~~~~~~~~}^{U_1}$};
\node at (.4,.9) {\small $\overbrace{~~~~~~~~~~~~~}^{U_2}$};
\node at (-1,.2) {\small $M$};
\node at (1,.2) {\small $M$};
\node at (-1,-.2) {\small $\alpha$};
\node at (1,-.2) {\small $\beta$};
\end{tikzpicture}\caption{$M\times [0,1]$. At the two ends, trivializations $\alpha$ and $\beta$ of $2T_M$ are chosen.}\label{fig_pinch}
\end{center}
\end{figure}
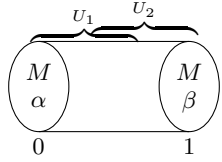

The degree of $\phi_{\ga\gb}$ can be presented in a different way. First recall that if a bundle $E\to X$ is trivial over a subspace $Y$ of $X$, then one can pinch $E|_Y$ together and get a new bundle over $X/Y$, and the new bundle will depend on the choice of trivialization on $Y$. Especially, even if $E$ was trivial to begin with, the new bundle can still be non-trivial over $X/Y$. Now consider the 4-manifold $M\times I$, and the bundle $2T_M$ thereon. Choose two different trivializations $\ga$ $\gb$ of $2T_M$ at $M\times \{0\}$ and $M\times\{1\}$, see fig.\ref{fig_pinch}.
Now pinch together $M\times(\{0\}\cup\{1\})$. Even though the bundle $2T_M$ on $M\times I$ is trivial, it becomes non-trivial after the pinching, we call this new bundle $E$. Its first Pontryagin class, denoted $\hat p_1$, is a \emph{relative} cohomology class
\bea \hat p_1\in H^{4}(M\times I; M\times \{0,1\}).\label{relative_cohomology}\eea
The degree of the map $\phi_{\ga\gb}$ is computed as
\bea \deg(\phi_{\ga\gb})=\frac{1}{2}\int_{M\times I}~\hat p_1.\label{deg_pontryagin}\eea
\begin{remark}
If Eq.\ref{deg_pontryagin} is not immediately clear to the reader, this remark serves as an explanation.

To compute the Pontryagin class, one needs first the connection. Let me pose the question, why do connections exist? Because one can always define the connection locally on patches by using the local trivialization and then use a partition of unity to obtain a connection defined globally (in contrast, analytic connections do not always exist \cite{AtiyahAnalyticConnection}, as the p.o.u is not holomorphic). To illustrate this in general, we cover a manifold with patches $U_I$, on which the bundle $E$ is trivialized. One can simply define the connection to be over each patch $U_i$, and glue them together as
\bea A_i=\sum_j\rho_jg^{-1}_{ji}dg_{ji},\nn\eea
where $g_{ij}$ is the transition function and $\{\rho_{i}\}$ is the partition of unity. By breaking $g_{ij}$ into $g_{ik}g_{kj}$, one easily checks the usual transformation property of the connections
\bea A_{i}=\sum_{j}\rho_{j}(g_{jk}g_{ki})^{-1}d(g_{jk}g_{ki})=g^{-1}_{ki}\big(\sum_{j}\rho_{j}g^{-1}_{jk}dg_{jk}\big)g^{-1}_{ki}+\sum_{j}\rho_{j}g^{-1}_{ki})d g_{ki}=g^{-1}_{ki}A_kg_{ki}+g^{-1}_{ki}d g_{ki}.\nn\eea

To apply this to our bundle, choose two trivializing patches $U_{1,2}$ as in fig.\ref{fig_pinch}. On $U_{1,2}$, the bundle is trivialized using $\ga$ and $\gb$ respectively, thus the transition function is $g=\phi^{-1}_{\ga\gb}$. The connection is calculated to be
\bea A_1=\rho_2g^{-1}dg,~~~A_2=\rho_1\,gdg^{-1},~~~~~\rho_1+\rho_2=1.\nn\eea
The curvature can be computed with either $A_1$ or $A_2$
\bea &&F_1=dA_1+A_1A_1=d\rho_2 g^{-1}dg+(\rho_2^2-\rho_2)(g^{-1}dg)^2~~\Rightarrow\nn\\
&&\hspace{1cm}\Tr[F^2]=-2d\rho_2(\rho_2-\rho_2^2)\Tr[(g^{-1}dg)^3].\nn\eea
Note that $\Tr[F^2]$ vanishes at $0$ and $1$, as a relative class should be, see Eq.\ref{relative_cohomology}.

Thus the integration of the Pontryagin class is
\bea \int_{I\times M}\,\hat p_1=-\frac{1}{4\pi^2}\int_{M\times I}~2d\rho_2(\rho_2-\rho_2^2)\Tr[(dg\cdot g^{-1})^3]
=\frac{1}{12\pi^2}\int_{M}\Tr[(\phi^{-1}_{\ga\gb}d\phi_{\ga\gb})^3].\nn\eea
The rhs is twice the degree of the map $\phi_{\ga\gb}$.
\end{remark}

\subsection{The Canonical Framing}
Back to the CS theory. The phase of the 1-loop determinant is not defined until one chooses a 2-framing for $M$, which would seem to suggest that CS theory only gave invariants for framed 3-manifolds. Yet, it is possible to choose the \emph{canonical framing}, and in this way, one can still interpret CS theory as giving 3-manifold invariants, so long as one performs his computation under the canonical framing.

The material here is taken from ref.\cite{AtiyahFraming}.
For an orientable closed 3-manifold $Y$, choose a 4-manifold $Z$ that is bounded by $Y$. Given a choice of trivialization $\ga$ of $2TY$ one can fix a trivialization of $2T_Z$ on the boundary $\partial Z=Y$ (since the normal bundle is always trivial). Using the construction reviewed in the previous section, one can construct a new bundle from $2T_Z$ by pinching together $Y$. Denote by $\hat p_1(2T_Z,\ga)$ the Pontryagin class of the new bundle, which is a relative cohomology class with an $\ga$-dependence. Define the integer
\bea \varphi(\ga)=\textrm{Sign}\,Z-\frac16\int_Z~\hat p_1(2T_Z,\ga),\label{canonical_framing}\eea
where $\textrm{Sign}\,Z$ is the Hirzebruch signature of the 4-manifold (the number of positive eigen-values minus the number of negative eigen-values in the pairing matrix $H^2(Z)\times H^2(Z)\to\BB{R}$). The canonical framing is chosen such that the lhs of Eq.\ref{canonical_framing} is zero.

Eq.\ref{canonical_framing} is quite similar to another formula that measures the difference between the signature of a 4-manifold to its first Pontryagin number
\bea \textrm{sign}\,Z=\frac13\int_Z p_1(\Gamma_g)-\eta_g,\label{APS}\eea
where $\Gamma_g$ is the Levi-Civita connection associated with the metric $g$ of the 4-manifold $Z$, and $\eta_g$ is the eta invariant of $Y$ computed with metric $g|_Y$.  Indeed, at the canonical framing, the eta invariant can be written simply as the gravitational Chern-Simons term of $T_Y$. Recall that, the gravitational CS term, like the CS term of the gauge theory, is not invariant under large gauge transformations. For the case of $T_Y$, a large gauge transformation exactly corresponds to the changing of framings. To relate Eq.\ref{canonical_framing} and \ref{APS}, we need to relate the relative class $1/2\hat p_1(2T_Z,\ga)$ and $p_1(\Gamma_g)$.
\begin{figure}[h]
\begin{center}
\begin{tikzpicture}
\draw[rotate = 0](0,0) ellipse (1 and 1.4);
\draw[rotate = 0, fill=gray!40](0,0) ellipse (.6 and 1);
\node at (0,0) {\small $(Z,\,\Gamma_g)$};
\node at (1.4,0) {\small $(Y,\,0)$};
\node at (0.6,0.8) {\small $\alpha$};
\end{tikzpicture}\caption{The gray bulk is the interior of $Z$, the white strip is the collar of the boundary $Y$. In the gray area, the connection is the Levi-Civita connection, and in the collar is connection is deformed smoothly to zero. The framing of $Y$ is the canonical framing $\ga$.}\label{fig_collar}
\end{center}
\end{figure}
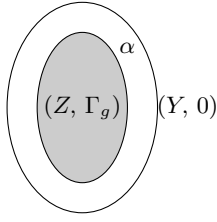
Referring to fig.\ref{fig_collar}, one can choose a connection that is Levi-Civita up to the boundary of the collar, and the connection becomes zero at the boundary. Thus the relative Pontryagin class $1/2\hat p_1(2T_Z;\ga)$ equals $p_1(\Gamma_g)$ in the gray area and is zero at the boundary. The integral in Eq.\ref{canonical_framing} can be broken into two parts
\bea \frac{1}{6}\int_Z\,\hat p_1(2T_Z;\ga)=\frac{1}{3}\int_{bulk}\,p_1(\Gamma_g)+\frac{1}{6}\int_{collar}\,\hat p_1(2T_Z;\ga).\nn\eea
Using the fact that the derivative of the CS term is the first Pontryagin class, the second integral is a surface term
\bea \frac12\int_{collar}\,\hat p_1(2T_Z;\ga)=-CS_{\ga},\nn\eea
where the subscript $\ga$ is to emphasize that the ambiguity of the CS term is now fixed by the framing. Combining the above two equations and Eq.\ref{APS}, one gets
\bea 3\eta_g=CS_{\ga}-3\varphi(\ga).\nn\eea
This means that, at the canonical framing, one can add to the Chern-Simons gauge theory action a \emph{local} counter term to offset the effect of $\eta_a$ .

In ref.\cite{FreedGompf}, the change of the 2-framing due to surgeries are worked out, but as this is quite specific to 2-framings, I will leave the discussion to the appendix.

%% file: Lecture_Notes_Appendix.tex
\section{}
\subsection{Computation of the Determinant}
The computation here is taken (almost word for word) from ref.\cite{RaySingerComplex}, but glossing over some mathematical paranoia such as the discussion about absolute convergence and the validity of exchanging summation with integration/differentiation, but expanding upon some technical details.

$\sbullet$ Flat $U(1)$-bundle $E$ over a torus, specified by the holonomy
\bea s(z+m\tau+n)=s(z)e^{2\pi i(mu+nv)},~~~u,v\in\BB{R}\backslash\BB{Z},\label{holonomy}\eea
where $\tau=\tau_1+i\tau_2,~\tau_2>0$ parameterizes the torus.

The eigen-modes can be chosen as
\bea \phi_{m,n}=\exp\Big(-\frac{\pi}{\tau_2}\big\{\bar z(m+u-\tau v)+z(n+v\bar\tau-u)\big\}\Big),~~\lambda_{m,n}=-\frac{4\pi^2}{\tau_2^2}|u+m-\tau(v+n)|^2.\label{eigen-mode}\eea
The determinant is $\log\det\Delta=-\zeta'(0)$
\bea \zeta(s)=\frac{1}{\Gamma(s)}\sum_{m,n}\int_0^{\infty}e^{t\lambda_{m,n}}t^{s-1}dt.\nn\eea

The following Poisson re-summation formula will be needed frequently in these notes
\bea \sum_{\ga\in\Lambda}\exp\big(-\eta\ga^2+2v\cdot\ga\big)=\Big(\frac{\textrm{vol}\Lambda^*}{\textrm{vol}\Lambda}\Big)^{1/2}
\Big(\frac{\pi}{\eta}\Big)^{d/2}\sum_{\gb\in\Lambda^*}\exp\Big(-\eta^{-1}\pi^2\big(\gb-\frac{iv}{\pi}\big)^2\Big)\label{Poisson_re_sum},\eea
where $\textrm{Re}\,\eta>0$, $\Lambda$ is a $d$ dimensional lattice embedded in $\BB{R}^d$, $\Lambda^*$ the dual lattice and $\cdot$ is the standard inner product in $\BB{R}^d$. This formula is derived from the simple fact
\bea \sum_{n\in\BB{Z}}e^{2i\pi nx}=\sum_{m\in\BB{Z}}\delta(x-m).\nn\eea

Apply this formula to $\zeta(s)$ to break up the term $|u+m-\tau(v+n)|^2$
\bea \sum_{m,n}\exp\Big(-\frac{4t\pi^2}{\tau_2^2}|u+m-\tau(v+n)|^2\Big)=\sum_{m,n}\frac{\tau_2}{4\pi t}\exp\Big(-\frac{1}{4t}|m\tau+n|^2+2\pi i(mu+nv)\Big),\nn\eea
compare with Eq.\ref{aymp_heat_kernel}.

Plugging this back into $\zeta(s)$, we have (the $m=n=0$ term drops)
\bea \zeta(s)&=&\frac{\tau_2}{4\pi\Gamma(s)}\sum_{m^2+n^2>0}e^{2\pi i(mu+nv)}\int_0^{\infty}~\exp\big(-\frac{|m\tau+n|^2}{4t}\big)t^{s-2}\nn\\
&=&\frac{\tau_2\Gamma(1-s)}{4\pi\Gamma(s)}\sum_{m^2+n^2>0}e^{2\pi i(mu+nv)}(\frac{4}{|m\tau+n|^2}\big)^{1-s}\nn\eea
The sum does not seem absolutely convergent at $s=0$ as it is, so in computing $\zeta'(0)$ we cannot set $s=0$ in the sum. However, if we first sum over $n$, it can be shown that the series is actually absolutely convergent. Thus we may simply set $s=0$ in the sum and get
\bea \zeta'(0)=I_1+I_2=\frac{2}{\pi}\tau_2\sum_1^{\infty}\frac{1}{n^2}\cos{2\pi nv}+\frac{1}{2\pi}\sum_{m\neq0}e^{2\pi imu}
\sum_{n=-\infty}^{\infty}e^{2\pi inv}\frac{2\tau_2}{|m\tau+n|^2}\nn\eea

For the first sum, we compute the following instead
\bea I'_1=\sum_{-\infty}^{\infty}\frac{1}{n^2}{\sin}^2{\pi nv}.\nn\eea
As usual apply once again the Poisson resummation
\bea I'_1=\int_{-\infty}^{\infty}\frac{1}{x^2}{\sin}^2(xv\pi)e^{inx}=\lim_{\epsilon\to 0}\int_{-\infty-i\epsilon}^{\infty-i\epsilon}\frac{1}{x^2}{\sin}^2(xv\pi)e^{inx},\nn\eea
where the second step is valid since the integral is absolutely convergent. Now assume that $-1<v<1$, and compute the integral by contour integral
\bea I'_1&=&-\frac{1}{4}\int_{-\infty-i\epsilon}^{\infty-i\epsilon}\frac{1}{x^2}\Big[e^{2\pi ix(n+v)}+e^{2\pi ix(n-v)}-2e^{2\pi ixn}\Big]\nn\\
&=&-\frac{2\pi i}{4}\Big[\sum_{n>-v}2\pi i(n+v)+\sum_{n>v}2\pi i(n-v)-2\sum_{n>0}2\pi in\Big]=\pi^2|v|\nn\eea
Thus
\bea \frac{\pi}{2\tau_2}I_1=\sum_1^{\infty}\frac{1}{n^2}-\sum_{n\neq0}\frac{{\sin}^2(n\pi v)}{n^2}=\frac{\pi^2}{6}-I_1'+(v\pi)^2=\frac{\pi^2}{6}-\pi^2|v|+\pi^2v^2.\nn\eea

As for $I_2$, we see that if the sum over $n$ is an integral, then we can change variable and free $m$
\bea \textrm{from}~\sum_n~1/|m\tau+n|^2(\cdots)~~\textrm{to}~~\frac{1}{m}\int dn\frac{1}{|\tau+n|^2}(\cdots),\nn\eea
then the sum over $m$ can be performed. To change summation to integration, we can apply the Poisson re-summation
\bea I'_2=\sum_{n=-\infty}^{\infty}e^{2\pi inv}\frac{1}{|m\tau+n|^2}=\sum_{n=-\infty}^{\infty}\int~e^{2\pi i(vx+nx)}\frac{1}{|m\tau+x|^2}\nn\eea
This is a trick that occurs over and over again in this type of calculations.

The denominator has simple zeros at $x=-\tau_1\pm i\tau_2$, thus depending on whether $n>v$ or $n<v$ (assume of course $0<v<1$) we have
\bea I_2'&=&\sum_{0}^{\infty}\frac{2\pi i}{2|m|\tau_2i}\exp\big(2\pi i(n+v)(-m\tau_1+i|m|\tau_2)\big)-\sum_{-\infty}^{-1}\frac{2\pi i}{-2|m|\tau_2i}\exp\big(2\pi i(n+v)(-m\tau_1-i|m|\tau_2)\big)\nn\\
&=&\frac{\pi}{|m|\tau_2}\sum_{n=-\infty}^{\infty}\exp\big(2\pi i(n+v)(-m\tau_1+i\epsilon_n|m|\tau_2)\big),~~~~~\epsilon_n=\big\{\mathop{}^{+1,~~n\geq0}_{-1,~~n<0}.\eea
Now the summation over $m$ simply gives a log
\bea I_2&=&\frac{\tau_2}{\pi}\sum_{m\neq0}e^{2\pi imu}\frac{\pi}{|m|\tau_2}\sum_{n=-\infty}^{\infty}\exp\big(2\pi i(n+v)(-m\tau_1+i\epsilon_n|m|\tau_2)\big)\nn\\
&=&-\sum_{n=-\infty}^{\infty}\log\Big|1-\exp\big(2\pi iu+2\pi i(n+v)(-\tau_1+i\epsilon_n\tau_2)\big)\Big|^2\nn\\
&=&-\log\Big(\big|1-e^{2\pi i(u-\tau v)}\big|^2\cdot
\prod_{n=1}^{\infty}|1-e^{2\pi i(u-\tau v+n\tau)}|^2\big|1-e^{2\pi i(-u+\tau v+n\tau)}|^2\Big)\nn\\
&=&-\log\prod_{n=-\infty}^{\infty}\Big|1-\exp2\pi i\big(|n|\tau-\epsilon_n(u-v\tau)\big)\Big|^2.\nn\eea

Finally, the determinant is ($w=u-\tau v$ and $0<v<1$)
\bea\log{\det}'\Delta_w=-\zeta'(0)=-2\tau_2\pi\big(\frac{1}{6}-|v|+v^2\big)+\log\prod_{n=-\infty}^{\infty}\Big|1-\exp2\pi i\big(|n|\tau-\epsilon_nw\big)\Big|^2.\label{det_non_zero}\eea

To compute ${\det}'\Delta_0$, one can recycle the above computation, but manually remove the mode
\bea -\frac{4\pi^2}{\tau_2^2}|u-\tau v|^2\nn\eea
from Eq.\ref{eigen-mode} and then let $w\to 0$
\bea {\det}'\Delta_0=\exp\bigg(-\frac{\tau_2\pi}{3}+\log\prod_{n\neq 0}\Big|1-\exp2\pi i\big(|n|\tau-\epsilon_nw\big)\Big|^2+\log\tau_2^2\bigg).\label{det_zero_prem}\eea
However this is not the full story, the determinant line bundle we are dealing with now is
\bea 0\to \Omega^{0,0}(T^2,E)\stackrel{\bar\partial_A}{\to} \Omega^{0,1}(T^2,E)\to 0,\nn\eea
where $E$ is the flat bundle whose holonomy is given by Eq.\ref{holonomy}. When the holonomy is non-trival, $\bar\partial_A$ has no zero modes, but by setting to zero the holonomy, the bundle $E$ becomes trivial and the above complex develops two zero modes $H^{0,0}(T^2)$, $H^{0,1}(T^2)$. So to compute the determinant, one needs to choose an explicit embedding of the two cohomology groups into $\Omega^{0,0}(T^2,E)$ and $\Omega^{0,1}(T^2,E)$ respectively. Or equivalently, what we have computed above should be regarded as (the absolute valued squared) of the section $(H^{0,0}(T^2))^*\otimes H^{0,1}(T^2)$ over the moduli space of the torus. As both cohomology groups are 1-dimensional, the section $|(H^{0,0}(T^2))^*\otimes H^{0,1}(T^2)|^2$ can be chosen as the integral $\int_{T^2}dz d\bar z\sim\tau_2$.
Thus dividing Eq.\ref{det_zero_prem} by $\tau_2$, one gets the final result for ${\det}'\Delta_{0}$
\bea {\det}'\Delta_0=\exp\bigg(-\frac{\tau_2\pi}{3}+\log\prod_{n\neq 0}\Big|1-\exp2\pi i\big(|n|\tau-\epsilon_nw\big)\Big|^2+\log\tau_2\bigg).\label{det_zero}\eea

\subsection{Knot Complement and the Alexander Polynomial}\label{KCatAP}
The treatment of this section is taken from ref.\cite{MilnorAlexander}, even though Milnor was not the first to prove the results below, his treatment is along the line of discussion of this note, and is an illustration of changing the coefficient ring of a complex so as to define the torsion.

Consider a knot inside of $K\hookrightarrow S^3$, let $\BS{K}$ be the tubular neighbourhood of $K$, then it is easy to see that the first homotopy group of the knot complement $\overline{S^3\backslash\BS K}$ has an infinite cyclic factor, generated by the cycle having linking number 1 with the knot $K$, we denote this generator by $t$. To deduce the homology of the knot complement, we can use the Mayer-Vietoris sequence (or simple intuition)
\bea &&\cdots \to H_i(\BS K\cap \overline{S^3\backslash\BS K})\stackrel{(i_*,-i_*)}{\longrightarrow} H_i(\BS K)\oplus H_i(\overline{S^3\backslash\BS K})\stackrel{+}{\longrightarrow } H_i(S^3)\stackrel{\partial_*}{\longrightarrow}\nn\\
&&\hspace{3cm}\to H_{i-1}(\BS K\cap \overline{S^3\backslash\BS K})\to H_{i-1}(\BS K)\oplus H_{i-1}(\overline{S^3\backslash\BS K})\to\cdots\label{MV_sequence}\eea
As the intersection $\BS{K}\cap\overline{S^3\backslash\BS K}=T^2$, one gets
\bea H_i(\overline{S^3\backslash\BS K})=\bigg\{\begin{array}{c}
                                                 \BB{Q}~~~i=0,1 \\ 0~~~i=2,3\end{array}.\nn\eea
The rational coefficient is used for safety, if one uses instead integer coefficient, there might be elements in the homology of finite order. Thus with $\BB{Z}$ coefficients, the simplicial complex of the knot complement is not acyclic. Let us try to change the coefficient ring. The intuitive idea is similar to the example on pg.\pageref{ex_flat_bundle}, where the circle has non-zero cohomology, but by adding a non-trivial holonomy, cohomology is killed.

To explain this, let me first add some basic facts about covering spaces. Let $X\stackrel{h}{\to}M$ be a covering, which can be more conveniently thought of as a bundle with discrete fibre. Then $h_*\pi_1(X)\subset \pi_1{M}$ is a normal subgroup. There is a fixed point free action of the quotient group $\pi_1(M)/h_*\pi_1(X)$ on $X$, called the \emph{deck transformation}. The action is constructed as follows, let $p\in X$ be a pre-image of $h^{-1}x$, and $g\in\pi_1(M,x)$, there is a unique lift of $g$ to $X$ starting from $p$ ending at, say $q\in h^{-1}x$. The action thus sends $p$ to $q$. Clearly, if $g\in h_*\pi_1(X)$, then $p=q$, the action is trivial.

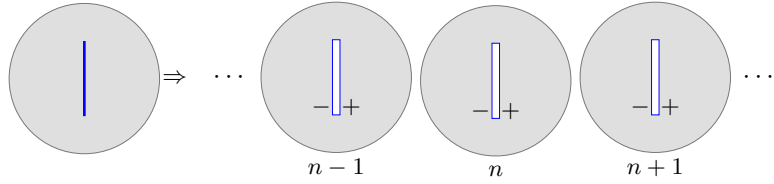
\begin{figure}[h]
\begin{center}
\begin{tikzpicture}[scale=1]
\draw [line width=0.1, fill=gray!50, opacity=.5] (0,0) circle (1);
\draw [line width=1, blue] (0,.5) -- (0, -0.5);
\node at (1.2,0) {\small$\Rightarrow$};
\node at (0,-1.3) {\small$~$};
\end{tikzpicture}
\begin{tikzpicture}[scale=1]
\draw [line width=0.1, fill=gray!50, opacity=.5] (0,0) circle (1);
\draw [line width=3, white] (0,.5) -- (0, -0.5);
\draw[blue, line width=.3] (-.05,-0.5) rectangle (.05,.5);
\node at (-0.2,-0.4) {\small$-$};
\node at (0.2,-0.4) {\small$+$};
\node at (0,-1.2) {\small$n-1$};
\node at (-1.4,0) {$\cdots$};
\end{tikzpicture}
\begin{tikzpicture}[scale=1]
\draw [line width=0.1, fill=gray!50, opacity=.5] (0,0) circle (1);
\draw [line width=3, white] (0,.5) -- (0, -0.5);
\draw[blue, line width=.3] (-.05,-0.5) rectangle (.05,.5);
\node at (-0.2,-0.4) {\small$-$};
\node at (0.2,-0.4) {\small$+$};
\node at (0,-1.2) {\small$n$};
\end{tikzpicture}
\begin{tikzpicture}[scale=1]
\draw [line width=0.1, fill=gray!50, opacity=.5] (0,0) circle (1);
\draw [line width=3, white] (0,.5) -- (0, -0.5);
\draw[blue, line width=.3] (-.05,-0.5) rectangle (.05,.5);
\node at (-0.2,-0.4) {\small$-$};
\node at (0.2,-0.4) {\small$+$};
\node at (0,-1.2) {\small$n+1$};
\node at (1.4,0) {$\cdots$};
\end{tikzpicture}\caption{The slit on the lhs is the disc bounded by the knot viewed from the side. One then removes a small tbn of this disc. To the right, the $-$ bank of the $(n-1)^{th}$ copy is glued to the + bank of the $n^{th}$ copy, whose $-$ bank is in turn glued to the + bank of the $(n+1)^{th}$ copy.}\label{fig_seifert_cover}
\end{center}
\end{figure}

\begin{figure}[h]
\begin{center}
\begin{tikzpicture}[scale=1]
\draw [line width=0.1, fill=gray!50, opacity=.5] (0,0) circle (1);
\draw [line width=1, blue] (0,.5) -- (0, -0.5);
\draw [->,line width=.3, blue] (0,-.75) arc (-90:-270:0.4);
\draw [->,line width=.3, blue] (0,.05) arc (90:-90:0.4);
\node at (-.5,-0.5) {\small$t$};
\node at (1.2,0) {\small$\Rightarrow$};
\end{tikzpicture}
\begin{tikzpicture}[scale=1]
\draw [line width=0.1, fill=gray!50, opacity=.5] (0,0) circle (1);
\draw [line width=3, white] (0,.5) -- (0, -0.5);
\draw[blue, line width=.3] (-.05,-0.5) rectangle (.05,.5);
\draw [<-,line width=.3, blue] (-0.05,.25) arc (95:150:0.6);
\draw [->,line width=.3, blue] (0.05,.25) arc (90:-96:0.5);
\draw [->,line width=.3, blue] (0,-.75) arc (-90:-263:0.4);
\draw [->,line width=.3, blue] (0.05,0.04) arc (90:-90:0.34);
\draw [->,line width=.3, blue] (0.05,-.64) arc (-90:-250:0.24);
\draw [-,line width=.3, blue] (0.05,-.16) arc (90:0:0.2);
\end{tikzpicture}\caption{What used to be a closed curve on the lhs is now un-wound and is traversing an infinite spiral staircase on the rhs}\label{fig_spiral}
\end{center}
\end{figure}
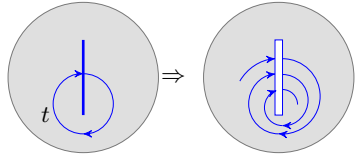

One can always construct a covering space of the knot complement $Xs\stackrel{h}{\to}\overline{S^3\backslash\BS{K}}$, on which the infinite cyclic factor called $t$ above acts by deck transformation. One way to construct the covering is the following (ref.\cite{Lickorish} ch.6). Let $D\subset S^3$ be the disc subtended by $K$ (called the Seifert surface), thicken this disc slightly and remove it from $S^3$, the remainder is depicted in fig.\ref{fig_seifert_cover} as a gray area with a white slit. Take infinitely many copies of this remainder and do according to the caption of fig.\ref{fig_seifert_cover}.
Then $t$, which used to be a closed curve (lhs of fig.\ref{fig_spiral}) is un-wound. And $t$ clearly acts by translation from $n$ to $n+1$.

Back to our problem, we have constructed a covering $X$ of $Y=\overline{S^3\backslash\BS{K}}$, there is a simplicial action of $t$ on $C_{\sbullet}(X,\BB{Z})$.
Consider the group ring $R=\BB{Z}[t,t^{-1}]$ and its field of fractions $F$, which is the field of rational functions in $t$ over $\BB{Q}$. One can show that, the complex $C_{\sbullet}(X,\BB{Z})$ considered as an $F$-module by means of the action of $t$, is acyclic. The combinitorial proof is found in ref.\cite{MilnorAlexander}, but the rough idea is already illustrated by the example on pg.\pageref{ex_flat_bundle}.

One needs to compute the torsion, but before doing so, one can first drastically simplify the complex $C_{\sbullet}(Y)$. Since $Y$ has now a boundary made up of some 2-cells, these 2-cells can be pushed in without changing the torsion (see the next remark). One can keep doing this, until one is left with a 2-dimensional complex (namely all the 3-cells can be gotten rid of this way, for otherwise, it would contradict $H_3(Y)=0$). Similarly, 1-cells with distinct end points can be shrunk, nor is this going to affect the torsion. The complex $Y$ after these simplifications will be called $\tilde Y$, it has only one zero cell. The covering of $\tilde Y$ is still called $X$.
\begin{remark}
Both assertions above about the invariance of torsion can be proved by using the property Eq.\ref{additivity_torsion}. Let me illustrate the first case, the second is left to the reader. Look at the middle picture of the cartoon fig.\ref{fig_push_in}. I want to push the red cell (2-dimensional, even though drawn as an arc) into the wedge. The sequence of complexes of Eq.\ref{additivity_torsion} is indicated in the figure.
\begin{figure}[h]
\begin{center}
\begin{tikzpicture}[scale=.6]
\draw [-,line width=0.1, fill=blue!30, opacity=.5] (0,0) circle (1);
\draw [white, fill=white] (30:0) -- (30:1)
arc (30:-30:1) -- cycle;
\node at (-1.5,0) {\small$0\to$};
\node at (1.3,0) {\small$\to$};
\node at (0,-1.2) {\small$A'$};
\end{tikzpicture}
\begin{tikzpicture}[scale=.6]
\draw [-,line width=0.1, fill=blue!30, opacity=.5] (0,0) circle (1);
\draw [-,line width=1, red] (.866,.5) arc (30:-30:1);
\draw [-,line width=1, blue] (0,0) -- (.866,.5) (0,0) -- (.866,-.5);
\node at (1.4,0) {\small$\to$};
\node at (0,-1.2) {\small$A$};
\end{tikzpicture}
\begin{tikzpicture}[scale=.6]
\draw [blue, fill=blue!30, opacity=.5] (30:0) -- (30:1)
arc (30:-30:1) -- cycle;
\draw [-,line width=1, red] (.866,.5) arc (30:-30:1);
\draw [dashed,line width=.8, blue] (0,0) -- (.866,.5) (0,0) -- (.866,-.5);
\node at (1.5,0) {\small$\to 0$};
\node at (.5,-1.2) {\small$A''$};
\end{tikzpicture}\caption{The 2-$d$ cartoon of the pushing-in. The 2-cell to be pushed in is in red, and the dashed lines are pinched to one point.}\label{fig_push_in}
\end{center}
\end{figure}
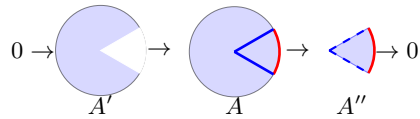
The space $A''$ is obtained from $A$ by pinching $A'$ together, and therefore is equivalent to a 3-ball bounded by a 2-sphere (the red 2-cell), and the torsion of this complex is 1 (lemma 7.2 \cite{MilnorTorsion}). Thus the torsion of $A'$ equals that of $A$.
\end{remark}

One can now compute the torsion of the complex
\bea 0\to C_2(X)\stackrel{\partial_2}{\to}C_1(X)\stackrel{\partial_1}{\to}C_0(X)\to 0.\label{complex_alexander}\eea
We first observe, as an $F$-module, $C_{\sbullet}(X)$ has a set of preferred bases. This is because for every cell of $\tilde Y$, its pre-image in $X$ are all related by the action of $t$. Any representative can be used as a basis, and different choices of the basis differ by a power of $t$, thus the torsion is well-defined up to some powers of $t$ too. We name the basis as
\bea \begin{array}{ccc}
       e & f_i & g_j \\
       \dim\,0 & \dim\,1 & \dim\,2\end{array},~~~j=1,\cdots r;~~~i=1,\cdots r+1.\nn\eea
For all 1-cells $f_i$, we have $\partial_1 f_i=(t^r-t^s)e$ and without loss of generality one can assume that the first 1-cell $f_1$ has boundary $\partial_1 f_1=(1-t)e$, thus in matrix form (where all the entries are divisible by $1-t$)
\bea \partial_1=\left[\begin{array}{c} 1-t \\ \vdots \end{array}\right]=(1-t)\left[\begin{array}{c} 1 \\ \vdots \end{array}\right].\label{matrix_del_1}\eea
The matrix corresponding to $\partial_2$ is $r\times (r+1)$, and we write it as
\bea \partial_2=\big[\begin{array}{cccc}
                  v_1 & v_2 & \cdots & v_{r+1} \end{array}\big],\label{matrix_del_2}\eea
where each $v$ is an $r$ dimensional vector, whose entries are in $R=\BB{Z}[t,t^{-1}]$ (obs. with \emph{integer} coefficients). The relation $\partial_2\partial_1=0$ tells us that $v_1$ is the linear cimbination of the rest of $v$'s. Let $A(t)$ be the determinant of the minor
\bea A(t)=\det\big[\begin{array}{cccc}
                  v_2 & \cdots & v_{r+1} \end{array}\big],\nn\eea
then it is straightforward to see that the torsion of the complex Eq.\ref{complex_alexander} is
\bea &&(f_1\cdots f_{r+1})/(g_1\cdots g_r)=\frac{1}{A(t)}f_1,\nn\\
&&\Delta=e/(f_1\cdots f_{r+1})/(g_1\cdots g_r)=\frac{A(t)}{t-1},\nn\eea
well-defined up to powers of $t$. Note that the torsion is in $F$, but the $A(t)$ is $R$ and it is called the \emph{Alexander polynomial}.

The Alexander polynomial has an important property (see ref.\cite{MilnorAlexander})
\bea A(t^{-1})=\pm t^iA(t),~~~A(1)=\pm1.\nn\eea
 One can normalize $A(t)$ and by multiply it with an appropriate power of $t$, achieve
\bea A(1)=1,~~~A(t^{-1})=A(t),~~~\partial_tA(t)\big|_{t=1}=0.\label{property_alexander}\eea
In fact, the first two imply the third.

\begin{remark}\emph{Geometric construction of the Alexander polynomial,} See ref.\cite{Lickorish} ch.6, ch.7\\
The Alexander polynomial is constructed as an invariant associated to the \emph{presentation} of the module $H_1(X,R)$, where as a reminder $R=\BB{Z}[t,t^{-1}]$.
The presentation of a module $\SF M$ is a sequence
\bea \SF{R}\stackrel{h}{\to} \SF F \to \SF M\to 0,\nn\eea
where $\SF R$ $\SF F$ are free modules and $\SF R$ is called 'relations'. If one writes the map $h$ as a matrix, then it can be shown that the ideal generated by the minors of $h$ is in a certain sense independent of the choice of the presentation or the choice of basis for $\SF F$, $\SF R$. Particularly, the determinant of the biggest minor is the Alexander polynomial.

Coming back to our problem, the module in question is $H_1(X;R)$, it can be presented as $C_2 \to \ker\partial_1\to H_1(X;R)\to 0$.
We again look at the complex Eq.\ref{complex_alexander}, \emph{except now we use $R$ as the coefficient}. We need to find out the kernel of $\partial_1$, but from Eq.\ref{matrix_del_1}, the kernel is $r$-dimensional, generated by
\bea -\frac{\partial_1^i}{\partial_1^1}f_1+f_i,~~~~i=2,\cdots r+1,\nn\eea
where $\partial_i^i$ denotes the $i^{th}$ entry of $\partial_1$ in Eq.\ref{matrix_del_1}. We take these as the generator of $\SF F$. As for $\SF{R}$, the generators are chosen as $g_i$ and the matrix form of $h$ is $h=[v_2,\cdots,v_{r+1}]$ as in Eq.\ref{matrix_del_2}. Clearly, the determinant of $h$ is $A(t)$.
\end{remark}

\subsection{Framing Change from Surgeries}
In ref.\cite{FreedGompf}, it was shown that if one performs an $SL(2,\BB{Z})$ surgery along a link inside of a 3-manifold with prescribed framing, then the new manifold inherits (unambiguously) a framing from the old one. However, the discussion there as well as the review given in ref.\cite{KirbyMelvin} are both laconic, I will try to insert some (hopefully not too verbose) explanations here and there. And by framing below, I will mean 2-framing.

\begin{figure}[h]
\begin{center}
\begin{tikzpicture}[scale=.8]
\draw [semithick,blue, fill=gray!20] (0,0) circle (1);
\draw [semithick,blue, fill=white] (0,0) circle (.6);
\draw [semithick, black, ] (0,0) circle (.8);
\draw [<-,blue] (1,0) to [out=25, in=-145] (2,-.3);
\draw [<-,blue] (.43,-.43) to [out=-45, in=-145] (2,-.3);
\draw [<-,blue] (.57,.57) to [out=45, in=145] (2,.3);
\node at (2,-.3) [right] {\small{original framing}};
\node at (2,.3) [right] {\small{standard framing}};
\end{tikzpicture}~~
\caption{The blue band is the thickened $T^2$, the original framing on the inner and outer $T^2$ are homotoped to the standard framing in the middle.} \label{fig_thickenT2}
\end{center}
\end{figure}
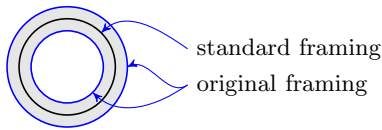

Let $L$ be a link in $S^3$ with a prescribed framing, and let $T^2$ be the boundary of the tbn of $L$. One can assume that this framing of $S^3$ restricted to $T^2$ is the standard one, specified as follows. Let the solid torus (the tbn of $L$) $S^1\times D^2$ be parameterized by $(\phi,r,\theta)$, then the framing restricted to $T^2=S^1\times S^1$ is given by two copies of $(\hat r,\hat\theta,\hat\phi)$. If the given framing of $S^3$ does not restrict to this standard form on $T^2$,
one can thicken $T^2$ a little, see fig.\ref{fig_thickenT2} and put the standard framing on the middle $T^2$. A homotopy of framing from that of the outer and inner torus to that of the middle one can always be found, essentially because $\pi_2(\textrm{spin}(6))=0$ (the reasoning is very close to the one used in the beginning of sec.\ref{2Fo3M}). One then does surgery by cutting along the middle torus instead. Even though there are a different homotopy classes of the homotopy above, it is easy to see that the framing change due to the homotopy on the inner and outer blue band cancels.

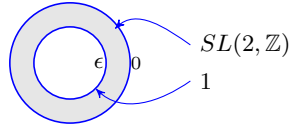
\begin{figure}[h]
\begin{center}
\begin{tikzpicture}[scale=.8]
\draw [semithick,blue, fill=gray!20] (0,0) circle (1);
\draw [semithick,blue, fill=white] (0,0) circle (.6);
\node at (1.1,0) {\scriptsize$0$};
\node at (.48,0) {\small$\epsilon$};
\draw [<-,blue] (.43,-.43) to [out=-45, in=-145] (2,-.3);
\draw [<-,blue] (.75,.57) to [out=45, in=145] (2,.3);
\node at (2,-.3) [right] {\small$1$};
\node at (2,.3) [right] {\small$SL(2,\BB{Z})$};
\end{tikzpicture}~~
\caption{The outer circle is the $T^2$ along which we cut the solid torus and perform the surgery. The framing on the inner and outer torus differ by an $SL(2,\BB{Z})$ which is interpolated to 1 from $0$ to $\epsilon$.} \label{fig_interpo}
\end{center}
\end{figure}

The solid torus that is taken out inherits the original framing from $S^3$, which is now assumed to be the standard one when restricted to $T^2$. When one reglues it back after an $SL(2,\BB{Z})$ twist, there is a discontinuity of framing along $T^2$: across $T^2$ the $\hat r$ direction remains unchanged but the $\hat\theta,\hat\phi$ directions undergo an $SL(2,\BB{Z})$ twist. One again shrinks the $T^2$ a little, and in this collar (which is $T^2 \times [0,\epsilon]$), the $SL(2,\BB{Z})$ twist is  homotoped to the identity as one go from 0 to $\epsilon$, see fig.\ref{fig_interpo}. This homotopy unambiguous if one demands that during the homotopy, the image of $T^2$ in $\textrm{spin}(6)$ is a point; thus the new manifold has an unambiguous inherited framing. The following example is not a good example in itself, but it is very concrete.
\begin{example}\emph{Framing Change from $T$}\\
Recall that for $T$, one takes out from $M$ a solid torus, adds a right hand twist to the solid torus and replace it into $M$. Parameterize the solid torus $S^1\times D^2$ as above and on $T^2$ it has the standard framing $2(\hat r,\hat\theta,\hat\phi)$.
We choose now an arbitrary extension of this framing into the interior of the solid torus. Since we are interested in obtaining the framing change, the choice of extension does not matter. Decompose the 2-framing at the torus as the 2-framing of $2T_{D^2}$ plus twice $\hat\phi$ (which remains unchanged throughout the inward extension).
Parameterize twice the tangent bundle $2T_{D^2}$ as
\bea \BB{H}=\BB{C}\oplus j\BB{C}.\nn\eea
Let $\lambda(r)$ be a monotonous function such that $\lambda(0)=0,~\lambda(1)=\pi/2$. Define an $\BB{H}$ valued function
\bea f(r,\theta)=e^{\frac i2\theta}e^{j\lambda(r)} e^{-\frac i2\theta}e^{-j\lambda(r)},\nn\eea
then the following defines four linearly independent sections of $2T_{D^2}$
\bea f(r,\theta),~~~f(r,\theta)i,~~~f(r,\theta)j,~~~f(r,\theta)k.\nn\eea
Note that $f(0,\theta)=1,~f(1,\theta)=e^{i\theta}$, thus the 2-framing is rotational invariant along $\theta$ and $\phi$ as was assumed in the beginning.

The $T$ surgery glues the solid torus back through a diffeomorphism $\tau$ on the torus
\bea\tau:~~\theta'=\theta+\phi,~~\phi'=\phi.\nn\eea
Note that this diffeomorphism can be extended to the interior of the solid torus
\bea \tau:~~\theta'=\theta+\phi,~~\phi'=\phi,~~r'=r.\nn\eea
This is an indication that $T$ surgery does not change the diffeomorphism type of the manifold. In general, $\tau$ cannot be extended.

The effect of $\tau$ on $\hat\phi$ is $\hat\phi\to \hat \phi+\hat\theta$, while its effect on $(\hat r,\hat\theta)$ can be worked out as follows.
Let $\zeta=f(r,\theta)\ga,~~\ga\in\BB{H}$ be any section of $2T_{D^2}$. By definition $\zeta$ acts on a function $h$ over $S^1\times D^2$ as
\bea \zeta\circ h(re^{i\theta},\phi)=\frac{d}{dt}h(re^{i\theta}+t\zeta,\phi)\big|_{t=0}.\nn\eea
Thus the pushforward acts as
\bea (\tau_*\zeta\circ h)(re^{i\theta'},\hat\phi)=\frac{d}{dt}h(e^{i\phi}(re^{i\theta}+t\zeta),\phi)\big|_{t=0}.\nn\eea
This gives the expression  for $\tau_*\zeta$
\bea \tau_*\zeta=e^{i\phi}f(r,\theta)\ga=e^{i\phi'}f(r',\theta'-\phi')\ga=e^{\frac i2(\theta'+\phi')}e^{j\lambda(r)} e^{-\frac i2(\theta'-\phi')}e^{-j\lambda(r)}\ga.\nn\eea
So the new framing of $2T_{D^2}$ differs from the old one by the function
\bea g^{-1}=e^{i\phi'}f(r',\theta'-\phi')f^{-1}(r',\theta')=e^{\frac i2(\theta'+\phi')}e^{j\lambda(r)}e^{\frac i2\phi'}e^{-j\lambda(r)}e^{-\frac i2\theta'},\nn\eea
One sees that at $r'=r=1$, $g=1$ and hence $\hat r,\hat \theta$ are not changed at $T^2$. We thus only need to homotope $\hat\phi+\hat\theta$ back to $\hat\phi$, this homotopy does not contribute to the essential change of framing. The only essential change of framing comes from the white interior in fig.\ref{fig_interpo}, which we compute next.

On the white interior, the old and new framing $2T_{D^2}$ differ by the function $g$ above, while $\hat\phi$ is unchanged.
Even though $g$ is map $S^1\times D^2\to SU(2)$, it is equal to identity at $\{\phi=0\}\cup \{r=1\}$, so $g$ can be still taken as a map $S^3\to SU(2)$.\footnote{Because $S^1\times D^2$ with $S^1\times S^1\cup \{*\}\times D^2$ pinched is equivalent to the suspension of $S^2$, which is $S^3$.} To compute the degree of this map, one can use the Pauli matrices to represent the quaternions $i=\sqrt{-1}\sigma^1$, $j=\sqrt{-1}\sigma^2$ and $k=-\sqrt{-1}\sigma^3$
and the degree is the integral
\bea \frac{1}{24\pi^2}\int \Tr\big[(g^{-1}dg)^3\big].\nn\eea
One can work out
\bea X=\Tr[g^{-1}(\partial_{\theta}g)\big(g^{-1}(\partial_{\phi}g)g^{-1}(\partial_{r}g)-g^{-1}(\partial_{r}g)g^{-1}(\partial_{\phi}g)\big)]
=16\dot{\lambda}\cos^3\lambda\sin\lambda\sin^2\frac{\phi}{2}\nn\eea
Thus
\bea \int\Tr(g^{-1}dg)^3=3\int~d\theta d\phi dr~X=48\int_0^{2\pi}d\phi\int_0^{2\pi}d\theta\int_0^{\pi/2}d\lambda~\cos^3\lambda\sin\lambda\sin^2\frac{\phi}{2}=24\pi^2,\nn\eea
verifying the claim.
\end{example}

Back from this long digression, we now focus on integer surgeries. Suppose the manifold $X_L$ is obtained from $S^3$ after some integer surgeries along the link, and that each component $L_i$ of the link is marked with the integer $p_i$ (recall that this means performing a surgery $T^{p_i}S$ along this component). Suppose also that $S^3$ is of canonical framing in the beginning, then in ref.\cite{FreedGompf} it was worked out that after the surgery, $X_L$ inherits a framing (constructed above) different from the canonical framing by
\bea \phi_L=-3\sigma_L+\sum_i p_i,\label{formula_framing}\eea
where $\sigma_L$ is the signature of the linking matrix of the link $L$.
\begin{remark}
$\sigma_L$ is also the signature of the 4-manifold $Z_L$ that is bounded by $X_L$ ($Z_L$ is presented as a handle body). Let us take a knot $L$ embedded in $S^3$, then the handle body is obtained as follows. Since $L$ has a tbn in $S^3$ equivalent to $S^1\times D^2$, the 2-handle is attached to the 4-ball $B^4$ in the following manner
\bea Z_{L}=B^4\cup^{f}_{S^1\times D^2} D^2\times D^2,\nn\eea
see fig.\ref{fig_handle_body}.
\begin{figure}[h]
\begin{center}
\begin{tikzpicture}[scale=.8]
\draw [blue, fill=blue!60,opacity=.4] (0,0) circle (1.4);
\draw [dashed,blue] (-1.4,0) to [out=90, in=90] (1.4,0);
\draw [-,blue] (-1.4,0) to [out=-90, in=-90] (1.4,0);
\node at (-0.6,.2) {\scriptsize$\sbullet$};
\node at (0.6,.2) {\scriptsize$\sbullet$};
\draw [red, line width =9pt, opacity=.6] (-.6,.2) to [out=110, in=180] (0,1.6);
\draw[rotate = 0, fill=red, opacity=.6 ](-0.6,0.2) ellipse (.18 and .12);
\draw [red, line width =9pt, opacity=.6] (.6,.2) to [out=70, in=0] (0,1.6);
\draw[rotate = 0, fill=red, opacity=.6 ](0.6,0.2) ellipse (.18 and .12);
\end{tikzpicture}
\caption{Handle-body. This picture is actually misleading: the 4-ball is drawn as the 3-ball, and hence the knot embedded in $S^3$, which should be like $S^1$, is drawn correspondingly as $S^0$, that is, two points.}\label{fig_handle_body}
\end{center}
\end{figure}
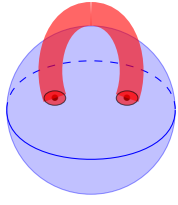
Concretely, the 2-handle $D^2\times D^2$ has boundary $S^1\times D^2\cup D^2\times S^1$, and the first part of the boundary is glued onto $B^4$ along the tbn of the knot (which is also $S^1\times D^2$) with an appropriate diffeomorphism\footnote{Generically such gluing gives manifolds with corners, so some smoothing procedure is needed, but I am not qualified to discuss about this point.} $f$. If, for example, the knot is marked with an integer $p$, then the gluing diffeomorphism consists of $p$ twists. To identify the 2-cycles, one can use short exact sequence (where $h$ denotes the handle $D^2\times D^2$)
\bea
0\to C_{\sbullet}(B^4\cap h) \stackrel{(i_*,-i_*)}{\longrightarrow} C_{\sbullet}(B^4)\oplus C_{\sbullet}(h)\stackrel{+}{\longrightarrow} C_{\sbullet}(B^4\cup^f h)\to 0\nn\eea
and the corresponding Mayer-Vietoris sequence
\bea 0\to H_2(B^4\cup^f h)\stackrel{\partial_*}{\longrightarrow} H_{1}(B^4\cap h)\to 0\nn.\eea
From this sequence, one can identify a representative for a 2-cycle in $H_2(B^4\cup^f h)$ as the sum $c_1+c_2$, with $c_{1,2}$ in $C_2(B^4)$ and $C_2(h)$ respectively, such that $\partial c_1=-\partial c_2$ and both are equal to the meridian $S^1\times\{0\}$. So $c_1$ can be taken as the Seifert surface of $L$ within $S^3=\partial B^4$ and $c_2=D^2\times\{0\}$. Construct similarly a representative 2-cycle for another link $L'$, then the intersection number of these two cycles is easily seen to be the number of times $\partial c_2=S^1\times \{0\}$ penetrates the Seifert surface of $L'$ (one can of course switch the role of $L$ and $L'$). The general situation with more complicated links is similar.
\end{remark}

I will not try to prove Eq.\ref{formula_framing} here, but rather give some examples so that the reader get a feeling of how things work.
The actual proof of Eq.\ref{formula_framing} is based on the examples plus an induction, see Thm 2.3 \cite{FreedGompf}.
\begin{example}\emph{Framing change from $TS$}\\
We observed in sec.\ref{sec_CGFwS} that an unknot with coefficient $\pm1$ that is unlinked with other links does not change the diffeomorphism type but merely shifts the framing by $\mp 2$. It is enough to check this for the case of $S^3$. We have a solid torus embedded in $S^3$, while the complement is also a solid torus. We use a line to represent the torus wall that separates these two solid tori, the one on the right is to be subject to a $T$ twist. Also note that, across the wall, the definition of $a$ and $b$ cycle is reversed. Let $\phi$ be the change of 2-framing, we have
\bea \phi_{1\big|TS}=\phi_{T^{-1}T\big|TS}=\phi_{T^{-1}\big|S^{-1}TSTS}=\phi_{T^{-1}\big|S^{-1}TSTSTT^{-1}}=\phi_{T^{-1}\big|T^{-1}}=-2.\nn\eea
where the relations $S^{-1}=-S$, $(ST)^3=-1$ are used. On the other hand, we have $\sigma_{+1}=1$, Eq.\ref{formula_framing} gives $\phi_{TS}=-3+1=-2$, in agreement.
\end{example}

\begin{example}\emph{Framing change from $T^pS\to T^{p+1}S$}\\
Let the manifold $X_L$ be obtained from $S^3$ from surgeries along one single unknot $L^p$, labelled with the integer $p$.
The next check is what happens if we add 1 to the surgery coefficient $p$.

We know that $X_{L^p}$ bounds a 4-manifold $Z_{L^p}$, which is obtained by attaching a handle to $B^4$ along the link $L$. The handle is attached to $B^4$ as
\bea Z_{L^p}=B^4\cup^{p}_{S^1\times D^2} D^2\times D^2.\nn\eea
Concretely, the meridian $S^1\times \{*\}\subset S^1\times S^1\subset S^1\times D^2$ is sent by the gluing map to the $b+pa$ cycle of the tbn of $L$.
Adding $+1$ to the surgery coefficient merely adds one more twist to this gluing map. Using Atiyah's formula Eq.\ref{canonical_framing} for the canonical framing, we have
\bea \phi_{X_{L^p}}=-3\sigma_{L}+\int_{Z_{L^p}}\hat p_1;~~~~\phi_{X_{L^{p+1}}}=-3\sigma_{L^{p+1}}+\int_{Z_{L^{p+1}}}\hat p_1.\nn\eea
Let $\lambda_p$ and $\lambda_{p+1}$ denote the two integrals of $\hat p_1$ above. The difference $\delta\lambda=\lambda_{p+1}-\lambda_p$ is likely to be calculable using the method of the example of $T$ earlier, but at any rate, what matters to us is that the difference does not depend on $p$, because $\delta\lambda$ can be written as an integral over the two handles glued back to back with \emph{one} twist
\bea \delta\lambda=\int_{h\cup^1h}\hat p_1.\nn\eea
This shows $\delta\lambda$ can be inferred from the known cases of framing change. For surgeries $TS$ and $T^{-1}S$ we have
\bea \phi_{TS}-\phi_{S}=-3(1-0)+\delta\lambda,~~~\phi_{S}-\phi_{T^{-1}S}=-3(0-(-1))+\delta\lambda,\nn\eea
We remark that, if we perform $T^pS$ surgery along an unknot, the self-linking number of this unknot is $p$.
Adding the above two equations together, we get
\bea -6+2\delta\lambda=\phi_{+1}-\phi_{-1}=-4~\Rightarrow ~ \delta\lambda=1.\nn\eea
With $\delta\lambda$ we conclude
\bea \phi_{X_{L^{p+1}}}-\phi_{X_{L^p}}=-3(\sigma_{L^{p+1}}-\sigma_{L^p})+1\label{framing_iterative}.\eea

For example, for a single unknot with coefficient $p>1$, we can use Eq.\ref{framing_iterative} iteratively reduce $p$
\bea \phi_p=\phi_{p-1}-3\times 0+1\nn\eea
until one reaches $p=1$ and $\phi_{1}=-2$. Thus
\bea \phi_p=p-3,~~~p>1.\nn\eea
We check this against the formula Eq.\ref{formula_framing}, in which case the signature of the self-linking matrix $\sigma_p=1$ and we have
\bea \phi_p=-3\times 1+p\nn\eea
in agreement.
\end{example}

\begin{example}\emph{Surgery along a Hopf link}\\
To complete the story, we look at a surgery along a Hopf link with integers $p$ and $q$ (assuming $p>1,\,q>1$), then the self-linking matrix is
\bea \sigma_{p,q}=\begin{array}{|cc|}
                    p & 1 \\
                    1 & q
                  \end{array}.\nn\eea
From the previous reasoning
\bea \phi_{p,q}&=&\phi_{p-1,q}-3\times 0+1=\cdots =\phi_{1,q}+p-1\nn\\
&=&\phi_{1,q-1}-3\times 0+1+p-1=\cdots =\phi_{1,2}+p-1+q-2\nn\\
&=&\phi_{1,1}-3\times (2-1)+1+p+q-3\nn\\
&=&\phi_{0,1}-3(1-0)+1+p+q-5\nn\\
&=&\phi_{0,0}+1+p+q-7=p+q-6,\nn\eea
where we have canceled the surgery $L_{0,0}$. Check this against the formula Eq.\ref{formula_framing}
\bea \phi_{p,q}=-3\times 2+p+q.\nn\eea
\end{example}

\subsection{QM Computation of the Heat Kernel}
Let $D:~E\to F$ be an elliptic differential operator mapping sections of the bundle $E$ to those of $F$, and also let $D^{\dagger}$ be its adjoint. The index of $D$ is the difference $\dim \ker D-\dim \ker D^{\dagger}$. The index can be computed using the heat kernel
\bea \Tr_E\big(e^{-tD^{\dagger}D}\big)-\Tr_F\big(e^{-tDD^{\dagger}}\big)\nn\eea
at the large $t$ limit, when clearly only the zero modes matter. This expression is actually independent of $t$, since the non-zero eigen-values of $D$ and $D^{\dagger}$ cancel out, thus one can also look at the small $t$ limit. If one reformulates the trace above as a path integral, then the small $t$ limit (high temperature limit) of the path integral will only involve the 1-loop determinant, making the derivation much simpler.

We are interested in the self-adjoint operator $D=d+d^{\dagger}$ defined on a manifold $X$ of dimension $2l$, as well as its twisted version $D=d_A+d^{\dagger}_A$, with $A$ being a connection of a bundle over $X$. Introduce an involution operator
\bea \tau:~\omega\to i^{l+p(p-1)}*\omega,~~~~\omega\in \Omega^p(X).\nn\eea
Since $D$ anti-commutes with $\tau$, it sends $\Omega_+(X)$ to $\Omega_-(X)$, where $\tau\Omega_{\pm}=\pm\Omega_{\pm}$. If $D_+$ is the restriction of $D$ to $\Omega_+$, the index problem above can be written as
\bea \textrm{ind}_{D_+}=\Tr\big((-1)^{\tau}e^{-tD^2}\big).\nn\eea
To see the significance of this index, take first $A=0$ for clarity, then the zero modes of $D$ are the harmonic forms and hence are representatives of the cohomology of $X$. Furthermore, the modes with definite $\tau$ eigenvalue can be written as $\omega_p\pm \tau\omega_p$ if $p\neq l$, and one sees that the zero modes of plus or minus $\tau$-eigenvalue cancel out pairwise except for the middle dimension. Now take $l$ even, the index thus gives the number of self-dual harmonic $l$-forms minus the number of anti self-dual ones. Since the self and anti self-dual harmonic $l$-forms are orthogonal w.r.t the Hodge pairing, one sees that the index of $D_+$ is the signature of the manifold $X$.

To compute the heat kernel so as to get the characteristic polynomial that gives the index density, we follow the method used by Alvarez-Gaume \cite{AlvarezGaume:1983rm} \cite{AlvarezGaume:1983wp}, who realized that heat kernel above is nothing but the calculation of the super-symmetric index of a susy quantum mechanics problem. Namely, if the super charge $Q$ corresponds to the operator $D$ above then the Hamiltonian of the system is $\{\bar Q,Q\}=H$ and the susy index $\Tr(-1)^Fe^{-\beta H}$ corresponds to the index of $D$. The key step is to identify the correct QM system such that $Q$ and $(-1)^F$ have the desired interpretation.

In fact, super-symmetry is not really essential so long as one can guarantee the cancelation between the non-zero modes. However, when susy is present, it is much easier to start from a Lagrangian and identify the Hamiltonian (this is not a trivial problem for curved target space, due to operator ordering ambiguity). On the other hand, as one is only interested in the small $t$ limit, certain ambiguity simply do not matter, after all, we are computing an index, which is supposed to be quite independent of the minute details of the system.

Too much talking already, we start with the Hamiltonian given by the operator $H=\Delta_A/2$ (to be defined presently), whose index we want to compute, then try to rewrite $Tr[e^{-TH}(-1)^F]$ as a path integral of an action
\bea \Tr[e^{-TH(\partial_q,q,\partial_{\theta},\theta)}]=\int DpDqD\bar\theta D\theta~ \exp\big(\int_0^Tdt~ (i\dot qp+\dot\theta\bar\theta-H)\big),\label{temp_12}\eea
where on the rhs one can integrate out $p$ in favour of $\dot q$ and thereby arrive at the Lagrangian formulation. Also notice that certain $i$'s are missing because we have already done the Wick rotation. In flat space, there is a well defined passageway from the lhs to the rhs, but for curved target space, a precise prescription is missing, again due to operator ordering problems. For the index problem, this ambiguity is not serious, we shall see soon that the ambiguous terms are suppressed as $T\to0$.

To start, introduce the fermionic coordinates $\theta^{\mu}$ to play the role of $dx^{\mu}$ on $X$, fermionic fibre coordinates $z^A$ of the bundle $E$ on $X$, as well as the operator
\bea \nabla_{\mu}=\partial_{\mu}-\Gamma^{\rho}_{\mu\nu}\theta^{\nu}\frac{\partial}{\partial\theta^{\rho}}-A^A_{\mu B}z^B\frac{\partial}{\partial z^A},\nn\eea
then $d_A^{\dagger}$ is written as
\bea d_A^{\dagger}=-g^{\mu\nu}\frac{\partial}{\partial\theta^{\mu}}\nabla_{\nu}\nn\eea
and it is easy to work out the Laplacian
\bea \Delta_A=\{d_A,d_A^{\dagger}\}=-g^{\mu\nu}\nabla_{\mu}\nabla_{\nu}+g^{\mu\nu}\Gamma^{\rho}_{\mu\nu}\nabla_{\rho}
+(F_{~\mu}^{\nu})^A_{~B}\theta^{\mu}\frac{\partial}{\partial\theta^{\nu}}z^B\frac{\partial}{\partial z^A}-R^{{\rho}~{\mu}}_{~\nu~{\rho}}\theta^{\nu}\frac{\partial}{\partial\theta^{\mu}}
+\frac12R^{\mu\nu}_{~~\rho\sigma}\theta^{\rho}\theta^{\sigma}\frac{\partial}{\partial\theta^{\mu}}\frac{\partial}{\partial\theta^{\nu}}.\nn\eea
By applying the recipe Eq.\ref{temp_12}, we arrive at the Lagrangian
\bea &&L^E=-\frac12g_{\mu\nu}\dot x^{\mu}\dot x^{\nu}-\bar\theta_{\mu}\nabla_t\theta^{\mu}-\bar z_A\nabla_t z^A
-\frac14R^{\rho\kappa}_{~~\sigma\lambda}\bar\theta_{\rho}\bar\theta_{\kappa}\theta^{\sigma}\theta^{\lambda},\nn\eea
\emph{with some ambiguous terms} like $R$ and the second last term in $\Delta_A$.

There is however another problem, we do not really want to compute the trace over all states, rather only over the 1-particle state of $z$ because a multi $z$-particle state gives the anti-symmetric product of the bundle $E$. There is a nice trick used in ref.\cite{Mostafazadeh:1994ma}, one inserts into the trace
\bea \exp(-cN),~~~~N=z^A\frac{\partial}{\partial z^A}=-\frac{\partial}{\partial z^A}z^A+\dim_E.\nn\eea
Then sending $c\to \infty$ gives the 0-particle state, while doing the following
\bea -\lim_{c\to\infty} e^c\frac{d}{dc} e^{-cN}= \lim_{c\to\infty} Ne^{-c(N-1)}\label{recipe_z}\eea
picks out the 1-particle state.
\begin{remark}
One wonders what is the point of writing $z^A\partial_{z^A}$ as $-\partial_{z^A}z^A+\dim_E$? The reason is that in the derivation of the correspondence Eq.\ref{temp_12}, one has to place the momenta (derivatives) operator to the left of the coordinates in $H(\partial_q,q,\partial_{\theta},\theta,\partial_z,z)$, then replace the operators with c-numbers (see for example the appendix of the book \cite{PolchinskiString}).
\end{remark}
We have yet to realize the involution operator $\tau$. Note that the relevant Hilbert space of the above system correspond to the section $\Omega^p(X)\otimes\Gamma(E)$, we pick the ground state $|\Omega\ket$ to be the state annihilated by all $\theta$, and a general state is
\bea \bar\theta^{\mu_1}\cdots \bar\theta^{\mu_p}|\Omega\ket,~~~\bar\theta^{\mu}=\bar\theta_{\nu}g^{\nu\mu},\nn\eea
which is a $p$-form. Now write $\theta,~\bar\theta$ in a real basis
\bea \bar\theta^{\mu}=\frac{1}{\sqrt2}(\psi-i\chi)^{\mu},~~~\theta^{\mu}=\frac{1}{\sqrt2}(\psi+i\chi)^{\mu}.\nn\eea
Switching $\chi\to-\chi$ swaps $\theta$ and $\bar\theta$ and the effect on a state is exactly $\tau$
\bea \bar\theta^{\mu_1}\cdots \bar\theta^{\mu_p}|\Omega\ket\to \frac{i^{l+p(p-1)}}{(2l-p)!}\epsilon^{\mu_1\cdots \mu_p}_{~~~~~~~\mu_{p+1}\cdots \mu_{2l}}
\bar\theta^{\mu_{p+1}}\cdots \bar\theta^{\mu_{p+1}}|\Omega\ket.\nn\eea
The eigenvalue of $\tau$ is determined by the number of $\chi$'s: $(-1)^{\tau}=(-1)^{N_{\chi}}$. We shall integrate over $\chi$ with periodic boundary conditions (which inserts $(-1)^{N_{\chi}}$ in the trace) while $\psi$, $z$ and $\bar z$ are integrated with anti-periodic boundary conditions.

The stage is set, now we pick a base point $x_0$ and perform the normal coordinate expansion
\bea x^{\mu}=x_0^{\mu}+\xi^{\mu}-\frac{1}{2}\xi^{\ga}\xi^{\gb}\Gamma^{\mu}_{\ga\gb}+\cdots\nn\eea
The action to quadratic order is
{\small\bea L_2^E=-\frac12g_{\mu\nu}\dot\xi^{\mu}\dot\xi^{\nu}+\dot\theta \bar\theta+\dot z^A\bar z_A+\frac{1}{2}\dot\xi^{\mu}\xi^{\nu}R_{\mu\nu~\sigma}^{~~\;\rho}\bar\theta_{\rho}\theta^{\sigma}+\frac12(F_{~\lambda}^{\kappa})^A_{~B}\bar\theta_{\kappa}\bar z_A\theta^{\lambda}z^B
-\frac14R^{\rho\kappa}_{~~\sigma\lambda}\bar\theta_{\rho}\bar\theta_{\kappa}\theta^{\sigma}\theta^{\lambda}+\frac{c}{T}(\bar z_Az^A-\dim_E).\nn\eea}
\noindent The mode expansion is
\bea \xi=\sum_{n\neq0}=\frac1{\sqrt T}\xi_ne^{2i\pi nt/T},~~\chi=\sum_{n\in\BB{Z}}=\frac1{\sqrt T}\chi_ne^{2i\pi nt/T},~~
\phi=\sum_{n\in\BB{Z}}\frac1{\sqrt T}\phi_ne^{i\pi(2n+1)t/T},~\phi=\{\psi, \bar z, z\}.\nn\eea
The action expanded to second order in small fluctuation is
\bea S^E_2&=&-\sum_{n>0}\frac{4\pi^2 n^2}{T^2}\xi_{-n}\cdot \xi_n-\sum_{n>0}\frac{2ni\pi}{T}\chi^{\mu}_{-n}\chi_n^{\nu}g_{\mu\nu}
-\sum_{n\geq0}\frac{i\pi(2n+1)}{T}\psi^{\mu}_{-n}\psi_n^{\nu}g_{\mu\nu}-\sum_{n\in\BB{Z}}\frac{i\pi(2n+1)}{T}z^A_{-n}\bar z_{An}\nn\\
&&-\frac{2i\pi n}{2T^2}\sum_{n>0}R_{\mu\nu\rho\sigma}\xi^{\mu}_{-n}\xi^{\nu}_{n}\chi^{\rho}_0\chi_0^{\sigma}+\frac1{4T}\sum_{n\in\BB{Z}}(F_{\kappa\lambda})^A_{~B}\chi_0^{\kappa}\bar z_{An}\chi_0^{\lambda}z_{-n}^B
-\frac{1}{2T}\sum_{n\geq0}R_{\rho\kappa\sigma\lambda}\chi^{\rho}_0\chi^{\kappa}_0\psi_{-n}^{\sigma}\psi_n^{\lambda}\nn\\
&&+\frac{c}{T}\sum_{n\in\BB{Z}}\bar z_{An}z^A_{-n}-c\dim_E.\nn\eea
Note if one tries to expand the term $R^{\rho~\nu}_{~\mu~\rho}\bar\theta_{\nu}\theta^{\mu}$ similarly, one gets a term of order $T^0$, and hence can be dropped.

The computation of the 1-loop determinant is standard. The zero modes $\chi_0^{\mu}$ will play the role of $dx^{\mu}$ on $X$ and we let $\mu_i$ be the skew eigenvalue of the Riemann curvature, i.e. the anti-symmetric matrix
$1/2R_{\mu\nu\rho\sigma}\chi_0^{\rho}\chi_0^{\sigma}$ is put into a block diagonal form with blocks
\bea \begin{array}{|cc|} 0 & \mu_i \\  -\mu_i & 0  \end{array},~~~i=1,\cdots \dim_X/2,\nn\eea
then the integration over $\xi,~\chi$ and $\psi$ gives
\bea \prod_{i=1}^{\dim_X/2}\frac{\mu_i/2}{\tan \mu_i/2}.\nn\eea
the integration involving $z$ give
\bea e^{-c\dim_E}\prod_{i=1}^{\dim_E}\cosh^2\big(\frac{\lambda_i}{4}-\frac{c}{2}\big),\nn\eea
where $\lambda_i$ is the eigenvalue of $1/2\chi_0^{\mu}\chi_0^{\nu}(F_{\mu\nu})^A_{~B}$. One can then apply the procedure Eq.\ref{recipe_z} (remembering $\sum\lambda_i=0$) and get for the final answer
\bea \textrm{index}\sim\int_X \Tr [e^{F/2}] \prod_{i=1}^{\dim_X/2}\frac{\mu_i/2}{\tan \mu_i/2}.\nn\eea
It is not easy to get the correct overall normalization for the index density using path integral. Instead, one can take a simple 4-manifold $\BB{C}P^2$ (whose signature is 1) and put a trivial bundle on it. The rhs of the above equation can be computed explicitly, and by comparing, one can fix the normalization.